%% file: main.tex
\documentclass[a4paper, titlepage,12pt,twoside]{book}

\usepackage[applemac]{inputenc} 
\usepackage[lmargin=3cm,rmargin=2.5cm,tmargin=2.5cm,bmargin=2.5cm]{geometry}
\usepackage[francais,english]{babel}  

\usepackage{fancyhdr}
\usepackage{graphicx}
\usepackage{epsfig}
\usepackage{hyperref}
\usepackage{epstopdf}
\usepackage{pstricks}

\fancypagestyle{plain}{ 
\fancyhead{}
}

\usepackage{amstext,amsmath,amssymb,amsfonts,bbm}

\include{macro}

\begin{document}

\include{pdg}

\frontmatter

\include{abstract}
\include{preface}

\include{acknowledgements}

\tableofcontents

\mainmatter

\include{intro} 
\include{part1}

\include{part2}

\include{part3}
\include{conclusion}

\backmatter                   

\appendix                     
\include{appendices}


\bibliographystyle{amsplain}
\bibliography{library}

\end{document}

%% file: macro.tex
\def\beq{\begin{equation}}
\def\be{\begin{equation}}
\def\ee{\end{equation}}
\def\bes{\begin{eqnarray}}
\def\ees{\end{eqnarray}}

\DeclareMathOperator{\im}{im}
\DeclareMathOperator{\Ad}{Ad}
\DeclareMathOperator{\Hom}{Hom}
\DeclareMathOperator{\rk}{rk}

\DeclareMathOperator{\SU}{SU}

\DeclareMathOperator{\Vol}{Vol}
\DeclareMathOperator{\codim}{codim}
\DeclareMathOperator{\corank}{corank}
\DeclareMathOperator{\Col}{Col}

\def\bra{\langle}
\def\ket{\rangle}
\def\f{\frac}

\def\pp{\partial}

\def\arr{\rightarrow}
 
\def\la{\langle} \def\ra{\rangle}
\def\f{\frac}

\newcommand{\alg}{\mathfrak{g}}

\def\g{\gamma}
\def\d{\delta}

\def\r{\rho}

\def\l{\lambda}
\def\o{\omega}

\def\L{\Lambda}

\def\G{\Gamma}
\def\D{\Delta}

\newcommand{\unit}{\mathbbm{1}}

\def\bra{\langle}
\def\ket{\rangle}
\def\f{\frac}

\def\mone{^{-1}}

\def\pp{\partial}

\def\Z{\mathbb{Z}}

\def\calA{{\mathcal A}}

\def\calF{{\mathcal F}}

\def\calH{{\mathcal H}}

\def\calM{{\mathcal M}}

\def\calZ{{\mathcal Z}}

\def\cF{{\mathcal F}}
\def\cT{{\mathcal T}}
\def\cA{{\mathcal A}}

\def\arr{\rightarrow}

\def\u{\underset}
\def\k{K_{\tau}}
\def\La{\Lambda}

\newcommand{\su}{\mathrm{SU}(2)}
\newcommand{\cH}{{\cal{H}}}

%% file: pdg.tex
\thispagestyle{empty}
\sloppy\hbadness=5000\vbadness=100
{\center 

\fbox{%
\raisebox{1.3cm}{\parbox{11cm}{%
\large \sc \center Universit\'e de la M\'editerran\'ee\\ Aix-Marseille II\\ Facult\'e des Sciences}

}}\\
\vspace{1cm}
{\huge \sc Th\`ese\\} 
\vspace{1.5cm}
Pr\'esent\'ee par\\
\vspace{1cm}
{\large \sc Matteo Smerlak} \\
\vspace{1cm}
Pour obtenir le grade de \\
\vspace{.5cm}
{\large \sc Docteur de l'universit\'e de la M\'editerran\'ee}\\
\vspace{1cm}
\'Ecole doctorale : Physique et Sciences de la mati\`ere\\
Sp\'ecialit\'e : Physique des particules, Physique math\'ematique\\
\vspace{2cm}
{\LARGE  \sc Divergences des mousses de spins}\\

\vspace{0.6cm}

{\Large Comptage de puissances et resommation dans le mod\`ele plat}\\

\vspace{2cm}

Soutenue le 7 décembre 2011, devant la commission d'examen compos\'ee de\\
\vspace{.5cm}
\begin{center}
\begin{tabular}[c]{ll}
{Krzysztof \sc Gawedzki} & Pr\'esident du Jury\\
{Carlo \sc Rovelli} & Codirecteur de th\`ese\\
{Vincent \sc Rivasseau} & Codirecteur de th\`ese\\
{Alejandro \sc Perez} & Examinateur\\
{Karim \sc Noui} & Rapporteur\\
{Daniele \sc Oriti} & Rapporteur\\
\end{tabular}
\end{center}
}
\newpage
~\newpage

%% file: abstract.tex
\chapter{Abstract/Résumé}


\section*{Abstract}

In this thesis we study the \emph{flat model}, the main buidling block for the spinfoam approach to quantum gravity, with an emphasis on its \emph{divergences}. Besides a personal introduction to the problem of quantum gravity, the manuscript consists in two part. In the first one, we establish an exact powercounting formula for the \emph{bubble divergences} of the flat model, using tools from discrete gauge theory and twisted cohomology. In the second one, we address the issue of \emph{spinfoam continuum limit}, both from the lattice field theory and the group field theory perspectives. In particular, we put forward a new proof of the Borel summability of the Boulatov-Freidel-Louapre model, with an improved control over the large-spin scaling behaviour. We conclude with an outlook of the \emph{renormalization program} in spinfoam quantum gravity.  

\bigskip

\bigskip

\bigskip

\section*{Résumé}

\noindent
L'objet de cette thèse est l'étude du \emph{modèle plat}, l'ingrédient principal du programme de quantification de la gravité par les mousses de spins, avec un accent particulier sur ses \emph{divergences}. Outre une introduction personnelle au problème de la gravité quantique, le manuscrit se compose de deux parties. Dans la première, nous obtenons une formule exacte pour le comptage de puissances des \emph{divergences de bulles} dans le modèle plat, notamment grâce à des outils de théorie de jauge discrète et de cohomologie tordue. Dans la seconde partie, nous considérons le problème de la \emph{limite continue des mousses de spins}, tant du point de vue des théorie de jauge sur réseau que du point de vue de la \emph{group field theory}. Nous avan\c cons en particulier une nouvelle preuve de la sommabilité de Borel du modèle de Boulatov-Freidel-Louapre, permettant un contrôle accru du comportement d'échelle dans la limite de grands spins. Nous concluons par une discussion prospective du \emph{programme de renormalisation} pour les mousses de spins. 

\newpage

%% file: preface.tex
\chapter{Foreword}

\bigskip

\bigskip

This manuscript presents my results on the structure of divergences in spinfoam models. They were obtained in collaboration with my advisors Carlo Rovelli and Vincent Rivasseau, and with Karim Noui, Jacques Magnen and Valentin Bonzom.

\subsection*{List of publications}

Five papers were written and submitted during the completion of this thesis:

\begin{itemize}
\item
J. Magnen, K. Noui, V. Rivasseau, M. Smerlak, \emph{Scaling behaviour of three-dimensional group field theory}, Class. Quantum Grav. \textbf{26} (3), 185012 (2009) [0906.5477 (hep-th)]
\item
    V. Bonzom, M. Smerlak, \emph{Bubble divergences from cellular cohomology}, Lett. Math. Phys. \textbf{93} (3), 295-305 (2010)  [1004.5196 (gr-qc)]
\item
V. Bonzom, M. Smerlak, \emph{Bubble divergences from twisted cohomology} (2010) [1008.1476 (math-ph)], submitted to Comm. Math. Phys.
\item
V. Bonzom, M. Smerlak, \emph{Bubble divergences: sorting out topology from cell structure}, Ann. Henri Poincaré, \emph{online first}, 2 July 2011 [1103.3961 (gr-qc)]
\item
C. Rovelli, M. Smerlak, \emph{In quantum gravity, summing is refining}, (2010) [1010.5437 (gr-qc)]
\end{itemize}

In addition to these works in spinfoam quantum gravity, I studied during my PhD years various aspects of relativistic statistical physics: Tolman effect in a static gravitational field, Unruh effect and vacuum entanglement, random walks in curved spacetimes. Neither these investigations, 

\begin{itemize}
\item
C. Rovelli, M. Smerlak, \emph{Thermal time and the Tolman-Ehrenfest effect: 'temperature as the speed of time'}, Class. Quant. Grav. \textbf{28} (7) 075007 (2011) [1005.2985 (gr-qc)]
\item
M. Smerlak, \emph{Random walks in curved spacetimes} (2011)
[1108.0320 (gr-qc)], submitted to New. Journ. Phys.
\item
C. Rovelli, M. Smerlak, \emph{Unruh effect without entanglement} (2011) [1108.0320 (gr-qc)], submitted to Phys. Rev. D, 
\end{itemize}
nor the didactic discussion
\begin{itemize}
\item
M. Smerlak, \emph{A blackbody is not a blackbox}, Eur. J. Phys. \textbf{32} 1143 (2011), [1010.5696 (physics.class-ph)]
\end{itemize}
will be addressed in this manuscript. The same holds for my earlier papers

\begin{itemize}
\item
C. Rovelli, M. Smerlak, \emph{Relational EPR}, Found. Phys.  \textbf{37}, 427-445 (2007) [quant-ph/0604064]
\item
V. Bonzom, E. R. Livine, M. Smerlak, S. Speziale, \emph{Towards the graviton from spinfoams: the complete perturbative expansion of the 3d toy model}, Nucl. Phys. B  \textbf{804}, 507-526 (2008) [0802.3983 (gr-qc)]
\end{itemize}

\subsection*{On mathematical rigor}

The level of mathematical rigor of the considerations presented in this thesis is variable: while the chapters 3 to 5 and 7 are precise (in the sense of mathematical physics), the chapters 2 and 6 contain many formal or heuristic arguments. This variability is consistent with my conviction that, in physics, certain matters are more efficiently addressed at the formal level, while others do require a full-fledged mathematical analysis. 

\subsection*{Notations}

For more clarity, we will therefore reserve the sign $=$ for mathematical identites, and will use the sign $\doteq$ for formal identities. The symbol $:=$ will denote a definition, and $\sim$ an asymptotic equivalence. 

%% file: acknowledgements.tex
\chapter{Acknowledgements}

At the time of leaving Eden at CPT, I want to thank heartfully Carlo Rovelli for having been over the years an extraordinarily trustful and inspiring mentor. \emph{Per me, sei tu il primo scienzato.} 

\medskip
\noindent
I am also particularly grateful to Vincent Rivasseau, for this thesis, for AIMS, and for much more. I am delighted that his shining intelligence and outstanding generosity are becoming sources of inspiration for our African fellows as well. 

\medskip
\noindent
My warmest thanks extend to the CPT and Orsay groups, and in particular to Simone Speziale and Francesca Vidotto, for their unique friendship, and to Eugenio Bianchi, for knowing everything -- and sharing it. 
\medskip
\noindent

\medskip
\noindent
Many thanks also to Bianca Dittrich and Daniele Oriti, for making sure that the end of this thesis is also the beginning of a career. See you soon!

\medskip
\noindent
Dearest thanks to my collaborators and friends, Karim Noui, Valentin Bonzom, and also to the King of Lower and Upper Egypt, Ahmed Youssef.

\medskip
\noindent
Et bien sûr, merci à mes parents, pour la première cellule, et pour les autres.

\medskip
\noindent
Et à Maéva, toujours.

\newpage

%% file: intro.tex
\part{Introduction and preliminaries}

\vspace*{\fill}

\begin{flushright}
\emph{C'est quand c'est mou que c'est dur, mais... c'est quand c'est dur que c'est bon.}

\smallskip

Un Suisse anonyme.
\end{flushright}

\vspace*{\fill}

\chapter{Introduction}\label{}

\selectlanguage{french}

\section{Le problème de la gravité quantique}

Aux yeux de la majorité des physiciens, la \emph{gravité quantique}
est et demeure un oxymore. Comment croire qu'une théorie
quantique de l'interaction gravitationnelle soit seulement possible,
quand les méthodes de quantification perturbative sont vouées à
l'échec, en raison de la non-renormalisabilité du couplage entre
gravitons\footnote{Ceci n'excluant pas la possibilité de prédictions
quantitatives à basse énergie, dans le cadre de la théorie des champs
effective \cite{Burgess2003}.} \cite{Goroff:1985th} ? Quand les approches
non-perturbatives, théories des boucles
\cite{Rovelli:2004tv,Thiemann:2007zz} ou des cordes
\cite{Green:1987sp,Polchinski:1998rq} notamment, faillent à se montrer
prédictives ? Quand les apports expérimentaux sont inexistants, et
pour longtemps sans doute ? Quatre-vingts ans après les premiers
travaux de Rosenfeld et Bronstein \cite{Rovelli1997}, on est en
droit de se demander si le programme de quantification de la
relativité générale n'est pas obsolète. Des voix aujourd'hui s'élèvent
pour réclamer \cite{Jacobson:1995zr} -- et parfois proposer
\cite{Volovik2003} -- un changement de paradigme.

\bigskip

Quel est au fond le problème de la gravité quantique ? La physique
contemporaine, on le sait, donne à voir \emph{deux} images du monde.
L'une, soutenue par la théorie de la relativité générale \cite{Misner1973,Wald1984}, fait de l'univers un ``mollusque"
\cite{Einstein1990} dynamique au sein duquel l'espace, le temps
et la matière s'influencent mutuellement, selon les équations locales
et déterministes d'Einstein. L'autre, inspirée des règles
opérationnelles de la mécanique quantique \cite{Basdevant2002}, présente
la matière comme étant animée de singulières fluctuations, gouvernées
par des lois holistiques et aléatoires. Au plan épistémologique, le
problème de la gravité quantique est donc un problème
d'\emph{unification} : il s'agit d'élaborer une théorie capable de
concilier les caractères apparemment contradictoires de la relativité
générale et de la mécanique quantique -- quelque chose comme la
description d'un espace-temps-matière fluctuant. La nécessité de cette
unification se fait jour dès lors qu'on considère la structure globale
\cite{Hawking1973} d'un espace-temps relativiste. Si l'on s'en
tient à une description purement classique de la matière, par un
tenseur d'énergie-impulsion satisfaisant une condition d'énergie, les
théorèmes de singularité d'Hawking-Penrose
\cite{Hawking1973,Wald1984} montrent en effet qu'un tel
espace-temps contient nécessairement une \emph{singularité} : une
région où les géodésiques s'arrêtent, purement et simplement. Qu'elle
passe par des violations des conditions d'énergie \cite{Mattingly2001}
ou plus radicalement par le dépassement de la relativité générale, la
solution de cette ``catastrophe spatio-temporelle" -- comme la
divergence du potentiel de Coulomb en électrostatique ou de l'énergie
thermique de la radiation -- fera certainement intervenir la mécanique
quantique.

Le problème de la gravité quantique est aussi un problème de
\emph{relativisation}. On sait comment les relativités galiléenne,
restreinte et générale ont évacué de l'appareil conceptuel de la
physique ces ``idoles" \cite{Finkelstein2001} que sont,
respectivement, l'espace absolu, le temps absolu, et l'espace-temps
absolu. De la même façon, la mécanique quantique a déposé le concept
d'état absolu, le remplaçant par celui, plus participatif, d'opération
(ou action \cite{Finkelstein1999}, préparation/mesure \cite{VonNeumann1955}, interaction \cite{Rovelli:1996cr}, relation
\cite{Bitbol2010}...). L'enjeu d'une théorie de la gravité
quantique est de croiser et de prolonger ces relativisations. Pour
reprendre une image de Rovelli, avec la relativité générale (pour
laquelle l'occurence d'un événement est absolue), et la mécanique
quantique (dans laquelle l'espace-temps est donné \emph{a priori}),
nous sommes seulement ``half-way through the woods"
\cite{Rovelli1997}. Les résultats préliminaires de Hawking
\cite{HAWKING1974} et de Unruh \cite{Unruh1976}, dans le cadre
de la théorie quantique des champs sur espace-temps courbes \cite{Wald1994} (où l'on néglige la rétro-action des champs quantiques
sur l'espace-temps), suggèrent que cette nouvelle relativisation
affectera notamment les notions de particules, de température et
d'entropie.

Enfin, et surtout, le problème de la gravité quantique est un problème
\emph{physique}. Loin de se confiner à des spéculations extrapolant,
malmenant parfois, notre connaissance de l'univers,\footnote{Exemples
: la supersymétrie, le \emph{big bounce}, le scénario ekpyrotique, le
principe holographique, les dimensions supplémentaires, la discrétude
de l'espace-temps...} il perce dans de nombreuses questions issues de
l'expérience, à la signification opérationnelle claire, et dont nous
ignorons la réponse :

\medskip\begin{itemize}
\item
La pression de Casimir gravite-t-elle \cite{Fulling2007} ?
\item
Existe-t-il un \emph{Lamb shift} gravitationnel \cite{Wang2010} ?
\item
La propagation de la lumière dans le vide est-elle dispersive
\cite{Amelino-Camelia1998} ?
\item
La symétrie de Lorentz est-elle brisée \cite{Jacobson2005} ?
Déformée \cite{Amelino-Camelia2002} ?
\item
...
\end{itemize}

En un mot, \emph{nous ignorons la nature de la relation entre matière
et espace-temps}. A mon sens, c'est ce problème fondamental -- et pas
un autre\footnote{Parmi les problèmes qui \emph{ne sont pas}
directement relié au problème de la gravité quantique, citons :
l'interprétation de la mécanique quantique, le problème de la
constante cosmologique, les divergences ultraviolettes en théorie
quantique des champs, les conditions initiales en cosmologie, le
problème du temps, de la flèche du temps, etc.} -- qui devra être
résolu par une théorie de la gravité quantique. Qu'elle se présente
comme une quantification de la relativité générale ou pas, celle-ci
n'aura rien d'un oxymore.
%

\section{Quantification ou émergence ?}

Pour aborder le problème de la gravité quantique, les approches
théoriques s'organisent en deux categories, que je qualifierais de
\emph{réductionnistes} et \emph{émergentistes}. Les premières
proposent de réduire le problème de la gravité quantique à celui de la
description de la dynamique de l'espace-temps lui-même, dans un
premier temps sans tenir compte de la présence de la matière (la
gravité pure). Elles se fondent en cela sur l'antique dichotomie entre
matière et géométrie, préservée dans la cinématique relativiste :
l'action d'Einstein-Hilbert est fonction de deux types de champs
distincts -- la métrique lorentzienne d'espace-temps d'une part, et
les champs de matière d'autre part. La gravité quantique à boucles,
dont il sera question dans cette thèse, fait partie de cette catégorie
: elle s'attache à quantifier l'espace-temps, comme une champ
dynamique à part entière, avec ses degrés de liberté propres (dont les
ondes gravitationnelles sont la manifestation perturbative). Elle marche en cela dans les pas de Dirac,
Feynman, Wheeler, DeWitt etc. \cite{Rovelli2000} : pour la théorie
des boucles, l'espace-temps est un champ classique comme les autres,
qui doit être quantifié indépendamment de la matière.

\bigskip

Les approches émergentistes
\cite{Volovik2003,Hu2009,T.Padmanabhan:2010zr}, initiées par
Sakharov \cite{Sakharov1991}, prennent le point de vue inverse :
l'espace-temps n'existe qu'en présence d'un substrat microscopique,
matériel ou non, et apparaît comme un type original de propriété
collective. Les troublantes relations entre gravité et thermodynamique
découvertes dans le contexte de la mécanique des trous noirs, et
approfondies par Jacobson \cite{Jacobson:1995zr} et Padmanabhan
\cite{T.Padmanabhan:2010zr} grâce à l'interprétation des équations
d'Einstein comme des équations d'état, donne un crédit important à ce
point de vue. (On trouve également une forme d'émergentisme dans la
théorie des cordes, pour laquelle les équations d'Einstein semblent
apparaître comme condition de cohérence pour la dynamique de cordes
supersymétriques.) Face aux difficultés rencontrées par les tentatives
réductionnistes de quantification de l'espace-temps, cet émergentisme
apparaît comme un contrepoint salutaire. Cela étant, les résultats de
Jacobson, Padmanabhan et d'autres dans cette direction restent trop
fragmentaires pour constituer un véritable ``changement de paradigme",
et en particulier ne prouvent pas que le ``champ gravitationnel ne
doit pas être quantifié" \cite{Volovik2008}. Gageons simplement que
les relations entre thermodynamique et gravité seront naturellement
intégrées dans une théorie future de la gravité
quantique.\footnote{Certains chercheurs, à commencer par Einstein \cite{Einstein1938}, ont imaginé une forme d'émergence inverse
: la matière serait une structure interne à l'espace-temps. Un
scénario de ce type est étudié aujourd'hui en théorie des boucles
\cite{Bilson-Thompson2006}, ce qui montre que, dans ce contexte en
tous cas, réduction et émergence ne s'excluent pas nécessairement.}

\bigskip

Dans cette thèse, je m'inscrirai donc dans le cadre de la théorie des
boucles, conscient de l'\emph{hypothèse} réductionniste qui la
sous-tend. Rappelons aux sceptiques que la procédure de quantification
des champs classiques a été donnée pour morte à deux reprises (dans
les années 40, en raison des divergences ultraviolettes, puis dans les
années 60, à cause du fantôme de Landau) -- et qu'elle est toujours
bien vivante. La non-renormalisabilité du développement perturbatif,
dans le cas de la gravité, n'est peut-être pas davantage rédhibitoire
: elle indiquerait simplement qu'il faut non seulement quantifier,
mais aussi \emph{quantifier intelligemment}.

\section{Quantification et troncation}

Si l'on considère le champ gravitationnel comme un champ classique
fondamental, comme les champs de Maxwell ou de Dirac, le problème de
la gravité quantique consiste à élaborer une procédure de
quantification de l'espace-temps. On insiste souvent sur le fait
qu'une telle quantification est radicalement originale, en ceci
qu'elle doit s'opérer sans le cadre cinématique de la relativité
restreinte (référentiels inertiels et transformations de Lorentz) ; on
dit qu'elle doit être ``background independent". Les difficultés
propres à cette situation ont été soulignées par de nombreux auteurs,
parmi lesquelles Rovelli fait figure de référence. Je renvoie à son
livre \cite{Rovelli:2004tv} pour une excellente discussion de cet
aspect.

Il est pourtant un aspect de la quantification de la métrique
d'espace-temps qui est similaire à celle des champs de matière usuels,
au sein du modèle standard de la physique des particules. Il s'agit de
la nécessité de procéder à une \emph{troncation} de l'ensemble des
degrés de liberté dynamiques. Comme l'électrodynamique ou la
chromodynamique, la géométrodynamique (la relativité générale) est en
effet une théorie non-linéaire. Ceci signifie qu'il est impossible de
scinder les équations du mouvement en composantes indépendantes, comme
on peut le faire par l'analyse de Fourier dans le cas du champ libre
de Klein-Gordon. Tous les degrés de liberté du champ sont couplés
entre eux, sans hiérarchie apparente.

Nous ignorons comment quantifier une telle infinité de degrés de
liberté en interaction. Pour progresser vers une théorie quantique
prédictive, il faut donc procéder à une troncation de l'espace des
phases, par laquelle on isole certains degrés de liberté, identifiés
comme plus pertinents pour le régime expérimental d'intérêt. On
établit par cette opération un schéma d'approximation, qui est
d'autant plus fin que le nombre de degrés de libertés conservés est
important. Plusieurs type de troncation sont utilisées couramment en
physique des hautes énergies :

\begin{enumerate}
\item
\emph{Le développement perturbatif \cite{Peskin1995}.} Dans les théories
faiblement couplées, comme l'électrodynamique à basse énergie ou la
chromodynamique à haute énergie, ou peut utiliser le nombre de
particules pour tronquer l'espace des phases. Rappelons qu'une
particule (virtuelle) est un mode de Fourier du champ, représentant un
état dans l'espace de Fock du champ libre correspondant. Le
développement perturbatif consiste à organiser les amplitudes
quantiques en série de termes au nombre de particules croissants. On
tronque habituellement cette série à l'ordre 1 pour une première
estimation des corrections quantiques, et à l'ordre $\sim 10$ pour les
calculs les plus précis.
\item
\emph{La discrétisation sur réseau \cite{Rothe2005}.} Pour
étudier les phénomènes résultant d'interactions fortes entre
particules élémentaires, comme le confinement des quarks et des gluons
au sein des hadrons, le développement perturbatif est inopérant. Avec
l'avènement des calculateurs numériques puissants, la troncation la
plus utile dans ce contexte est la discrétisation sur réseau : on
réduit l'espace-temps à un nombre fini de points sur un graphe
régulier, séparés par une distance $a$ infime par rapport à la
dimension caractéristique du phénomène que l'on souhaite étudier (la
masse du hadron, par exemple). Dans cette approche, toutes les
fluctuations d'échelle inférieure à $a$ sont supprimées.
\item
\emph{Factorisation des symétries de jauge.} Il existe une classe de
théories de champs, telles les célèbres théories de champs
topologiques \cite{Atiyah1989}, dans lesquelles le nombre infini de
degrés de liberté est imputable à une symétrie de jauge. La réduction
de cette symétrie de jauge réduit le système à un nombre fini de
degrés de libertés, qu'on peut quantifier par les méthodes usuelles de
mécanique quantique. A proprement parler, cette opération n'est pas
une troncation de l'espace des phases, puisque les degrés de libertés
éliminés sont redondants. Aucune information physique n'est donc
perdue.
\end{enumerate}

Dans le développement perturbatif et la mise sur réseau, la troncation
d'un nombre infini de degrés de liberté ne peut se faire sans tenir
compte de leur résultante effective sur les degrés de libertés
préservés. Cette opération, qualifiée traditionnellement de
\emph{renormalisation} \cite{Rivasseau:2003zy} dans le premier cas et
de \emph{groupe de renormalisation} \cite{Wilson1983} dans le
second, est à l'origine de l'essentiel des difficultés techniques et
conceptuelles liées à la troncation de l'espace des phases. Son
résultat, la définition de constante de couplages ``habillées"  ou
``d'actions parfaites", dépendant de l'échelle, doit s'interpréter
comme la condition de cohérence d'un schéma de troncation associant
échelles spatiales et force des interactions.

Comme tout schéma d'approximation, la troncation possède une limite
exacte, dans laquelle le couplage entre tous les degrés de libertés
est rétabli. Dans le cas du développement pertubatif, la définition et
l'étude de cette limite est le sujet de la \emph{théorie constructive}
\cite{Riv} ; dans les approches sur réseau, on parle de \emph{limite
continue}. 

Un point important à noter concernant cette limite exacte est la forme
qu'elle prend dans les cas 1 et 2, respectivement. La limite
constructive du développement pertubatif est essentiellement une somme
: pour caricaturer un processus en vérité très subtil (et sur lequel
nous reviendrons plus tard), on peut dire qu'il s'agit d'\emph{ajouter} la
valeur de tous les diagrammes de Feynman, à tous les ordres.
La limite continue des théorie sur réseau, à l'inverse, s'obtient en
\emph{raffinant} le graphe. Les deux schémas de troncation apparaissent donc
techniquement très différents, même s'ils correspondent tous deux à
l'élargissement de la section d'espace des phases préservée par la
troncation.

\bigskip

En gravité quantique comme dans le modèle standard, il est nécessaire
d'opérer une troncation des degrés de liberté contenu dans la métrique
pour réaliser la quantification des équations d'Einstein. On peut
considérer que la singulière difficulté de la gravité quantique réside
dans le fait qu'aucun des schémas cités précédemment ne s'applique au
cas de la relativité générale.

\begin{enumerate}
\item
\emph{Non-renormalisabilité du développement perturbatif.} La
quantification perturbative de la relativité générale, fondée sur le
concept de graviton, consiste à supposer que la métrique
d'espace-temps fluctue peu autour de la métrique de Minkowski. Ceci
conduit à un développement des amplitudes en puissances de la
constante de Newton $G$ pour lequel la procédure de renormalisation
n'est pas close. Techniquement, on montre que l'interaction des
gravitons entre eux est non-renormalisable. En présence d'un cutoff en
énergie $\Lambda$, on peut toutefois construire une théorie de champs
effective \cite{Burgess2003}, consistant en un double développement
en $G$ et $\Lambda$. Celle-ci permet de calculer notamment les
premières corrections quantiques au potentiel gravitationnel de
Newton, mais ne possède pas de limite ultraviolette exacte.
\item
\emph{Absence d'une métrique de fond.} La mise sur réseau de la
relativité générale est rendue impossible par le fait qu'aucune
métrique de fond ne permet de définir le pas du réseau $a$. Cette voie
de troncation est donc exclue pour la gravité quantique.
\item
\emph{Présence d'une infinité de degrés de liberté physiques.} Même
s'il est vrai que quatre des six degrés de liberté par point de la
métrique (en quatre dimensions) sont redondants, du fait de la
symétrie par difféomorphisme de la relativité générale, deux degrés de
liberté par point sont physiques et doivent être quantifiés. C'est
dire que la relativité générale n'est pas à l'image de la théorie de
Chern-Simons \cite{Witten1988} ou de la théorie BF
\cite{Horowitz1989} : son espace des phases réduit est de dimension
infinie.
\end{enumerate}

Une contribution essentielle à l'élaboration d'un schéma de troncation
de la relativité générale qui dépasse ces difficultés est due à Regge
\cite{Regge1961}. S'appuyant sur une triangulation $\Delta$ de la
variété d'espace-temps, il a en effet découvert une approximation
simple de l'action d'Einstein-Hilbert, s'exprimant en fonction de la
longueur $l_e$ des arêtes des simplex : pour un espace-temps de
dimension $d$, notant $t\in\Delta_{d-2}$ un $(d-2)$-simplex (un ``os"
ou ``charnière") de $\Delta$, $A_t(l_e)$ son volume euclidien et
$\epsilon_t(l_e)$ l'angle de déficit autour de $t$, l'action de Regge
s'écrit\footnote{On n'inclut pas ici de constante cosmologique.}
\be\label{reggeaction}
S_R(l_e)=\sum_{t\in\Delta_{d-2}}A_t(l_e)\epsilon_t(l_e).
\ee
et approche l'action de Einstein-Hilbert d'autant mieux que le nombre
de $d$-simplexes est grand. Cette observation est aujourd'hui au
fondement de plusieurs approches à la gravité quantique, parmi
lesquelles on peut citer la théorie des triangulations dynamiques
causales \cite{Ambjorn2010} et le calcul de Regge quantique
\cite{Williams2009}.

Un aspect remarquable de l'approche de Regge est sa capacité à décrire
une géométrie quadrimensionnelle sans devoir recourir à un choix de
jauge, c'est-à-dire de coordonnées sur l'espace-temps
\cite{Rovelli2011b}. En localisant la courbure sur les ``os" d'une
triangulation, elle offre en effet la possibilité d'exprimer la
dynamique en termes de longueurs \emph{physiques} -- la dimension de
ces ``os''. Ainsi, le schéma de troncation de Regge est également une
solution (alternative à celle de Komar et Bergmann
\cite{Bergmann1960}) au problème fondamental de la gravité pure :
la difficulté d'identifier des observables invariantes par
difféomorphismes en l'absence de champs de matière.

\section{Réseaux et mousses de spins}

La théorie des boucles
\cite{Rovelli:2004tv,Ashtekar:2004eh,Thiemann:2007zz} développée à
l'origine par Rovelli et Smolin \cite{Rovelli1988,Rovelli1990} sur la base du
travail d'Ashtekar \cite{Ashtekar1986}, promeut un schéma de
troncation de la relativité générale d'un genre différent, mêlant
discrétisation sur réseau et factorisation des symétries de jauge
(exemples 2 et 3 ci-dessus). Elle se présente sous deux formes,
canonique et covariante, correspondant pour le cas gravitationnel aux
quantifications de Schrödinger et de Feynman utilisées en mécanique
quantique.

Chronologiquement, la première version de la théorie des boucles est
la version canonique. S'inscrivant dans le sillage de Arnowitt, Deser
et Misner \cite{Arnowitt1960}, celle-ci s'appuie sur une
décomposition de l'espace-temps en feuilles d'espaces indexées par une
coordonnées de temps, et reformule les équations d'Einstein comme des
équations de contraintes. Sur chacune de ces feuilles d'espaces, elle
utilise comme variables canoniques -- c'est là son originalité --  la
connexion $\textrm{SU(2)}$ d'Ashtekar $A$ et son champ électrique
conjugué $E$, et fait l'hypothèse que l'ensemble des \emph{holonomies}
$H$ de $A$ -- les boucles -- et des \emph{flux} bidimensionnels $F$ de
$E$ forme un jeu de variables adaptées pour une troncation des degrés
de liberté du champ gravitationnel compatible avec sa symétrie sous
les difféomorphismes.

Cette hypothèse est motivée par la théorie des noeuds
\cite{Rovelli1988,Gambini1996,Baez1994}, qui sont précisément des structures invariantes
par difféomorphismes dans un espace tridimensionnel \cite{Baez1994}.
En ne considérant que les holonomies supportées sur une classe
d'équivalence $\gamma$ de graphes dirigés et plongés dans une feuille
d'espace donnée, on obtient ainsi un troncation finie de la relativité
générale, qu'il est possible de quantifier à l'instar du champ de
Yang-Mills \cite{Gambini1996}. Ceci conduit à la représentation
des holonomies de $A$ et des flux de $E$ par des opérateurs sur
l'espace de Hilbert des fonctions de carré sommable pour la mesure de
Haar
\be
\mathcal{H}_\gamma=L^2\big(\textrm{SU(2)}^L\big)/\textrm{SU(2)}^N
\ee
où $N$ et $L$ désignent respectivement le nombre de noeuds et de liens
de $\gamma$, et où le quotient correspond à la symétrie de jauge sur
réseau traditionnelle (action aux noeuds du graphe).  Les états
quantiques générant cet espace par la décomposition de Peter-Weyl sont
appelés \emph{réseaux de spins} \cite{Rovelli1995}. On peut
considérer qu'ils décrivent un ``espace quantique tronqué", dont les
aires et volumes, essentiellement données par les opérateurs de
Penrose \cite{Penrose1971,Moussouris1983}, sont quantifiées
\cite{Rovelli:1994ge}. Pour résumer, la théorie des boucles fait du
champ gravitationnel un \emph{champ de jauge sur un réseau dynamique}.
Son point commun avec le calcul de Regge est qu'elle s'appuie sur des
structures discrètes pour définir des \emph{observables relationnelles
internes au champs gravitationnel}.

Dans cette approche, on obtient l'espace de Hilbert (cinématique,
c'est-à-dire permettant de représenter la contrainte hamiltonienne)
total par une limite inductive sur $\gamma$, qui est aussi une somme
directe de sous-espaces
$\mathcal{H}^*_\gamma\subset\mathcal{H}_\gamma$ sur lesquels tous les
spins sont-nuls:\footnote{Du point du vue de l'analyse fonctionnelle,
cet espace $\mathcal{H}$ accompagné de sa représentation de l'algèbre
des holonomies et flux de la connection d'Ashtekar est isomorphe à un
espace de fonctionnelles de connections distributionnelles, de carré
intégrable pour une mesure dite d'Ashtekar-Lewandowski
\cite{Ashtekar1995,Ashtekar1995a}. (Cette dernière présentation
est analogue à la formulation de Schrödinger de la mécanique
quantique.)
}
\be\label{inductivelimithilbert}
\mathcal{H}=\underset{\longrightarrow}{\lim}\
H_\gamma=\bigoplus_\gamma\mathcal{H}_\gamma^*.
\ee
Précisons ce point. Pour obtenir l'espace de Hilbert cinématique
total, il faut considérer des graphes ``arbitrairement denses". Pour
ce faire, on peut observer que l'ensemble de tous les graphes, équipé
de la relation d'ordre $\preceq$ correspondant à l'inclusion d'un graphe
dans un autre, forme un \emph{ensemble ordonné filtrant}
(\emph{directed set} en anglais) : pour toute paire de graphes
$\gamma_1$ et $\gamma_2$, il existe un graphe $\gamma_3$ contenant
$\gamma_1$ et $\gamma_2$ comme sous-graphes. Grâce à cette structure,
ainsi que les injections
$\mathcal{H}_\gamma\hookrightarrow\mathcal{H}_{\gamma'}$ quand
$\gamma\preceq\gamma'$ obtenues en coloriant les arêtes de $\gamma'$ non
contenues dans $\gamma$ par un spin nul, on peut construire la limite
inductive \eqref{inductivelimithilbert}. Nous reviendrons dans le
paragraphe suivant sur le fait que, dans le schéma de troncation en
boucles, la limite continue est aussi une somme. 

\bigskip

La limite de la formulation canonique de la théorie des boucles est la
difficulté de résoudre la contrainte hamiltonienne scalaire, qui
génère les reparamatrisations de la coordonnée temporelle dans
l'espace-temps \cite{Thiemann:2007zz}. Pour contourner ce problème,
Reisenberger et Rovelli  \cite{Reisenberger1996} ont introduit une
intégrale de chemins pour les réseaux de spins. Dans ce schéma, une
``trajectoire" d'un réseau de spins est représenteée par un
$2$-complexe colorées par des spins, appelé \emph{mousse de spins}. On
définira précisément ces $2$-complexes, ou \emph{mousses}, dans le
chapitre 2. Pour notre propos présent, il suffit de se représenter une
mousse $\Gamma$ comme un ensemble de vertex, arêtes et faces,
attachées le long de leurs bords, et possédant un graphe $\gamma$
comme frontière ($\gamma$ pouvant être le graphe vide). De même que le
graphe $\gamma$ représente une troncation de la géométrie spatiale
dans l'approche canonique, la mousse $\Gamma$ représente une
troncation de la géometrie spatio-temporelle. Suivant la logique de
l'intégrale de chemin, on peut définir une dynamique pour
l'espace-temps quantique en associant à chacune de ces mousses une
\emph{amplitude} complexe $\mathcal{Z}(\Gamma)$. En principe, cette
amplitude devrait pouvoir être calculée à partir des éléments de
matrice de la contrainte hamiltonienne. Ceci a pu être réalisé en
trois dimensions \cite{Noui2004}, mais semble hors de portée en
quatre dimensions. La stratégié utilisée en théorie des boucles est
donc de faire un ansatz pour cette amplitude : on parle de
\emph{modèles} de mousses de spins.

\bigskip

Ces dernières années ont vu l'introduction d'une nouvelle classe de
tels modèles en quatre dimensions, par Engle, Pereira, Rovelli
\cite{Engle2007,Engle:2007wy} mais aussi par Freidel et Krasnov
\cite{Freidel:2007py} sur la base de travaux de Livine et Speziale
\cite{Livine:2007ya}, qui semblent prometteurs. Ainsi, Barrett et ses
collaborateurs \cite{Barrett:2009gg} ont montré que, pour une mousse
$\Gamma$ contenant un seul vertex et dont la frontière $\gamma$ est le
graphe complet sur $5$ noeuds, correspondant aux $5$ tétraèdres qui
bordent un $4$-simplex, la limite de grand spin de
$\mathcal{Z}(\Gamma)$ est proportionnelle à l'exponentielle de
l'action de Regge pour un $4$-simplex. Ceci montre que le schéma de
troncation en mousses de spins est lié au calcul de Regge, dont on
sait qu'il est lui-même une troncation de la relativité générale. Un
autre résultat remarquable concerne l'identification d'un régime
perturbatif \cite{Modesto2005} : en utilisant un état cohérent sur
$\gamma$ piquant la géométrie quantique sur l'espace-temps plat de
Minkowski, Rovelli \emph{et al.} ont pu montrer que les fluctuations
quantiques sur $\Gamma$ ont la même structure que le graviton en
relativité générale linéarisée. Leur fonction de corrélation à $2$
\cite{Bianchi2009} et $3$ points \cite{Rovelli2011a}, en
particulier, ont la même structure tensorielle que celles du graviton.

\bigskip

Dans cette thèse, j'étudierai un modèle plus simple, le \emph{modèle
plat}, qui forme le squelette de ces ``nouveaux modèles" et en partage
certaines propriétés, comme la limite Regge
\cite{PR,Dowdall2009,Barrett:2009gg}. Les motivations pour étudier
dans un premier temps ce modèle-ci plutôt que les nouveaux modèles
sont nombreuses :

\medskip\begin{itemize}
\item
l'essentiel des résultats supportant l'approche par mousses de spins
portent sur le modèle plat
\item
tous les modèles de mousses de spins étudiés dans la littérature sont
des modifications du modèle plat
\item
techniquement, il est considérablement plus simple que les nouveaux modèles
\item
il est lié à la topologie quantique (et en est en quelque sorte à l'origine)
\item
son interprétation géométrique est claire : il décrit un système de
connections de jauge discrètes \emph{plates}
\end{itemize}

\bigskip

J'étudierai en profondeur un aspect du modèle plat qui a été peu
considéré jusqu'à présent : celui de ses \emph{divergences}. Le
premier problème spécifique sur lequel je concentrerai mes efforts est
ainsi celui des \emph{divergences de bulles}. Certains résultats
préliminaires dûs à Perez et Rovelli \cite{Perez:2000fs} et Freidel
\emph{et al.} \cite{Freidel2002,Freidel2004a,Freidel2004}
suggèrent en effet que pour une mousse $\Gamma$ contenant des
``bulles" (à définir précisément), la quantité $\mathcal{Z}(\Gamma)$
est divergente. Cette situation, qui rappelle les divergences
ultraviolettes des amplitudes de Feynman en théorie quantique des
champs, réclame une analyse quantitative : quel est le \emph{degré de
divergence} d'une mousse $\Gamma$ ? De quoi dépend-il ? De la
topologie de $\Gamma$ ? D'autre chose ? Ces questions apparaissent
essentielle pour une compréhension à la fois rigoureuse et générale du
modèle plat et, partant, des modèles de mousses de spins en général.
La seconde partie de ce manuscrit (chapitres \ref{Laplace} à \ref{applications}) fournit une solution détaillée à ce
problème.

\section{Limite continue des mousses de spins}

L'autre problème qui se pose quand on étudie le modèle plat est bien
sûr celui de la limite continue : de l'ensemble des amplitudes
tronquées $\mathcal{Z}(\Gamma)$, comment obtient-on une amplitude
complète $\mathcal{Z}$ contenant tous les degrés de liberté du champ
gravitationnel quantique ? On a vu précédemment qu'au niveau
cinématique, la limite continue, consistant à raffiner les graphes
jusqu'à l'infini, prend aussi la forme d'une somme directe d'espace de
Hilbert. Qu'en est-il au niveau dynamique ? Faut-il définir
$\mathcal{Z}$ en sommant sur les mousses, formellement
$\mathcal{Z}=\sum_\Gamma\mathcal{Z}(\Gamma)$, comme dans un
développement perturbatif à la Feynman (cas 1) ? Ou raffiner $\Gamma$
infiniment, $\mathcal{Z}=\lim_\Gamma\mathcal{Z}(\Gamma)$, comme en
théorie de jauge sur réseau (cas 2) ? Cette question a été la source
de nombreux débats entre spécialistes des mousses de spins
\cite{Freidel:2005qe}.

\bigskip

L'implémentation la plus populaire de la \emph{somme} sur les mousses
est due à Boulatov \cite{Boulatov:1992vp} et Ooguri
\cite{Ooguri:1992eb}, et est connue sous le nom \emph{théorie de
champs sur le groupe} (\emph{group field theory})
\cite{Freidel:2005qe,Oriti:2006se}. Dans ce schéma, les amplitudes
$\mathcal{Z}(\Gamma)$ apparaissent comme d'authentiques amplitudes de
Feynman \cite{Reisenberger:2000fy} pour une théorie de champs
auxiliaires, définies sur une certaine puissance cartésienne d'un
groupe de Lie compact, $\textrm{SU}(2)$ le plus souvent. C'est dire
que, pour la théorie de champs sur le groupe, la divergence de
$\mathcal{Z}(\Gamma)$ est un problème de renormalisation, et la somme
sur $\Gamma$ un problème constructif. Le calcul du degré de divergence
d'une mousse $\Gamma$ prend ainsi tout son sens, puisqu'il permet
d'étudier la renormalisabilité de la théorie de champs sur le groupe
de Boulatov-Ooguri. Rivasseau a également souligné que ce point de vue
présente l'avantage remarquable de fixer le coefficient de chaque
mousse $\Gamma$ dans la somme \cite{Rivasseau2011} : c'est le
coefficient combinatoire du développement de Feynamn de la théorie de
champs sur le groupe. Enfin, Oriti soutient que cette perspective
permet de rapprocher la théorie des boucles de la matière condensée
\cite{Oriti2010}, et ainsi de réduire la fracture entre
réductionnisme et émergentisme en gravité quantique.

Dans le chapitre \ref{resummation} de cette thèse, je prendrai ce point de vue pour
étudier la limite continue des mousses de spins : je montrerai qu'il
est possible, dans le cadre de la théorie de champs sur le groupe de
Boulatov (en trois dimensions), de réaliser la somme sur les mousses
dans un sens rigoureux, celui de la resommation de Borel. J'utiliserai pour cela l'outil le plus récent
de la théorie constructive, développés par Rivasseau et Magnen : le
développement en cactus \cite{Rivasseau:2007fr,Magnen2008}. Ce
résultat donne un crédit supplémentaire à la viabilité de la théorie
de champs sur le groupe.

\bigskip

Mais on peut également définir une forme de \emph{limite inductive}
pour les mousses de spins \cite{Zapata2002,Rovelli2010}. Celle-ci forme une
alternative à la théorie de champs sur le groupe, dont l'avantage est
de préserver la relation entre somme et limite suggérée par l'idée,
fondamentale en théorie des boucles, qu'un espace-temps est un
processus virtuel, et non une structure \emph{a priori} : une mousse
est donc à la fois un diagramme de Feynman (cas 1) \emph{et} un réseau
(cas 2). Cet argument sera présenté en détail au
chapitre \ref{spincont}.

%

%% file: part1.tex
\selectlanguage{english}

\chapter{The flat spinfoam model}\label{}


In this preparatory chapter, we introduce the Ponzano-Regge model, give a first account of its \emph{bubble divergences}, discuss its relationship with three-dimensional quantum gravity, and introduce a cellular generalization coined the \emph{flat spinfoam model}.  

\section{The Ponzano-Regge model}

The \emph{Ponzano-Regge model} was put forward in 1968 as a simplicial model of quantum gravity in three dimensions. Its influence over the later developments of theoretical and mathematical physics cannot be overestimated: the spinfoam and group field theory approaches to quantum gravity, as well as the mathematical theory of state-sum topological invariants,\footnote{Part of my work with Bonzom has been concerned with the topological features of the (generalized) Ponzano-Regge model, and in particular with its relationship to twisted Reidemeister torsion. This aspect of our work will not be addressed in this thesis.} are direct offsprings of the Ponzano-Regge model. 


\subsection{The Ponzano-Regge asymptotic formula}

At the roots of these developments is a most surprising result: the Wigner $\{6j\}$ symbol (a key object in $\textrm{SU}(2)$ recoupling theory \cite{Varshalovich1988}, with applications in nuclear, atomic and molecular physics) contains, in the large spin asymptotics, the three-dimensional Regge action for a tetrahedron with edge lengths $l_e=j_e+1/2$, where $j_e$ are half-integers. Explicitly, if $V(l_e)$ is the volume of a Euclidean tetrahedron with edge lengths $l_e$, and $S_R(l_e)$ the corresponding Regge action \eqref{reggeaction}, one has in the uniform $j_e\rightarrow\infty$ limit\footnote{Meaning that the spins are scaled as $j_e\mapsto\lambda j_e$ with a single $\lambda\rightarrow\infty$. In this regime, the ``shape" of the tetrahedron, defined by the ratios $j_e/j_{e'}$ with $e\neq e'$, is preserved in the limit.}


\be \label{6jasympPR}
\begin{Bmatrix} j_1 &j_2 &j_3 \\ j_4 &j_5 &j_6 \end{Bmatrix} \sim
\f1{\sqrt{12\pi V(l_e)}}\cos\left( S_{\rm R}(l_e) + \f\pi4\right).
\ee
Moreover, when the spins $j_i$ do not define a Euclidean tetrahedron, the $\{6j\}$ symbol is exponentially suppressed in this limit. 

Ponzano and Regge conjectured this asymptotic behavior on the basis of numerical simulations \cite{PR}. An influential proof thereof, using geometric quantization, was given by Roberts in \cite{Roberts1998}; see \cite{Aquilanti2010,Bonzom2011} for the most recent advances on the semiclassics of Wigner symbols. (Higher orders in the large-spin asymptotic expansion were also investigated using saddle point techniques in my paper \cite{Bonzom:2008xd} and the subsequent \cite{Dupuis2009}.)

The remarkable character of the asymptotic formula \eqref{6jasympPR} lies in the fact that, not only it contains the Regge action $S_R$ (hence it is related to three-dimensional gravity), but contains it \emph{within a cosine}. This is reminiscent of the semiclassical (WKB) approximation of quantum amplitudes (with the only difference that the eikonal exponential $e^{iS}$, where $S$ is the Hamilton function, is replaced by the cosine.\footnote{This difference can be traced back to the fact that $6$ edge lengths define a tetrahedron up to rotation \emph{and reflection} \cite{Baez:1999sr}.}) In other words, the $\{6j\}$ symbol appears to be related to semiclassical \emph{quantum} gravity.

This striking connection between $\textrm{SU}(2)$ representation theory and three-dimensional quantum gravity prompted Ponzano and Regge to put forward the first spinfoam model ever written.

\subsection{A simplicial partition function}

Since a $\{6j\}$ symbol appears to represent a \emph{quantum tetrahedron},\footnote{The notion of \emph{quantum tetrahedron} was later expounded by Barbieri \cite{Barbieri1997} and Baez and Barrett \cite{Barrett1999} with applications in \emph{four-dimensional} spinfoam gravity \cite{Barrett1997}.} with quantized edge lengths and Regge semiclassics, it is not unreasonable to surmise that a model of  (Riemannian) quantum gravity on a triangulated $3$-manifold $\Delta$ can be obtained by 

\medskip\begin{itemize}
\item
associating one $\{6j\}$ symbol to each tetrahedron, and
\item
summing over the spins, in the spirit of Feynman's path integral.
\end{itemize}\medskip
Indeed, Ponzano and Regge defined in \cite{PR} a partition function $\mathcal{Z}_{\textrm{PR}}(\Delta)$ for the triangulation $\Delta$ by
\be\label{PRpartition}
\mathcal{Z}_{\textrm{PR}}(\Delta) \doteq\sum_{(j_e)_{e\in\Delta_1}} (-1)^{\chi} \prod_{e\in\Delta_1} (2j_e+1) \prod_{t\in\Delta_3}\{6j\}_t.
\ee
Here, $\Delta_i$ denotes the set of $i$-simplices of $\Delta$, $\{6j\}_t$ is the $\{6j\}$ symbol based on the spins coloring the $6$ edges of the tetrahedron $t\in\Delta_3$, and $\chi$ is a linear function of the spins $j_e$, left unspecified by Ponzano and Regge (except for very special cases) but fixed in \cite{Barrett:2009ys}. Note that, if $\Delta$ has a boundary, this expression defines a \emph{transition amplitude} for a two-dimensional Regge geometry with fixed edge lengths. 


The key input in this heuristic quantization of three-dimensional geometry is the \emph{quantization of edge lengths}. This follows from the quantization of irreducible representations (spins) $j_i$ of $\textrm{SU}(2)$, and the identification $l_i=j_i+1/2$ suggested by the Ponzano-Regge asymptotic formula \eqref{6jasympPR}. As pointed out by Rovelli \cite{Rovelli1993}, this quantization hypothesis has a compelling justification in loop quantum gravity. We refer to Rovelli's book \cite{Rovelli:2004tv} for more details on this aspect.

\subsection{Bubble divergences}\label{bubbles}

It did not escape Ponzano and Regge's attention that the multi-series in the definition \eqref{PRpartition} is unlikely to converge for a general triangulation $\Delta$ \cite{PR}. Unfortunately, it is difficult to turn this concern into a precise mathematical statement. Even absolute convergence of \eqref{PRpartition} is hard to check, mainly for two reasons.

\begin{itemize}
\item
First, not all the values of the spins $j_e$ actually contribute to the sum, because of the \emph{triangle constraints} which enter the definition of $\{6j\}$ symbols. For instance, a necessary condition for the $\{6j\}$ symbol in \eqref{6jasympPR} is that $\vert j_{2}-j_{3}\leq j_{1}\leq j_{2}+j_{3}$. Describing the domain $D_\Delta$ of admissible spins for a general triangulation $\Delta$ is by no means straightforward. 
\item
Even if $D_{\Delta}$ could be described explicitly, as in the elementary cases studied in \cite{Perini:2008pd}, we would not know how the $\{6j\}$ symbols behave there: the Ponzano-Regge asymptotic formula only prescribes its behavior in the homogeneous limit, where all the spins grow at the same rate.\footnote{Recent work has improved this situation: we now know the asymptotic behavior of the $\{6j\}$ symbol with small angular momenta as well \cite{Yu2011,Bonzom2011}.} This is only a very restricted subdomain of $D_\Delta$. 
\end{itemize}
Because of these difficulties, simple estimates based on a balance between the divergent edge factors $(2j_e+1)$ and the convergent tetrahedra factors $V(j_e)^{-1/2}\sim j_e^{-3/2}$ are not reliable: the absolute convergence of the Ponzano-Regge partition function is an open problem.

Worse still, even assuming that we can resolve these issues satisfactorily, the most likely outcome is be that the Ponzano-Regge sum \eqref{PRpartition} is \emph{not} absolutely convergent. One would then ask whether it is perhaps \emph{semi-convergent}, like the alternating series $\sum(-1)^n/n$. Unfortunately, the complicated oscillation pattern induced by the factors $(-1)^\chi$ and the Ponzano-Regge cosines $\cos\left( S_{\rm R}(l_i) + \f\pi4\right)$ make this question almost impossible to answer in full generality. 

This notwithstanding, one observation made in an attempt by Perez and Rovelli \cite{Perez:2000fs} to describe the domain $D_\Delta$ deserves to be mentioned. First, consider the \emph{dual cell complex} $\Delta^*$ of the triangulation $\Delta$, whose $k$-cells are in one-to-one correspondence with the $(3-k)$-simplices of $\Delta$. (Since the partition function \eqref{PRpartition} only involves the edges and tetrahedra of $\Delta$, and not its vertices, we can actually restrict to the $2$-skeleton of $\Delta^*$.) Within $\Delta^*$, the triangle constraints can be localized on the edges $e^*$, where three spin-colored faces $f^*$ meet. A moment of reflection then shows that, for one the spins to be free, namely to have admissible values from zero to infinity, a necessary condition is that there exists a \emph{closed} set of faces within $\Delta^*$: a \emph{bubble}.

This insight has two immediate consequences. First, it suggests that the \emph{number of bubbles} of $\Delta^*$ is somehow the driving parameter in the divergence degree of $\mathcal{Z}_{\textrm{PR}}(\Delta)$, if such a thing can be defined. Second, it gives a field-theoretic flavor to the issue of divergences in the Ponzano-Regge model: \emph{bubbles} are higher-dimensional analog of \emph{loops} -- and we know that loops support the divergences of Feynman amplitudes. This circumstance calls for an interpretation of the Ponzano-Regge model as a ``higher categorical'' version of perturbative quantum field theory \cite{Baez:1999sr}.

We shall see in this thesis that both intuitions are relevant.

\section{A discrete gauge theory}

So far, the only evidence we have given for the connection between the Ponzano-Regge ansatz and three-dimensional quantum gravity is the Ponzano-Regge asymptotic formula. There is however another argument to this effect, relying on the Palatini-Cartan formulation of general relativity. The latter sheds a new light on the Ponzano-Regge model: it is a \emph{discrete gauge theory}. This new perspective turns out to be the key for understanding its bubble divergences, as we will see in the second part of this thesis. 

\subsection{Palatini-Cartan gravity}

Recall that, in a ``first order" formulation of (Riemannian) three-dimensional general relativity, the gravitational field can be represented by a pair of fields $(e,\omega)$, where $\omega$ is an $\su$ connection $1$-form on spacetime $M$, and $e$ is an $\mathfrak{su}(2)$-valued $1$-form over $M$, or equivalently a triad field on $M$.\footnote{This is identification is possible due to the fact that the adjoint representation of $\su$ is also its vector representation.} The so-called \emph{Palatini-Cartan}\footnote{The name given to various gravitational actions is very author-dependent. We use "Cartan" to mean that a vielbein is involved, and "Palatini" to express the fact that the connection is an independent variable.} action is then  \cite{Baez1994}
\be\label{palatinicartan}
S_{\textrm{PC}}(e,\omega):=\int_M\langle e\wedge F(\omega)\rangle.
\ee
Here the bracket is the Killing form in the Lie algebra $\mathfrak{su}(2)$. It is not hard to see that the Euler-Lagrange equations for this action, namely
\be
F(\o)=0 \qquad\textrm{and}\qquad d_\o(e)=0,
\ee
are indeed the vacuum Einstein equation $R_{ab}=0$ in disguise: the first states that $\o$ is flat, and the second that it is the unique spin connection associated to $e$, see \cite{Baez1994}.

The Palatini-Cartan action is an instance of the \emph{BF models} considered by Horowicz \cite{Horowitz1989}, defined in dimension $d$ by the analogue of \eqref{palatinicartan} with $e$ replaced by a $(d-2)$-form $B$. Like the latter, it is invariant under an extended gauge group which, besides the diffeomorphisms and $\su$ gauge transformations, includes a \emph{shift symmetry} generated by
\be\label{shiftsymmetry}
e\mapsto e+d_\o\l,
\ee
where $\l$ is an $\mathfrak{su}(2)$-valued $0$-form over $M$ and $\o$ is a flat connection. 

Let us now perform a formal path-integral ``quantization'' of the Palatini-Cartan, according to
\be\label{pathintegralBF1}
\widetilde{Z}_{\textrm{PC}}(M):\doteq\int\mathcal{D}e\mathcal{D}\omega\ e^{iS_{\textrm{PC}}(e,\omega)}.
\ee
Integrating out the $e$ field gives
\be\label{pathintegralBF2}
\widetilde{Z}_{\textrm{PC}}(M)\doteq\int\mathcal{D}\omega\ \delta\big(F(\omega)\big).
\ee
The tilde and quotes here are meant to remind us that \emph{no gauge-fixing has been performed} for the on-shell gauge symmetry \eqref{shiftsymmetry} \cite{Freidel2002,Freidel2004}, and so we cannot speak of a genuine quantization in the usual sense of gauge theory \cite{Witten1988,Horowitz1989}. For clarity, the operation performed in \eqref{pathintegralBF1}-\eqref{pathintegralBF2} should perhaps be called a ``half-quantization" of Palatini-Cartan gravity. 


Before we move on to explain the connection between this procedure and the Ponzano-Regge model, one more comment is in order. In quantizing a classical theory \`a la Feynman, it is not uncommon that the space of histories has to be enlarged with respect to its classical counterpart. For instance, the quantum histories of a free particle contains rough (viz. non-differentiable) paths in addition to the smooth ones considered in the classical theory. Similarly, the quantum histories of a free field include distributional fields as well as smooth ones. Now, the path-integral half-quantization of the Palatini-Cartan action relies on a similar extension of the classical notion of a gravitational history, including \emph{torsion} and \emph{degenerate metrics}. Witten has argued \cite{Witten1988} that such configurations are essential in three-dimensional quantum gravity (showing that they permit a renormalizable perturbative extension), but later changed his mind \cite{Witten2007}. All in all, it is fair to say that the status of torsion and degenerate metrics is quantum gravity remains to be understood.\footnote{I thank Simone Speziale for drawing my attention to this point.}

\subsection{Discrete gauge theory}


Let us now consider the half-quantized expression \eqref{pathintegralBF2}, and \emph{discretize it}. To this aim, we must
\begin{enumerate}
\item
introduce a triangulation $\Delta$ of $M$, and its dual cell complex $\Delta^*$, and
\item
discretize the connection $\o$ (resp. the curvature $F$) on the edges (resp. the faces) of the $2$-skeleton $\Gamma$ of $\Delta^*$
\item
Choose orientations for the edges and faces of $\G$.
\end{enumerate}
Recall that, given this data, discrete gauge theory is formulated in terms of \emph{parallel transport operators} and \emph{holonomies}. For a given cell decomposition $\Delta^*$ of $M$, these are defined respectively as assignments of group elements $g_e\in\su$ to each edge $e$, and of the products 
\be\label{defhol}
 H_f=\prod_{e\in\partial f} g_e^{\epsilon_{fe}}
\ee
to each face $f$, with $\epsilon_{fe}=\pm 1$ depending on the relative orientation of $e$ and $f$. Thus, a \emph{discrete connection} $A=(g_e)_e$ is an element of $\su^{E}$ and the \emph{curvature} of $A$ is the family of holonomies  $\big(H_f\big)_f$ defined by \eqref{defhol}. 

Using these definitions, we can make the following ansatz for the discretization of \eqref{pathintegralBF2} on $\Gamma$:
\be\label{discreteBF}
\widetilde{z}_{\textrm{PC}}(\G):\doteq\prod_{e\in \G_1}\int_{\textrm{SU}(2)}dg_e\ \prod_{f\in \G_2}\delta(H_f).
\ee
Here, $dg$ is the Haar measure on $\su$ and $\Gamma_i$ denotes the set of $i$-cells of $\Gamma$. Just like the formal path integral \eqref{pathintegralBF2} described a system of flat $\su$ connections, the expression \eqref{discreteBF} can be seen as the path-integral for a system of flat discrete connections.

Now comes the main result of this section: \emph{the half-quantized, discretized path integral \eqref{discreteBF} is formally identical to the Ponzano-Regge partition function \eqref{PRpartition}.} To understand how this comes about, we must recall the basis of the Peter-Weyl theory of Fourier analysis on compact groups. 

Let $f:G\rightarrow\mathbb{C}$ be a smooth function on a compact group $G$, and $\rho_j:G\rightarrow\textrm{GL}(V^j)$ be the sequence of unitary irreducible representations of $G$, labelled by a discrete index $j$.\footnote{More precisely, $j$ ranges over the set of equivalence classes of unitary irreps of $G$, and $\r^j$ is an element of $j$.} For each $j$, pick a basis $(v^j_m)_{m=0,\dots,\dim V_j}$ of $V_j$, and define the representative functions $D^j_{mn}:V_j\rightarrow\mathbb{C}$ by \be D^j_{mn}(g)=\langle v_m^j\vert\r^j(g)v^j_n\rangle.\ee Then the \emph{Peter-Weyl theorem} states that $f$ can be decomposed over the representative functions as 
\be
f(g)=\sum_j\sum_{mn}{f^j_{mn}}D^j_{mn}(g),
\ee
with
\be
f^j_{mn}:=\int_Gdg\ D^j_{mn}(g)^*f(g).
\ee
When $f$ is a class function, viz. when it is conjugation invariant, this decomposition reduces to
\be
f(g)=\sum_{j}f^j_{mn}\chi^j(g).
\ee
where $\chi^j=\sum_mD^j_{mm}$ is the \emph{character} of the representation $\r^j$. These decompositions extend to square-integrable functions and Schwartz distributions. In particular, one computes that
\be\label{fourierdelta}
\delta(g)=\sum_j(\dim V_j)\chi^j(g).
\ee

The second ingredient we need to derive the Ponzano-Regge model from the discrete path-integral \eqref{discreteBF} is the recoupling identity
\be\label{recoupling}
\int_Gdg\ \bigotimes_{a=1}^k\rho^{j_a}(g)=\sum_{\iota}\vert\iota\rangle\langle\iota\vert,
\ee
in which $\iota$ ranges over an orthonormal basis of the $G$-invariant subspace of the tensor product representation $\bigotimes_{a=1}^kV^{j_a}$. For the case $G=\su$ and $k=3$, there is only one such $\iota$: the Wigner $(3jn)$-symbol. 

Let us now come back to the integral \eqref{discreteBF}, and use \eqref{fourierdelta} to decompose the delta functions, noting that $j$ is now a spin. This gives
\be
\widetilde{z}_{\textrm{PC}}(\G)\doteq\sum_{(j_f)_{f\in\Delta^*_2}}\prod_{e\in\Delta^*_1}\int_{\su}dg\ \prod_{f\in\Delta^*_2}(2j_f+1)\chi^{j_f}(H_f).
\ee
Now, since $\Delta$ is the triangulation of a three-dimensional manifold, every edge $e$ of the dual decomposition is shared by exactly three faces. Therefore the recoupling identity \eqref{recoupling} can be applied along each edge $e$. Associating the vectors $\vert\iota\rangle$ and $\langle\iota\vert$ to the source and target of each $e$, this results in a contraction pattern whereby four invariant vectors $\vert\iota\rangle$ are contracted at each vertex of $\Delta^*$. Examination of this contraction pattern shows that the corresponding invariant is nothing but the Wigner $\{6j\}$ symbol. All in all, we find that 
\be
\widetilde{z}_{\textrm{PC}}(\Delta)=\mathcal{Z}_{\textrm{PR}}(\Delta)
\ee

This result clearly vindicates the relevance of the Ponzano-Regge for quantum gravity. But it also provides us with a new interpretation of the bubble divergences discussed in sec. \ref{bubbles} \cite{Freidel2004}: they are the discrete remnants of the (non-compact) shift symmetry of the Palatini-Cartan action. Building on this insight, these authors developed a regularization scheme based on a tentative gauge-fixing of this symmetry. It appears, however, that their ansatz is successful only in limited cases. Recent work by Barrett and Naish-Guzman \cite{Barrett:2009ys}, and also by Bonzom and myself \cite{Bonzom}, has clarified the reasons of this failure and the way to overcome it.

\subsection{Boulatov's group field theory}

The presentation of the Ponzano-Regge model as a discrete gauge theory, with \emph{group elements} instead of spins as basic variables, has another interesting payoff: it puts us on the right track to understand its bubble divergences as genuine field-theoretic divergences.

To proceed in this direction, let us follow Boulatov \cite{Boulatov:1992vp} and introduce an auxiliary field $\phi$, defined over the configuration space $\SU(2)\times\SU(2)\times\SU(2)$ and invariant under the right diagonal action of $\su$ and cyclic permutations $c$ of its arguments, i.e.:
\be\label{invariances}
\phi(g_1h,g_2h,g_3h)=\phi(g_1,g_2,g_3) \;\;\;\;\;\;\textrm{and} \;\;\;\;\phi(g_{c(1)},g_{c(2)},g_{c(3)})=\phi(g_1,g_2,g_3).
\ee
Let us assume furthermore that the ``dynamics'' of this field is given by the \emph{non-local} action
\begin{multline}
S_B(\phi):=\f{1}{2}\int\prod_{i=1}^3dg_i\ \phi^2(g_1,g_2,g_3)\\+\f{\lambda}{8}\int\prod_{i=1}^{6}dg_i\ \phi(g_1,g_2,g_3)\phi(g_3,g_4,g_5)\phi(g_5,g_2,g_6)\phi(g_6,g_4,g_1),
\end{multline}
where $dg_i$ denotes the $\su$ Haar measure. Note that, in the quartic term, the six integration variables are repeated twice, following the pattern of the edges of a tetrahedron. (The factor $8$ is only for simpler final formulas.)

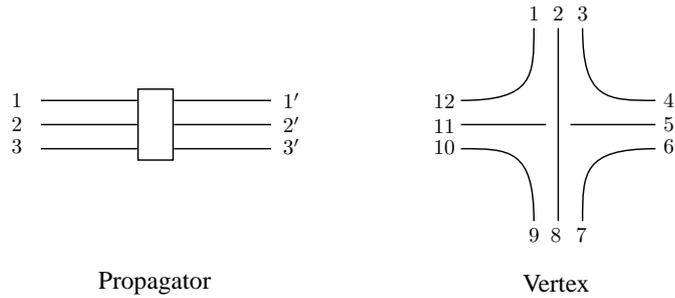
\begin{figure}[t]\label{propavertexfig}
\begin{center}
\scalebox{0.8} 
{
\begin{pspicture}(0,-2.4336915)(11.207031,2.4336915)
\psbezier[linewidth=0.024](8.703125,2.024004)(8.703125,1.2240039)(8.703125,0.82400393)(7.5031247,0.82400393)
\psbezier[linewidth=0.024](9.503125,-1.1759961)(9.503125,-0.37599608)(9.503125,0.024003906)(10.703124,0.024003906)
\psbezier[linewidth=0.024](10.703124,0.82400393)(9.903126,0.82400393)(9.503125,0.82400393)(9.503125,2.024004)
\psbezier[linewidth=0.024](7.5031247,0.024003906)(8.303125,0.024003906)(8.703125,0.024003906)(8.703125,-1.1759961)
\psline[linewidth=0.024cm](9.103125,2.024004)(9.103125,-1.1759961)
\psline[linewidth=0.024cm](7.5031247,0.4240039)(8.903125,0.4240039)
\psline[linewidth=0.024cm](9.303125,0.4240039)(10.703124,0.4240039)
\usefont{T1}{ptm}{m}{n}
\rput(9.069346,-2.2109962){Vertex}
\usefont{T1}{ptm}{m}{n}
\rput(8.706641,2.264004){\footnotesize $1$}
\usefont{T1}{ptm}{m}{n}
\rput(9.10664,2.264004){\footnotesize $2$}
\usefont{T1}{ptm}{m}{n}
\rput(9.506641,2.264004){\footnotesize $3$}
\usefont{T1}{ptm}{m}{n}
\rput(10.9266405,0.82400393){\footnotesize $4$}
\usefont{T1}{ptm}{m}{n}
\rput(10.9266405,0.4240039){\footnotesize $5$}
\usefont{T1}{ptm}{m}{n}
\rput(10.9266405,0.06400391){\footnotesize $6$}
\usefont{T1}{ptm}{m}{n}
\rput(9.48664,-1.3959961){\footnotesize $7$}
\usefont{T1}{ptm}{m}{n}
\rput(9.06664,-1.3959961){\footnotesize $8$}
\usefont{T1}{ptm}{m}{n}
\rput(8.706641,-1.3959961){\footnotesize $9$}
\usefont{T1}{ptm}{m}{n}
\rput(7.2266407,0.8040039){\footnotesize $12$}
\usefont{T1}{ptm}{m}{n}
\rput(7.2266407,0.40400392){\footnotesize $11$}
\usefont{T1}{ptm}{m}{n}
\rput(7.2266407,0.044003908){\footnotesize $10$}
\psline[linewidth=0.024cm](0.5983423,0.82400393)(2.183125,0.82400393)
\psline[linewidth=0.024cm](0.5983423,0.4240039)(2.183125,0.4240039)
\psline[linewidth=0.024cm](0.5983423,0.024003906)(2.183125,0.024003906)
\psframe[linewidth=0.024,dimen=outer](2.783125,1.0240039)(2.183125,-0.1759961)
\psline[linewidth=0.024cm](2.783125,0.82400393)(4.383125,0.82400393)
\psline[linewidth=0.024cm](2.783125,0.4240039)(4.383125,0.4240039)
\psline[linewidth=0.024cm](2.783125,0.024003906)(4.383125,0.024003906)
\usefont{T1}{ptm}{m}{n}
\rput(2.4712205,-2.2109962){Propagator}
\usefont{T1}{ptm}{m}{n}
\rput(0.20664063,0.82400393){\footnotesize $1$}
\usefont{T1}{ptm}{m}{n}
\rput(0.20664063,0.4240039){\footnotesize $2$}
\usefont{T1}{ptm}{m}{n}
\rput(0.20664063,0.06400391){\footnotesize $3$}
\usefont{T1}{ptm}{m}{n}
\rput(4.7166405,0.82400393){\footnotesize $1'$}
\usefont{T1}{ptm}{m}{n}
\rput(4.7166405,0.4240039){\footnotesize $2'$}
\usefont{T1}{ptm}{m}{n}
\rput(4.7166405,0.06400391){\footnotesize $3'$}
\end{pspicture} 
}
\caption{The propagator $C$ and the interaction vertex $T$ of the Boulatov model. The box in the propagator represents the group averaging $\int_{\SU(2)}dh$.}
\end{center}
\end{figure}

Consider now the perturbative quantization of this model, by mean of the formal partition function 
\be
\calZ_{B}:\doteq\int D\phi\ e^{-S_{B}(\phi)}.
\ee
Examination of the corresponding Feynman rules reveals that the propagator is three-stranded, with integral kernel
\be
C(g_{1},g_{2},g_{3};g'_{1},g'_{2},g'_{3})=\int_{\SU(2)}dh\ \delta(g_{1}^{-1}hg'_{1})\delta(g_{2}^{-1}hg'_{2})\delta(g_{3}^{-1}hg'_{3}).
\ee
Likewise, the interaction vertex can be written a product of six delta functions, matching twelve group elements pair-wise, as in fig. 2.1:
\be
T(g_{1},\dots, g_{12})=\delta(g_{1}g_{12}^{-1})\delta(g_{2}g_{8}^{-1})\delta(g_{3}g_{4}^{-1})\delta(g_{5}g_{11}^{-1})\delta(g_{6}g_{7}^{-1})\delta(g_{9}g_{10}^{-1}).
\ee

For a given stranded graph $\G$, these Feynman rules correspond to a contraction pattern which gives, after integration over all $h$ variables, the Feynman amplitude (for a vacuum diagram, viz. with no external leg)
\be
\mathcal{A}_{B}(\Gamma)\doteq\Big(\f{\l}{8}\Big)^{V}\prod_{e\in \G_1}\int_{\textrm{SU}(2)}dg_e\ \prod_{f\in \G_2}\delta(H_f),
\ee
where $V$ is the number of vertices of the stranded graph $\Gamma$, $e$ denotes an edge $\G$, and $f$ a face, i.e. a closed strand. Up to a multiplicative constant, this is nothing but the Ponzano-Regge partition function \eqref{discreteBF} evaluated on the $2$-complex defined by the edges and faces of the stranded Feynman graph. If we interpret furthermore this $2$-complex as dual to a three-dimensional space, as in \cite{DePietri:2000ii}, the connection with the Ponzano-Regge model becomes transparent: Boulatov's model is a \emph{generating function} for Ponzano-Regge amplitudes. (Note however that not all such $2$-complexes define a three-dimensional \emph{manifold}: there can be point-like singularities, yielding what is known as a \emph{pseudo-manifold}. We refer the interested reader to \cite{Gurau2010,Smerlak2011,Gurau2011} for a discussion on this point.)


\section{The flat spinfoam model}


The upshot of the above discussion of the Ponzano-Regge model can be summarized as follows:
\begin{enumerate}
\item
Although originally defined on a triangulation $\Delta$ of the spacetime manifold, the structure of the partition function (and in particular its divergence) is better understood in terms of the dual cell complex $\D^*$, and in fact of the $2$-skeleton $\G$ of $\D^*$. 
\item
The analytic structure of the Ponzano-Regge partition function is more transparent when written as a group integral \eqref{discreteBF}; its gauge-theoretic underpinning, and in particular its connection to the Palatini-Cartan formulation of gravity, its then manifest.
\end{enumerate}
This motivates a generalization of the Ponzano-Regge model defined on more general $2$-complexes: a kind of \emph{cellular gauge theory}, with no reference to a particular spacetime manifold. This is concomitant with Boulatov's approach, which, as we already mentioned, departs from the strict setup of three-dimensional manifolds.

\subsection{Foams}\label{deffoams}

Before we introduce this generalization, let us spell out what we mean by the expression ``$2$-complex'' above. 

The most general possible definition one can think of is a \emph{finite, connected two-dimensional CW complex}. Recall that a $d$-dimensional \emph{CW-complex} $X$ is the last of a nested sequence of topological spaces $X_{1} \subset X_{2}\subset\cdots\subset X_{d} = X$ defined inductively as follows. The $0$-skeleton $X_{0}$ is just a discrete set of points. For each integer $k > 0$, the $k$-skeleton $X_{k}$ is the result of attaching a set of $k$-dimensional balls $B^{k}$ to $X_{k-1}$ by gluing maps of the form $\sigma: \pp B^{k} \arr X_{k-1}$. For each $k$, the subspace $X_{k}$ is called the $k$-skeleton of $X$, and the interiors of the balls $B^k$ attached to $X_{k-1}$ are the $k$-cells of $X$. 

Contrary to what one might expect (perhaps on the basis of the easy classification of two-dimensional \emph{manifolds}), two-dimensional CW complexes are very subtle objects, about which several outstanding conjectures remain open.\footnote{One of them, the Zeeman conjecture, actually implies the Poincar\'e conjecture.} In order to stay away from such mathematical difficulties, in this thesis we will stick to a more restricted context, where the gluing maps are themselves \emph{cellular}: we will assume that each ball $B^{k}$ has a cellular decomposition such that the $(k-1)$-cells on the boundary of $B^{k}$ are glued onto $(k-1)$-cells of $X_{k-1}$. This condition introduces a form of rigidity in the gluing patterns, thereby removing most of the topological difficulties associated to general CW complexes.

The most useful examples of such CW complexes can be found in the polyhedral category, where the balls $B^{k}$ are in fact convex polyhedra. Explicitly, a \emph{polyhedral complex}, or \emph{piecewise-linear cell complex}, is a CW complex where each ball $B^k$ is in fact a $k$-dimensional convex polytope $P^k$ in some $\mathbb{R}^n$, and the gluing maps $\sigma: \pp P^{k} \rightarrow X_{k-1}$ are cellular homeomorphisms onto their images. When all the polytopes are simplices, one speaks of a \emph{$\D$-complex} \cite{hatcher}.

One reason which makes two-dimensional polyhedral complexes, from now on called \emph{foams}, particularly convenient to deal with is that their structure can be described purely combinatorially: attaching a polygon onto the $1$-skeleton $X_{1}$ amounts to choosing a \emph{cycle} in $X_{1}$, i.e. a finite  sequence of vertices ($0$-cells) such that, from each of its vertices there is an edge ($1$-cell) to the next vertex in the sequence, and the start and end vertices are the same. In other words, a two-dimensional polyhedral complex $\G$ can be equivalently defined as a triple $\G=(V_\G,E_\G, F_\G)$ with $V_\G$ a finite set of \emph{vertices}, $E_\G$ a set of ordered pairs of vertices $e=(v,v')$, or \emph{edges}, and $F_\G$ a finite set of cycles, or \emph{faces}. This data uniquely defines a two-dimensional polyhedral complex up to homeomorphism. 

Note that this definition includes the stranded Feynman diagrams of the Boulatov model, but is also much more general: unlike the former, the number of faces adjacent to any edge can be arbitrary. It is in this sense that the flat spinfoam model, being defined on general foams, generalizes the Ponzano-Regge model to ``arbitrary dimensions''. 

Finally, observe that there is a notion of \emph{boundary} for foams. Indeed, let us call the edges of $E_\G$ appearing exactly once in only one face its \emph{links}, and the other ones its \emph{interior edges}. Similarly, let us call the vertices appearing exactly once in an interior edge its \emph{nodes}, and the other ones its \emph{interior vertices}. 
The sets of nodes and links of a foam $\G$ generally do not form a graph, but when they do, and moreover the orientation of each link matches the one induced by the unique face passing through it, we say that $\G$ is a \emph{proper foam}. We then define the \emph{boundary} $\pp\G$ as the subcomplex of $\G$ defined by those faces of $\G$ which contain at least one link. The underlying graph of $\pp\G$ is the \emph{boundary graph} of $\G$. In this language, foams with the same boundary graphs can still have different boundaries, because the boundary faces can be different.



\subsection{Definition of the model}

We are now in the position to define the \emph{flat spinfoam model}. Let $\Gamma$ be a foam, and $G$ be a compact Lie group. Denote $\calA(\G,G):=G^{E}$ the space of discrete $G$-connections $A=(g_{e})_{e\in\G_{1}}$ over $\G$, and $H=(H_{f}\big)_{f\in\G_{2}}$ its curvature as in \eqref{defhol}. The flat model is then defined formally by the expression
\begin{equation}\label{flatpartition}
\calZ(\Gamma,G):\doteq\int_{\calA(\G,G)}dA\ \prod_{f\in \G_2}\delta(H_f),
\end{equation}
where $dA=\prod_{e\in\G_{1}}dg_{e}$ is the Haar measure on $\calA(\G,G)$.

When $\G$ is the $2$-skeleton of the cell complex dual to a triangulated $3$-manifold, this expression of course coincides with the Ponzano-Regge partition function. But this is only a very special case of \eqref{flatpartition}: there are many foams which cannot be constructed in this way. Note in particular that, when there are more than three faces adjacent to each edge, the Peter-Weyl transform of \eqref{flatpartition} does not consist of $\{6j\}$-symbols. For instance, if $\G$ is the $2$-skeleton of the dual cell complex of a triangulated $4$-manifold, these are replaced by $\{15j\}$-symbols: this is the so-called \emph{Ooguri model} \cite{Ooguri:1992eb}.


Another particular case of the flat spinfoam model is when $\G$ is the cellular decomposition of a closed, orientable \emph{surface}: the partition function \eqref{flatpartition} then corresponds to the weak coupling limit of \emph{two-dimensional Yang-Mills theory}, considered notably by Witten in \cite{Witten1991}. When the genus of $\G$ is greater or equal to $2$, extensive studies of \eqref{flatpartition} after \cite{Witten1991} have showed that $Z(\G)$ is finite.\footnote{It is then equal to $\zeta(-\chi(\G))$, with $\zeta$ the Riemann zeta function and $\chi(\G)$ the Euler characteristic of $\G$.} We will re-derive this result in sec. \ref{surfaceexamples}.

\subsection{Heat kernel regularization}
Just like its Ponzano-Regge specialization, the flat spinfoam model is ill-defined. In the definition \eqref{flatpartition}, this is manifest from the fact that delta distributions supported at coincident points are \emph{multiplied} -- a famously illicit operation.


Very roughly speaking, we can say that a troublesome situation will arise every time the number of delta functions ($F$, the number of faces of $\G$) does not match the number of integrals in \eqref{flatpartition} ($E$, the number of $edges$). Thus, when $F>E$, we might expect that the a possible regularization of $\calZ(\G,G)$ consists in \emph{removing} some \emph{redundant} delta functions in \eqref{flatpartition}. Indeed, this is the idea followed by Freidel and Louapre in \cite{Freidel2002,Freidel2004}, and later by Barrett and Naish-Guzman \cite{Barrett:2009ys}. We will comment on this approach in sec. \ref{firstresults}. 

For now, we will content ourselves with a more straightforward regularization, consisting simply in replacing the delta functions by smooth functions, peaked on the unit $\unit\in G$. One way to construct such functions is by letting the delta \emph{diffuse} over $G$ for some small time $\tau>0$. This gives the so-called \emph{heat kernel}\footnote{Explicit expression of the heat kernel for simply connected and connected semi-simple groups are known, see \cite{hkcompact}.} $K_{\tau}$, viz. the solution of the heat equation 
\be
\pp_{\tau}K_{\tau}=\Delta K_{\tau},
\ee
with initial condition $\lim_{\tau\rightarrow0}\ K_{\tau}=\delta$. Here $\Delta$ is the Laplace-Beltrami operator, defined using some bi-invariant metric on $G$ (e.g. the one induced by the Killing form on its Lie algebra, when $G$ is semi-simple). 

Using the heat kernel on $G$, we thus define the \emph{regularized flat spinfoam model} by 
\be\label{regularized}
\calZ_{\tau}(\Gamma,G):=\int_{\calA(\G,G)}dA\ \prod_{f\in \G_2}K_{\tau}(H_f).
\ee
This ansatz turns the problem of understanding the bubble divergences of the flat spinfoam model (and in particular of the Ponzano-Regge and Ooguri models) into a well-posed analytical question: what is the $\tau\arr0$ asymptotic behavior of $\calZ_{\tau}(\Gamma,G)$? Answering this question is the purpose of the next chapters of this thesis.

%% file: part2.tex
\part{Homological powercounting of bubble divergences}

\vspace*{\fill}

%

In this part of the thesis, we address the problem of computing
the \emph{divergence degree} of the regularized flat spinfoam model $\calZ_{\tau}(\G,G)$ in the $\tau\arr0$ limit. The approach we take has two ingredients, detailed in chapters \ref{Laplace} and \ref{cohomology} respectively:
\medskip\begin{enumerate}
\item
study the $\tau\rightarrow0$ asymptotics of \eqref{regularized} by means of a \emph{(generalized) Laplace approximation}
\item 
use \emph{(twisted) cellular homology} to disentangle the interplay between the topology of $\Gamma$ and the non-Abelian structure of $G$ within $\Omega(\Gamma,G)$ 
\end{enumerate}

\bigskip

We will see that the divergence degree is given by the number of ``bubbles" of $\Gamma$ only in very special cases; in general, the combinatorial structure of the foam does not fix $\Omega(\Gamma,G)$ alone -- the r\^ole played by the non-Abelian structure of $G$ is crucial. Applications of our result will be presented in chapter \ref{applications}.

\vspace*{\fill}

\chapter{Powercounting by Laplace's method}\label{Laplace}

\section{The amplitude as a Laplace integral}\label{}

In this section, we define the divergence degree of a foam, and show that, in principle, it can be computed by means of the Laplace method.

\subsection{Divergence degree}

The small-time asymptotic behaviour of the heat kernel on $G$ is well-known, and easy to understand: when $\tau\rightarrow0$, a Brownian particle on $G$ has no time to explore the geometry of $G$ and confuses the latter with the Euclidean space $T_{\unit}G$. In this limit, the heat kernel therefore takes a Euclidean form:
\be\label{heatkernelasymptotics}
K_{\tau}(g)\u{\tau\rightarrow 0}{\sim}(4\pi\tau)^{-\dim G/2}e^{-\vert g\vert^2/4\tau},
\ee
where $\vert g\vert$ is the Riemannian distance between $g$ and the unit $\unit$ of $G$. In particular, we have 
\be
K_{\tau}(1)\underset{\tau\rightarrow0}{\sim}\L_{\tau}^{\dim G},
\ee
with
\be
\L_{\tau}:=(4\pi\tau)^{-1/2}
\ee
playing the r\^ole of a large-spin cutoff. This motivates the definition of the \emph{divergence degree} of $\Gamma$ as the number $\Omega(\Gamma,G)$ such that the limit
\be
\mathcal{Z}'(\Gamma,G):=\u{\tau\rightarrow 0}{\lim}\ \L_{\tau}^{-\Omega(\Gamma,G)} \mathcal{Z}_{\tau}(\Gamma,G)
\ee
is finite and non-vanishing. Let us emphasize that we do not assume \emph{a priori} that $\Omega(\G,G)$ is a multiple of $\dim G$, as would be implied by the notion that to regularize the flat spinfoam model, one should remove redundant delta functions \cite{Freidel2004}.

\subsection{A Laplace integral}

Denote $V,E,F$ the number of vertices, edges and faces of the foam $\G$. Using the asymptotics \eqref{heatkernelasymptotics}, we can write the regularized amplitude $\calZ_{\tau}(\G,G)$ in the $\tau\rightarrow 0$ limit as the integral
\be\label{partitionlaplace}
\calZ_{\tau}(\Gamma,G)\u{\tau\rightarrow 0}{\sim}\Lambda_\tau^{(\dim G)F}\int_{\mathcal{A}(\Gamma,G)}dA\ e^{-S(A)/\tau},
\ee
where
\be\label{action}
S(A):=\f{1}{4}\sum_{f\in\Gamma_2}\vert H_f(A)\vert^2.
\ee
The function $S:\mathcal{A}(\Gamma,G)\rightarrow\mathbb{R}$ thus defined is smooth, non-negative, and vanishes on flat connections where $H_f(A)=\unit$ for every face $f$ of $\Gamma$. This suggests that the integral on the right-hand side of \eqref{partitionlaplace} can be estimated by means of Laplace's method, which we now review.

\section{Degenerate Laplace asymptotics}\label{}

Let $M$ be a (closed) Riemannian manifold, $dx$ the canonical volume form on $M$, and $f$ a smooth function. A point $y\in M$ is a \emph{critical point} of $f$ if the differential of $f$ vanishes at $y$; it is \emph{non-degenerate} if the Riemannian Hessian $\nabla^2f(y):T_{y}M\arr T_{y}M$ is non-degenerate, i.e. if $\det\nabla^2f(y)\neq0$. If $f$ has only non-degenerate critical points (viz. it is a \emph{Morse function}), Laplace's method states that
\be\label{laplace}
\int_Mdx\ e^{-f(x)/\tau}\underset{\tau\rightarrow0}{\sim}(\pi\tau)^{\dim M/2}\sum_i\f{e^{-f(y_i^*)/\tau}}{\sqrt{\det\nabla^2f(y_i^*)}}.
\ee
Here $y_i^*$ runs over the minima of $f$. This formula can be proved by Taylor-expanding $f$ to second order about $y^*_i$, and performing the resulting Gaussian integral. 

It is tempting to try to apply this well-know asymptotic formula to the amplitude \eqref{partitionlaplace}. Unfortunately, one sees easily that the function $f$ defined by \eqref{action} does have the required property. Indeed, the Morse lemma\footnote{If $f:M\rightarrow\mathbb{R}$ is a Morse function and $y^*$ a critical point of $f$, there is a coordinate patch $(x_1,\dots,x_n)$ around $y^*$ in which $f-f(y^*)=\sum_{j=1}^p x_j^2-\sum_{j=p+1}^n x_j^2$ for some $p$.} implies that non-degenerate critical points are isolated. But we know that topologically non-trivial spaces admit a \emph{continuum} of flat connections, hence that $\mathcal{F}(\Gamma,G)$ consists of \emph{degenerate} critical points.  To deal with this situation, we need a \emph{generalized Laplace method}. 

\subsection{The generalized Laplace method}

Let us consider the case where the critical set of $f$ is a disjoint union of \emph{submanifolds} of $M$. Denote $C$ the (possibly disconnected) critical manifold with \emph{maximum dimension}. Using the metric on $M$, we can split the tangent bundle of $M$ along $C$ into its tangential and normal sub-bundles, as 
\be\label{normalbundle}
TM_{\vert C}:=\bigsqcup_{y\in C} T_{y}M =TC\oplus NC.
\ee
Let us then define the \emph{normal Hessian} $\nabla_\perp^2f$ of $f$ as the restriction of $\nabla^2f$ to the normal fibers $N_yC$, and assume that $\nabla_\perp^2f$ is non-degenerate (one says that $f$ is a \emph{Morse-Bott} function). Laplace's formula \eqref{laplace} then generalizes to 
\be\label{generalizedlaplace}
\int_Mdx\ e^{-f(x)/\tau}\underset{\tau\rightarrow0}{\sim}(\pi\tau)^{(\dim M-\dim C)/2}\int_{C}dy\ \f{e^{-f(y)/\tau}}{\sqrt{\det\nabla_\perp^2f(y)}},
\ee
where $dy$ is the induced Riemannian volume form on $C$. Indeed, the orthogonal decomposition \eqref{normalbundle} implies that, along $C$, the volume form on $M$ factorizes as 
\be
dx=dy\wedge dz,
\ee 
where $dz$ is the Riemannian volume form on the fibers normal to $C$. Using the exponential decay $e^{-f/\tau}$ away from $C$ to restrict the integral over $M$ to a tubular neighborhood $U$ of $C$, in which $x\in U$ is parametrized by $(y,z)\in C\times N_{y}C$, and then applying Fubini's theorem, we obtain
\be
\int_Mdx\ e^{-f(x)/\tau}\underset{\tau\rightarrow0}{\sim}\int_{U}dx\ e^{-f(x)/\tau}\underset{\tau\rightarrow0}{\sim}\int_{C}dy\int_{N_yC}dz\ e^{-f(y,z)/\tau}.
\ee
The usual Laplace formula \eqref{laplace} can then be applied to the normal integrals $\int_{N_yC}dz_y\ e^{-f(y,z_y)/\tau}$; this gives immediately \eqref{generalizedlaplace}.

\subsection{Tame singularities}\label{tamesingularities}

What if $f$ is not a Morse-Bott function? Define a \emph{singularity} of $f$ as a critical point were either $C$ is not a manifold, or the normal Hessian $\nabla_\perp^2f$ is degenerate. Following \cite{Frohman2011}, say that  the singularities of $f$ \emph{tame} if for every $\epsilon$-neighborhood $U_\epsilon$ of the set $D$ of singularities,
\be\label{tame}
\underset{\epsilon\rightarrow0}{\lim}\ \underset{\tau\rightarrow0}{\lim}\ (\pi\tau)^{-(\dim M-p)/2}\int_{U_\epsilon}dx\ e^{-f(x)/\tau}=0.
\ee
(Unfortunately, there is no useful sufficient condition on $f$ for this relation to hold; it must be checked explicitly case by case.)

Assuming \eqref{tame}, we can apply the generalized Laplace method by simply excluding the singularities from the integral:

\be\label{generalizedlaplace2}
\int_Mdx\ e^{-f(x)/\tau}\underset{\tau\rightarrow0}{\sim}(\pi\tau)^{(\dim M-\dim C)/2}\int_{C\setminus D}dy\ \f{e^{-f(y)/\tau}}{\sqrt{\det\nabla_\perp^2f(y)}}.
\ee
In other words, for what concerns the $\tau\rightarrow0$ asymptotic behaviour of a Laplace-type integral, tame singularities are like no singularities at all.\footnote{That not all singularities are tame, however, can be seen on the two-dimensional example

\be
z_\tau:=\int_{\mathbb{R}^2}dxdy\ e^{-\f{(xy)^2}{\tau}}.
\ee
Here, the critical set is the `cross' $\{x=0\}\cup\{y=0\}$, and has a singularity at $(x,y)=(0,0)$. A naive application of the generalized Laplace method would give $z_\tau\propto\tau^{1/2}$ as $\tau\rightarrow0$, since the critical set is of codimension one except at the origin. However, this is not the correct estimate, which turns out to be $z_\tau\propto\tau^{1/2}\ln\tau$. In this case, indeed, integrating along the normal fibers yields a Hessian which is not integrable at the singularity: formally,
\be
z_\tau=\int_{\mathbb{R}}dx\left(\int_{\mathbb{R}}dy\ e^{-\f{(xy)^2}{\tau}}\right)=\sqrt{\pi\tau}\int_{\mathbb{R}}\f{dx}{\vert x\vert}.
\ee
It can be shown that isolated degenerate critical points always yield an asymptotic behaviour of the form $\tau^{a/2}\ln^b\tau$ with $a,b$ two integers, although computing the power of the logarithm is a difficult task, which requires to expand $f$ to orders higher than two. For extended degenerate critical sets, there is no general result.} 


\section{The divergence degree: first results}\label{firstresults}

\subsection{A powercounting formula}\label{veryfirstresult}
Let us now come back to the regularized amplitude
\be\label{bingo}
\calZ_{\tau}(\Gamma,G)\u{\tau\rightarrow 0}{\sim}\Lambda_\tau^{(\dim G)F}\int_{\mathcal{A}(\Gamma,G)}dA\ e^{-S(A)/\tau},
\ee 
and \emph{assume that $S$ has tame singularities}. Applying the generalized Laplace method described above then gives
\be\label{binga}
\calZ_{\tau}(\Gamma,G)\u{\tau\rightarrow 0}{\sim}\Lambda_\tau^{(\dim G)(F-E)+\dim\mathcal{F}_0(\Gamma,G)}\int_{\mathcal{F}_0(\Gamma,G)}\f{d\phi}{\sqrt{\det\nabla_\perp^2S(\phi)}},
\ee 
where $\mathcal{F}_0(\Gamma,G)$ is the set of \emph{flat connections} $\mathcal{F}(\Gamma,G)$ minus its singularities, and $d\phi$ is the induced Riemannian volume form on $\mathcal{F}_0(\Gamma,G)$. Since the remaining integral does not contain $\tau$ anymore, we can read off from \eqref{binga} an explicit for the sought-for divergence degree: 
\be\label{firstresult}
\Omega(\Gamma,G)=(\dim G)(F-E)+\dim\mathcal{F}_0(\Gamma,G).
\ee

Recall that from the formal expression \eqref{flatpartition}, we inferred that a foam is potentially divergent if it has more faces than edges, because then some of the delta functions are redundant. This is confirmed by the first term in the above formula. But the latter also shows how to sharpen this naive hunch, namely by \emph{studying the geometry of the space of flat connections} $\mathcal{F}(\Gamma,G)$. We will address this problem in some detail in the next section. For now, let us ask: is the intuition of Perez and Rovelli \cite{Perez:2000fs} and Freidel and Louapre \cite{Freidel2002,Freidel2004} that the divergence of a foam can be traced back to the presence of \emph{bubbles} vindicated by the formula \eqref{firstresult}?


\subsection{Bubbles from cellular homology: the simply connected case}\label{simplyconnected}

The notion of \emph{bubble} has a precise mathematical definition: it is a \emph{$2$-cycle} in cellular homology (see the appendix for a review of cellular homology). On a $2$-complex $\Gamma$, the space of $2$-cycles coincides with the second homology group $H_2(\Gamma)$. Thus, a tentative definition of the ``number of bubbles" of $\Gamma$ is $b^2(\Gamma):=\dim H_2(\Gamma)$, the \emph{second Betti number} of $\G$. Is the divergence degree $\Omega(\Gamma,G)$ related to $b^2(\Gamma)$?


Consider the case when the foam is simply connected, i.e. when its fundamental group $\pi_1(\Gamma)$ is trivial. Then we have that\medskip \begin{itemize}
\item
the flat connections on $\Gamma$ are all gauge-equivalent to the trivial one, hence
\be\label{flattrivial}
\dim\mathcal{F}_0(\Gamma,G)=(\dim G)(V-1),
\ee
\item
the first homology group of $\Gamma$ vanishes\footnote{By virtue of the Hurewicz theorem, according to which $H_1(\Gamma)$ is the abelianization of $\pi_1(\Gamma)$.}, \be H_1(\Gamma)=0.\ee
\end{itemize}
In this case, the result \eqref{result} gives therefore $\Omega(\Gamma,G)=(\dim G)(F-E+V-1)$. Furthermore, the equations $b_0(\Gamma)=1$ ($\Gamma$ connected) and $b_1(\Gamma)=0$ ($\Gamma$ simply connected), together with the Euler-Poincar\'e theorem, give $F-E+V=1+b^2(\Gamma)$.  Hence \be
\Omega(\Gamma,G)=(\dim G)\ b^2(\Gamma).
\ee
That is, \emph{for simply connected foams, the divergence degree is -- indeed -- given by the number of bubbles of $\Gamma$}, times the dimension of $G$.
\chapter{The (co)homology of flat connections}\label{cohomology}

Unfortunately, this identity between divergence degree and number of bubbles (second Betti number) breaks down when $\G$ is not simply connected, because then the term $\dim\mathcal{F}_0(\Gamma,G)$ in the divergence degree intertwines the topological structure of $\Gamma$ with the non-Abelian structure of $G$. (In particular, it does not factorize as $\dim G$ times a function of $\Gamma$ as in \eqref{flattrivial}.) We will see in this section that, this notwithstanding, the divergence degree can still be seen as counting the number of bubbles -- only these are not subsets of $\Gamma$. Rather, they are \emph{twisted bubbles}, in the sense of twisted (co)homology. To develop this perspective, we need to introduce more concepts from discrete gauge theory. This is the goal of this chapter.  

\section{More on discrete gauge theory}\label{moreondiscretegauge}

Discrete gauge theory has been studied extensively in two dimensions, especially after the seminal works of Atiyah and Bott \cite{Atiyah2007}, Goldman \cite{Goldman1984} and Witten \cite{Witten1991}. In this preliminary section, we extend this framework from triangulated $2$-manifold to general foams.

\subsection{Further definitions}

In the previous chapter, we called an element of $\mathcal{A}(\Gamma,G):=G^E$ a (discrete) \emph{G-connection} on $\Gamma$, and defined the \emph{curvature map} $H:\mathcal{A}(\Gamma,G)\rightarrow G^F$ by 
\be
H(A)\,:=\, \Bigl( H_f(A)=\prod_{e\in\partial f} g_e^{\epsilon_{fe}} \Bigr)_{f\in\Gamma_2}, \qquad \epsilon_{fe}=\pm1.
\ee 
Let us now complete these definitions. 

\paragraph{Flat connections}

Let us say that $A\in\mathcal{A}(\Gamma,G)$ is \emph{flat} if $H(A)=\unit$, and denote $\mathcal{F}(\Gamma,G):=H^{-1}(\unit)$ the \emph{space of flat connections}. (From now on, we shall write $\phi$ for a generic element of $\mathcal{F}(\Gamma,G)$.) Moreover, in differential-geometric parlance we say that a connection $A$ is \emph{regular} if the differential $dH_A$ has maximal rank, i.e. 
\be
\rk dH_A=\dim G\cdot\min\{E,F\};
\ee
else we say that $A$ is \emph{critical}.

\paragraph{Parallel transport}
Let $A=(g_{e})_{e\in\G_{1}}$ be a connection on $\G$. If $v$ and $w$ are two vertices of $\Gamma$, and $\g=(e_{1}^{\eta_{1}},\dots,e_{n}^{\eta_{n}})$ is an edge-path connecting them, with $n_{i}=\pm1$ to take into account the orientations of the edges, we define the \emph{parallel transport operator} from $v$ to $w$ by
\be
P_{A}(v,w;\g):=\prod_{i=1}^{n}g_{e_{i}}^{\eta_{i}}.
\ee
If $A$ is flat, two paths in the same homotopy class yield the same parallel transport operator. 
\paragraph{Gauge transformations}

Connections are by nature acted upon by \emph{gauge transformations}. In this discrete setup, these correspond to the action of $G$ at the vertices of $\Gamma$, according to 
\be\label{gauge}
h\cdot A:= \bigl(h_{t(e)}\,g_e\,h^{-1}_{s(e)}\bigr)_{e\in\Gamma_1},
\ee
where $h=(h_v)_{v\in\Gamma_1}\in G^V$ and $s(e)$ and $t(e)$ denote the source and target vertices of $e$. For each connection $A$, the relation \eqref{gauge} defines a map $\gamma_A$ from the gauge group $\mathcal{G}:=G^V$ to $\mathcal{A}(\Gamma,G)$, the \emph{local gauge map}. The set of gauge transformations $\zeta_A:=\gamma_A^{-1}(A)$ fixing $A$ is the \emph{stabilizer} of $A$, and the image $\mathcal{O}_A:=\gamma_A(\mathcal{G})$ of $\gamma_A$ is the \emph{orbit} of $A$; they are subgroups of $\mathcal{G}$ and $\mathcal{A}$ respectively. A connection $A$ such that $\zeta_A$ is only the center $Z(\mathcal{G})$ of $\mathcal{G}$ is called \emph{irreducible}; else it is \emph{reducible}. Finally, two connections $A$ and $A'$ belonging to the same orbit are called \emph{gauge equivalent}.

\paragraph{Moduli space}

The space of gauge orbits of \emph{flat} connections is a very important object in two-dimensional gauge theory: it is the semiclassical limit of the Yang-Mills phase space. This space is usually called the \emph{moduli space of flat connections}. Here, we denote it
\be
\mathcal{M}(\Gamma,G):=\mathcal{F}(\Gamma,G)/G^V=\{\mathcal{O}_\phi\subset\mathcal{F}(\Gamma,G), \phi\in\mathcal{F}(\Gamma,G)\}.
\ee


\subsection{Reducing gauge symmetries}\label{reducing}

The moduli space of flat connections is best described after gauge transformations have been \emph{reduced}, in the following sense. 

Since gauge transformations multiply the elements $g_e$ of a connection $A=(g_e)_{e\in\Gamma_1}$ independently at each vertex, every connection $A$ is gauge equivalent to one which has $g_e=\unit$ on a subset $T$ of $\Gamma_1$ touching all the vertices of $\Gamma$ without forming loops -- a \emph{spanning tree}. In particular, the orbit of a flat connection on $\Gamma$ is the same as the orbit of a flat connection on $\Gamma/T$, and we can show that\footnote{Note however $\mathcal{F}(\Gamma/T,G)\neq\mathcal{F}(\Gamma,G)$. The relation \eqref{result} and the gauge-invariance of the amplitude shows that, unless the singularities are wild, $\dim\mathcal{F}(\Gamma/T,G)=\dim\mathcal{F}(\Gamma,G)-(\dim G)(V-1)$, or equivalently $\codim\mathcal{F}(\Gamma/T,G)=\codim\mathcal{F}(\Gamma/T,G)$.}
\be
\mathcal{M}(\Gamma/T,G)=\mathcal{M}(\Gamma,G).
\ee
The advantage of this reduction is that the action of gauge transformations on $\Gamma/T$ is simply the conjugation of each element $g_e$ of $A\in\mathcal{A}(\Gamma/T,G)$ by a single group element $h\in G$:
\be
h\cdot A:= \bigl(hg_e\,h^{-1}\bigr)_{e\in(\Gamma/T)_1}.
\ee
Understanding the orbits of this action is usually much simpler than that of $G^V$ with $V>1$.

Moreover, since the integrand in $\calZ_\tau(\Gamma,G)$ is gauge-invariant, we can safely replace $\Gamma$ by the foam $\Gamma/T$ to compute the amplitude:
\be
\calZ_\tau(\Gamma/T,G)=\calZ_\tau(\Gamma,G).
\ee
For this reason, hereafter (and unless explicitly stated) we shall only consider foams with one vertex, understanding that the reduction of the gauge symmetry has already been performed.

\subsection{Representations of the fundamental group}

There is another benefit to this reduction of gauge transformations: it provides a very useful algebraic interpretation of the space of flat connections. The latter is based on the one-to-one relationship between foams with a single vertex and group presentations, as follows.

If $\Gamma$ is a foam with a single vertex, one can immediately read off a presentation of the fundamental group $\pi_1(\Gamma)$ by associating a generator $a_e$ to each edge of $\Gamma$ and one relation per face:
\be\label{standardpresentation}
\pi_{1}(\Gamma)=\bra(a_{e})_{e\in\Gamma_{1}}\
\vert\ (\prod_{e}a_{e}^{\epsilon_{fe}})_{f\in\Gamma_{2}}=\unit \ket.
\ee
Reciprocally, a finite presentation of a group $\pi$ unambiguously determines a foam $\Gamma$. From a single vertex, draw an edge for each generator, and attach the faces according to the relators.\footnote{Note that trivial relations such as $aa^{-1}=\unit$ must not be eliminated from the presentation of $\pi$ for this duality to hold. An example of this issue is provided by the `dunce hat': while $\bra a\vert a^2a^{-1}=\unit\ket$ is obviously equivalent to $\bra a\vert a=\unit\ket$ as a group presentation, the corresponding $2$-complexes, the dunce hat and the disc respectively, are not.}

Furthermore, the obvious translation of the flatness conditions $\prod_{e\in\partial f} g_e^{\epsilon_{fe}}=\unit$ into relations in a presentation of $\pi_1(\Gamma)$ shows that a flat connection on $\Gamma$ can be seen as a homomorphism from $\pi_1(\Gamma)$ to $G$, hence that
\be \label{representation space}
\mathcal{F}(\Gamma,G)\simeq \Hom\bigl(\pi_1(\Gamma),G\bigr).
\ee
Equipped with this algebraic interpretation, the space of flat connections is referred to as the \emph{representation variety} of $\pi_1(\Gamma)$ into $G$. Moreover, the moduli space of flat connections 
\be \label{representation space}
\mathcal{M}(\Gamma,G)\simeq \Hom\bigl(\pi_1(\Gamma),G\bigr)/G.
\ee
is called the \emph{character variety} of $\pi_1(\Gamma)$.

\subsection{The geometry of $\mathcal{F}(\Gamma,G)$ and $\mathcal{M}(\Gamma,G)$}

Unless all flat connections are regular, in which case the inverse function theorem proves that $\mathcal{F}(\Gamma,G)=H^{-1}(\unit)$ is a smooth submanifold of $\mathcal{A}(\Gamma,G)$, the space of flat connections has singularities. Although little is known about the detailed structure of these singularities in the general case, it is easy to see that $\mathcal{F}(\Gamma,G)$ has the structure of an \emph{affine algebraic set} whenever $G$ is algebraic: the flatness relations $H_f(\phi)=\unit$ are then polynomial equations in some affine space. This has various consequences.

First, an algebraic set consists of several \emph{irreducible components}, which by definition do not decompose as the union of smaller algebraic sets. Each irreducible component, also called an \emph{algebraic variety} $\mathcal{V}$, has a well-defined dimension, in spite of the fact that it is not a manifold, viz. contains singular points: $\dim\mathcal{V}$ is simply the dimension of a suitable neighborhood of any point where it \emph{is} a manifold. 

Second, the singularities of an algebraic variety $\mathcal{V}$ can be identified by computing the \emph{Zariski tangent space} $T_v\mathcal{V}$ at $v\in\mathcal{V}$. This tangent space is defined as the joint kernel of the differentials of all the polynomials which vanish on $\mathcal{V}$. It is naturally a vector space, and we have in general 
\be
\dim\mathcal{V}\leq\dim T_v\mathcal{V}.
\ee
The \emph{smooth}, or \emph{non-singular} points of $\mathcal{V}$ are those saturating this inequality.\footnote{And in fact we can define $\dim\mathcal{V}:={\min}_{y\in\mathcal{V}}\dim T_v\mathcal{V}$.} To get a feel for this definition, one can think of the cross $\{xy=0\}\subset\mathbb{R}^2$ in the plane considered in sec. \ref{tamesingularities}. At each point different from $(0,0)$, there is one tangent direction to the variety: the Zariski tangent space is one-dimensional. But at the origin, there are \emph{two} independent tangent vectors, spanning a \emph{two}-dimensional Zariski tangent space: $(0,0)$ is the unique singular point of the cross. The set of smooth points $\mathcal{V}_0$ of a variety $\mathcal{V}$, its \emph{smooth locus}, always forms a manifold that is dense\footnote{In the Zariski topology; we do not elaborate on this aspect here.} in $\mathcal{V}$. For instance, the smooth locus of the cross in $\mathbb{R}^2$ is the (disconnected) union of four half-lines. 

In the case of the space of flat connections $\mathcal{F}(\Gamma,G)$, one sees immediately that the smooth locus $\mathcal{F}_0(\Gamma,G)$ contains all the regular connections $\phi$: the map $H$ is submersive at $\phi$, hence by the inverse function theorem $\phi$ has a manifold neighborhood. In fact, the smooth locus is precisely the subset of $\mathcal{F}(\Gamma,G)$ where $H$ has maximal rank, and the Zariski tangent space is just $T_\phi\mathcal{F}(\Gamma,G)=\ker dH_\phi$.

Note also that the singularities of $\mathcal{F}(\Gamma,G)$ in the algebraic-geometric sense coincide with the analytic singularities of the function $S$ defined in sec. \ref{tamesingularities}. To see this, it is enough to check that
\be
\nabla^2S(\phi)=\f{1}{2}dH_\phi^\dagger dH_\phi,
\ee
hence that the degeneracy of $\nabla_\perp^2S$ corresponds to points where $\dim\ker dH_\phi$, the dimension of the Zariski tangent space, is larger than expected. 

Observe that the smooth locus is stable under the action of $G$ by conjugation: if $\phi$ is non-singular and $\phi'$ is gauge-equivalent to $\phi$, then $\phi$ is also non-singular. This fact turns $\mathcal{F}_0(\Gamma,G)$ into a $G$-manifold, and allows to describe the quotient $\mathcal{M}_0(\Gamma,G):=\mathcal{F}_0(\Gamma,G)/G$ as consisting of a disjoint union of manifolds characterized by their orbit type $H=\zeta_\phi$. Each such \emph{orbit stratum} $\mathcal{M}_0(\Gamma,G)_{(H)}$ has dimension $\dim\mathcal{M}_0(\Gamma,G)_{(H)}=\dim\mathcal{F}_0(\Gamma,G)-\dim G+\dim H$. In particular, if there are irreducible flat connections, they form the stratum with maximal dimensional dimension, or \emph{top stratum} of $\mathcal{M}_0(\Gamma,G)$.

\section{Twisted (co)homology}\label{}

We are now ready to introduce the theory of \emph{twisted cohomology}.\footnote{Although Bonzom and I were unaware of their work when we started using this tool, the relevance of twisted cohomology for the Ponzano-Regge model was first pointed out by Barrett and Naish-Guzman in \cite{Barrett:2009ys}.}

\subsection{Curvature as a coboundary operator}\label{curvatureascoboundary}

To understand the relationship between the space of flat connections and cohomology in more general terms than in sec. \ref{simplyconnected}, it is illuminating to compute explicitly the differential of the curvature map $H$ at $\phi=\unit$, the trivial connection. For $X=(X_e)_{e\in\Gamma_1}\in T_\unit\mathcal{A}(\Gamma,G)$, we find
\be\label{diffunit}
dH_\unit(X)=\Big(\sum_{e\in\pp f}\epsilon_{ef}X_e\Big)_{f\in\Gamma_2},
\ee
where $\pp f$ denotes the family of edges of $\Gamma$ lying on the boundary of the face $f$, and as before $\epsilon_{ef}$ is a sign encoding the relative orientation of $e$ and $f$. Now, \emph{this map $dH_{\unit}$ coincides with the first coboundary operator $\delta^1$ in the cellular cohomology $C^*(\Gamma,\alg)$ of $\Gamma$ with coefficients in $\mathfrak{g}$}, the Lie algebra of $G$ (see the appendix for a reminder on cellular cohomology):
\be\label{cochainwithcoeffs}
0\longrightarrow\mathfrak{g}\overset{0}{\longrightarrow}\mathfrak{g}^E\overset{\delta^1}{\longrightarrow}\mathfrak{g}^F\longrightarrow 0.
\ee
Here $\delta^0$ vanishes because $\Gamma$ has a single vertex. 

This observation provides further insight into the fact that for simply connected foams (which after gauge reduction admit a single flat connection, the trivial one), we have
\be
\Omega(\Gamma,G)=(\dim G)b^2(\Gamma)=b^2(\Gamma,\alg).
 \ee
Indeed, from $H_1(\Gamma)=\ker\delta^1/\im \delta^0=0$ we get $\ker\delta^1=\ker dH_\unit=0$, hence the flat connection $\unit$ is non-degenerate. The standard Laplace method then gives 
\be
\Omega(\Gamma,G)=(\dim G)F-\rk dH_\unit=(\dim G)F-\rk\delta^1=b^2(\Gamma,\alg).
\ee
That is, \emph{the corank of the curvature map $H$ is the second Betti number of the cochain complex} \eqref{cochainwithcoeffs}, hence counts the number of ``bubbles with coefficients in $\alg$". This explains the homological nature of the divergence degree $\Omega(\Gamma,G)$ for simply connected foams. 

\subsection{Interlude: twisted deRham cohomology}

If $M$ is a smooth $d$-manifold, it is well known that the exterior derivative $d^i:\Omega^i(M)\rightarrow\Omega^{i+1}(M)$ defines a cochain complex
\be
0\longrightarrow\Omega^0(M)\xrightarrow{d^0}\Omega^1(M)\xrightarrow{d^1}\dots\xrightarrow{d^{n-1}}\Omega^n(M)\xrightarrow{d^n}0,
\ee
whose cochain groups $\Omega^i(M)$ are the space of $i$-forms on $M$. Its cohomology $H_\textrm{dR}^*(M)$ is called the de Rham cohomology of $M$.\footnote{Physically, the first de Rham cohomology group $H^1_\textrm{dR}(M)$ is the space of Maxwell fields up to gauge; more generally,  $H^i_\textrm{dR}(M)$ is the solution space of `$i$-form electrodynamics' \cite{Henneaux1986}.}

Now, if $\phi$ is flat connection on a principal $G$-bundle $P\rightarrow M$, this construction extends to forms over $M$ valued in the adjoint bundle $\Ad(P)=P\times_{\Ad}\alg$, by means of the covariant exterior derivative $d^i_ \phi:\Omega^i\big(M,\Ad(P)\big)\rightarrow\Omega^{i+1}\big(M,\Ad(P)\big)$. Indeed, the flatness of $\phi $ entails $d_ \phi ^{i+1}d_\phi^{i}=[F(A),\ \cdot\ ]=0$, which means that
\be
0\longrightarrow\Omega^0\big(M,\Ad(P)\big)\xrightarrow{d^0_\phi}\Omega^1\big(M,\Ad(P)\big)\xrightarrow{d^1_\phi}\dots\xrightarrow{d^{n-1}_\phi}\Omega^n\big(M,\Ad(P)\big)\xrightarrow{d^n}0,
\ee
is also a cochain complex. The corresponding cohomology $H^*_ \phi\big(M,\Ad(P)\big)$ is the \emph{twisted} de Rham cohomology of $M$.\footnote{Twisted deRham cohomology describes non-Abelian gauge theory in a background field $\phi$. For instance, $H_ \phi ^{d-2}(M)$ is the space of solutions  up to gauge of the equation of motion $d_ \phi ^{d-2}B=0$ of BF theory \cite{Horowitz1989}.}


\subsection{Twisted cellular cohomology}

Just like de Rham cohomology, there is a twisted version of cellular cohomology. Here is one possible description of it. (The mathematically-oriented reader can consult Turaev's book \cite{Turaev2001} for a more algebraic presentation.)

Let $K$ be a finite-dimensional CW complex, with cochain groups $C^i(K)$, and let 
\be
C^i(K,\alg):=C^i(K)\otimes\alg\simeq\alg^{c_i(K)}.
\ee 
Here $c_i(K)$ is the number of $i$-cells of $K$. 

This space is the discrete counterpart of $\Omega^i\big(M,\Ad(P)\big)$: it consists linear combinations of $i$-cells of $K$ with coefficients in $\alg$. 

To describe the corresponding analogue of the covariant exterior derivative, first pick a reference vertex $v^i_\alpha$ on the boundary of each $i$-cell $e^i_\alpha$ of $K$. Then choose for each $(i+1)$-cell $e^{i+1}_{\beta}$ adjacent to $e^i_\alpha$ an edge-path connecting $v^i_{\alpha}$ and $v^{i+1}_{\beta}$ on the boundary of $e^{i+1}_{\beta}$. 

For each flat connection $\phi$, consider the parallel transport operator $P_\phi(v^i_\alpha,v^{i+1}_\beta)$ along a path of edges connecting $v^i_\alpha$ and $v^{i+1}_\beta$ on the boundary of $e^{i+1}_\beta$. (Since the connection is flat, the operator is independent of the chosen path.) Via the adjoint representation of $G$, these operators act on the cochain spaces $C^i(K,\alg)$. Use them to define the twisted coboundary operators $\delta_\phi^i:C^i(K,\alg)\rightarrow C^{i+1}(K,\alg)$ by
\beq
\delta_\phi^i(e^i_\alpha\otimes X) := \sum_{\beta}[e_\beta^{i+1},e_\alpha^i]\ \Big(e^{i+1}_\beta\otimes\Ad_{P_\phi(v^i_\alpha,v^{i+1}_\beta)}(X)\Big),
\ee
for $X\in\alg$. Here, $[e_\beta^{i+1},e_\alpha^i]$ is the incidence number of the pair of cells $(e_\beta^{i+1},e_\alpha^i)$ (the higher-dimensional analogue of the numbers $\epsilon_{ef}$ considered so far). 

It can then be checked that $\delta_\phi^{i+1}\circ\delta_\phi^i=0$, and that the \emph{$\phi$-twisted cellular cohomology groups} \be H_\phi(K,\alg):=\ker\delta_\phi^i/\im\delta_\phi^{i-1}\ee are well-defined, i.e. do not depend on the choice of reference vertices. Like the usual cohomology groups, they are homotopy invariants of $K$; their dimensions $b^i_\phi(K,\alg):=\dim H_\phi(K,\alg)$ are the \emph{$\phi$-twisted Betti numbers} of $K$. 

Let us emphasize that, in general, these are not simply related to the standard Betti numbers $b^i(K)$. There is such a simple relationship, however, for the \emph{twisted Euler characteristic}, which satisfies
\be\label{characteristic}
\chi_\phi:=\sum_{i=0}^d(-1)^i\dim C^i(K,\alg)=(\dim G) \chi(K).
\ee
By the Euler-Poincar\'e theorem, it follows that
\be
\sum_{i=0}^d(-1)^ib^i_\phi=(\dim G) \chi(K).
\ee

\subsection{Twisted cohomology in the flat model}\label{twistedinflat}

Twisted cellular cohomology arises naturally in the flat spinfoam model. Thus, a moment of reflection shows that the $0$-th twisted coboundary operator $\delta^0_\phi$ is nothing but the differential of the local gauge map $\gamma_\phi$ at the unit,
\be
\delta^0_\phi=(d\gamma_\phi)_\unit.
\ee
That is to say that the space of $1$-coboundaries $ \im \delta^0_\phi$ is the tangent space of the gauge orbit of $\phi$, and the space of $0$-cocycles $\ker\delta^0_\phi$ is the Lie algebra of its stabilizer $\zeta(\Gamma,G)_\phi$. In particular, $H^0_\phi$ vanishes if and only if $\phi$ is irreducible. 

Similarly, we can check that
\be
\delta^1_\phi=dH_\phi.
\ee
Hence, the space of $1$-cocycles $\ker\delta_\phi^1$ is the tangent space to $\mathcal{F}(\Gamma,G)$ at $\phi$, and the first cohomology group $H_\phi^1(\G,\alg)$ is the tangent space to $\mathcal{M}(\Gamma,G)$ at $\mathcal{O}(\Gamma,G)_\phi$. Moreover, $H^2_\phi=0$ means that $\rk dH_\phi=(\dim G)F$, i.e. that $\phi$ is regular.\footnote{The reader might wonder whether the higher coboundary operators also have a gauge-theoretic interpretation. From the perspective of discrete connections on a foam, the answer is negative. However, the flat spinfoam model can be analyzed as a discrete BF model, with an additional variable in $\alg$ on each face of $\G$; the higher coboundary operators turn out to describe the (reducible) shift gauge symmetry of the discrete $B$ field. We will detail this point of view in \cite{Bonzom}.}

Note that, by definition of non-singular connections, the second twisted Betti number $b^2_\phi$ has a constant value 
\be
b^2_0:=\min_{\phi\in\mathcal{F}(\Gamma,G)}b^2_\phi
\ee 
on $\mathcal{F}_0(\Gamma,G)$, given by
\be
b^2_0=(\dim G)F-\rk dH_\phi=(\dim G)(F-E)+\dim\mathcal{F}_0(\Gamma,G).
\ee
Comparing with the result \eqref{firstresult}, we see that
\be
\Omega(\Gamma,G)=b^2_0.
\ee
In words: \emph{(provided the singularities of the curvature map are tame) the divergence degree of a foam is given by the value of its second twisted Betti number on non-singular connections}. This is the main result of this chapter, and the precise sense in which divergences in the flat model are counted by ``bubbles". 

\section{Examples}\label{}

Let us now illustrate this result with several examples of interest. 

\subsection{Abelian structure group}

Consider first the case where the structure group $G$ is \emph{Abelian}. In this case the curvature map $H$ is a group homomorphism $G^E\rightarrow G^F$, hence
 \begin{itemize}
\item
the space of flat connections $\mathcal{F}(\Gamma,G)=\ker H$ is a Lie subgroup of $\mathcal{A}(\Gamma,G)=G^E$, and in particular has no singularity,
\item
the relation \eqref{diffunit} holds for \emph{any} flat connection $\phi\in\mathcal{F}(\Gamma,G)$, i.e. \be \delta^1_\phi(X)=dH_\phi(X)=\Big(\sum_{e\in\pp f}\epsilon_{ef}\theta(X_e)\Big)_{f\in\Gamma_2},\ee where $X=(X_e)_{e\in\Gamma_1}\in T_\phi G^E$ and $\mu:T_\phi G\rightarrow\alg$ is the Maurer-Cartan form on $G$. In other words,
\be
\delta_\phi^1(X)=\delta^1_\unit\big(\mu(X_{1}),\dots,\mu(X_{E})\big).
\ee
\end{itemize}\medskip
Both observations imply that $b^2_\phi=\corank\delta^1_\phi$ is constant on $\mathcal{F}(\Gamma,G)$, equal to its value on the trivial connection $\unit$. But we saw in the previous section that the twisted cochain complex of $\Gamma$ at $\phi=\unit$ is just the cellular cochain complex of $\Gamma$ with coefficients in $\alg$, hence
\be
\Omega(\Gamma,G)=b^2_0=b^2(\Gamma,G)=(\dim G)b^2(\Gamma). 
\ee
Thus, for Abelian groups like for simply connected foams, the relevant ``bubbles" are the cellular $2$-cycles of $\Gamma$. The initial intuition that, in the flat spinfoam model, powercounting amounts to the chasing of ``closed surfaces'' within $\G$ is correct is this case too.

\subsection{Surface groups with $G=\SU(2)$}\label{surfaceexamples}

Assume now that $\Gamma$ is the cellular decomposition $\Gamma_k$ of a (closed, orientable) \emph{surface} with genus $k$, and that $G=\SU(2)$. This is the case studied by Witten in the context of two-dimensional Yang-Mills theory \cite{Witten1991}, and by Goldman \cite{Goldman1984} and Atiyah and Bott \cite{Atiyah2007} from the mathematical perspective. 

As well-known, $\Gamma_k$ can always be reduced to a flower graph with $2k$ edges supporting only one face. This corresponds to the following presentation of the fundamental group:
\be
\pi_1(\Gamma_k) = \bra a_1,\,b_1,\dotsc,a_k,\,b_k\ \vert\ [a_1,b_1]\dotsm [a_k,b_k]=\unit \ket.
\ee
Here the square brackets denote the group commutator in $\pi_1(\Gamma_g)$, $[a,b] := aba^{-1}b^{-1}$. 

The space ${\calF}_k := \calF(\Gamma_k,\SU(2))$ of discrete $\SU(2)$-connections on $\G_{k}$ can therefore be described as
\be\label{flatsurface}
{\calF}_k =\bigl\{ (a_1,\,b_{1},\dotsc,a_k,\,b_k)\ \in\,\SU(2)^{2k},\ [a_1,b_1]\dotsm [a_k,b_k]=\unit \bigr\}.
\ee
Geometrically, the group elements $a_i$ and $b_i$ can be thought of as rotations with the same axis of rotation $\hat{n}_i\in S^2$, but $a_i$ and $b_j$ can have different axes for $i\neq j$. Computing the action of $\SU(2)$ on $\calF_{k}$ by conjugation, we find that a gauge transformation is a rotation of these axes.

It is not difficult to convince oneself that there are singular connections in ${\calF}_k$, as the differential of the flatness relation is clearly not of constant rank. Yet, one circumstance greatly simplifies the description of $\calF_k$ with respect to the general case: its singularities are classified by the orbit type $\zeta_\phi$. This follows from Poincar\'e duality, which states that $H^2_\phi\simeq H^0_\phi$. Hence, in this case a flat connection $\phi$ is non-singular ($H^2_\phi=0$) if and only if it is irreducible ($H^0_\phi=0$). 

To make this observation more concrete, we now describe explicitly the Zariski tangent space $\ker dH_\phi$, for $\phi\in\calF_{k}$. We treat the cases $k=1$ (the torus) and $k\geq2$ separately (the sphere, which is simply connected, was already treated in sec. \ref{curvatureascoboundary}).


\paragraph{Higher genera ($k\geq2$)} Identifying the tangent space $T_\phi\mathcal{A}(\Gamma,G)$ with its Lie algebra $\alg^E$ by means of the Maurer-Cartan form, we have in this case
\be\label{diffhigher}
dH_\phi(u,v)=\delta^1_\phi(u,v) = \sum_{i=1}^g \bigl(1-\Ad_{b_i}\bigr)\,u_i - \bigl(1-\Ad_{a_i}\bigr)\,v_i.
\ee
for algebra elements $u:=(u_i)_{i=1,\dotsc,k}\in\alg^E$ and $v:=(v_i)_{i=1,\dotsc,k}\in\alg^E$. Considering $u_i$ and $v_i$ as vectors in three-dimensional space, the adjoint action $\Ad_{a_i}$ is a rotation around the axis $\hat{n}_i$. Thus, for each $i$, it is natural to decompose $u_i$ and $v_i$ into orthogonal and parallel components to the direction $\hat{n}_i$, viz.
\be
u_i = u_{i}^\parallel +u_{i}^\perp,
\ee 
and similarly for $v_i$. 

All parallel components $u_{i}^\parallel$ drop out from \eqref{diffhigher}, since $(1-\Ad_{a_i})v_i = (1-\Ad_{a_i})v_{i}^\perp$. Moreover, the latter quantity belongs to the orthogonal plane to $\hat{n}_i$, where $(1-\Ad_{a_i})$ is invertible. This means that, varying $u_i$ and $v_i$, each term of the sum in \eqref{diffhigher} spans the orthogonal plane to $\hat{n}_i$. Finally, if at least two axes among the $k$ directions are distinct, then the span of $\delta^1_\phi(u_i,v_i)$ is the whole algebra $\alg$. 

Hence, there exist flat connections where the curvature map is submersive, i.e. there are regular flat connections. (Note that they are irreducible.) Consequently, by the implicit function theorem, the non-singular flat connections form a smooth manifold of dimension
\be
\dim \calF_0 = \dim \ker\delta^1_{\vert\calF_0} = 6k - 3,
\ee
and therefore
\be
\Omega\big(\Gamma_k,\SU(2)\big)=b^2_{0} = 0.
\ee 
Thus, we find that the regularized amplitude 
\be
\calZ_\tau(\Gamma_g,\SU(2)) = \int \prod_{i=1}^g da_i\,db_i\ \k\bigl([a_1,b_{1}]\dotsm [a_g,b_g]\bigr)
\ee
has a finite limit as $\tau\rightarrow0$. This is consistent with Witten's formula \cite{Witten1991}
\be\label{witten2d}
\underset{\tau\rightarrow0}{\lim}\ \calZ_\tau\big(\Gamma_k,\SU(2)\big)= \bigl(\Vol G\bigr)^{2k} \sum_{n\geq 1} n^{-(2k-2)}=\bigl(\Vol G\bigr)^{2k}\zeta(2k-2).
\ee
We refer to the work of Sengupta \cite{Sengupta2001} for details on why singularities are tame in this case.

Let us also say a word about the singular connections. In this case, they are of two kinds:
\medskip\begin{itemize}
\item
\emph{Abelian connections}. All group elements have the same axis $\hat{n}$. These form a submanifold of dimension $2k+2$, on which $\rk\delta^1_\phi = 2$. Such flat connections have a $\textrm{U}(1)$ stabilizer, corresponding to rotations around $\hat{n}$.
\item
\emph{Central connections}. All group elements are in the center of $\SU(2)$, i.e. $a_i,b_i=\pm\unit$. They are isolated points, where $\rk\delta^1_\phi = 0$. (Indeed, $\Ad_{\pm\unit}=1$, hence $dH_\phi=0$.) They are completely reducible, viz. $\zeta_\phi=\SU(2)$. 
\end{itemize}\medskip 
These observations are consistent Poincar\'e duality: the more a flat connection is reducible, the more it is singular.

\paragraph{The torus ($k=1$)} The case of the torus is more subtle, because all flat connections are critical, and also more interesting, because $\Omega\big(\Gamma_{k=1},\SU(2)\big)$ is not a multiple of $\dim G=3$.

To see this, recall first that the curvature map in this case is a single group commutator, $(a,b)\mapsto [a,b]=aba\mone b\mone$, hence flat connections $\phi$ consist in pairs of rotations with angles $0\leq\psi_{a},\psi_{b}<\pi$ and the same axis $\hat{n}\in S^2$. Moreover, the coboundary operator $\d_\phi^1$ (again pulled back to the Lie algebra $\alg^2$) reads
\be\label{difftorus}
\d_\phi^1= d\bigl([a,b]\bigr) = \bigl(1-\Ad_b\bigr)- \bigl(1-\Ad_a\bigr).
\ee
As before, let us distinguish between the Abelian and central connections:
\begin{itemize}
\item
\emph{Abelian connections.} Assume that either $a$ or $b$ (say $b$) is not $\pm\unit$. Then the operators $(1-\Ad_a)$ and $(1-\Ad_b)$ in \eqref{difftorus}, seen as linear maps on $\alg\simeq\mathbb{R}^3$, have the same one-dimensional kernel, namely the direction parallel to $\hat{n}$: this corresponds to variations of the angles $\psi_a$, $\psi_b$ for a fixed $\hat{n}$. Restricted to the orthogonal plane, the map $(1-\Ad_b)$ is invertible, hence the equation for the kernel of $\delta^1_\phi$ reads 
\be\label{difftorus}
u_{a}^\perp = (1-\Ad_b)^{-1} (1-\Ad_a) u_{b}^\perp,
\ee
and therefore fixes $2$ real components of $(u_a,u_b)\in\mathbb{R}^3\times\mathbb{R}^3$. It expresses the condition that the connection remains flat under variations of the directions of $a$ and $b$. It follows that $\rk\delta^1_\phi=2$, hence that $b^2_\phi=3-2=1$.
\item
\emph{Abelian connections.} If on the other hand  both $a$ and $b$ are $\pm\unit$, , the differential \eqref{difftorus} vanishes, hence $\rk\delta^1_\phi=0$, i.e. $b^2_\phi=3$. 
\end{itemize}
Thus, the singular connections are the central ones, and  
\be
\Omega\big(\Gamma_k,\SU(2)\big)=b^2_{0} = 1.
\ee
This confirms that the torus is divergent, as suggested by \eqref{witten2d}.

\subsection{A foam with wild singularities}\label{}

We close this section by an example of a foam presenting \emph{wild} singularities, still with $G=\SU(2)$. Consider the group\footnote{We do not know if this is the fundamental group of a $3$-manifold. (But it is that of a $4$-manifold, like any other finitely generated group.)}
\beq
\pi = \langle a,b,h\,\vert\,[a,h] = [b,h] =\unit \rangle.
\ee
The set of flat connections on the corresponding foam $\Gamma$ with three edges and two faces is determined by the relations
\beq
\calF = \left\{ (a,b,h)\in\SU(2)^3,\ [a,h] = [b,h] =\unit \right\}.
\ee
This set has two irreducible components. 

\begin{itemize}
\item
If $h$ is in the center of $\SU(2)$, i.e. $h=\pm \unit$, then, $a,b$ can be arbitrary:
\beq
\calF_{\operatorname{irred}} := \left\{ (a,b,\pm\unit),\ (a,b)\in\SU(2)^2\right\}.
\ee
These are the irreducible flat connections. 
\item
If $h$ is not in the center, then $a,b$ and $h$ have to lie in a common $\textrm{U}(1)$ subgroup of $\SU(2)$. These are the Abelian representations. If we write a generic element $g=\exp(i\psi \hat{n}\cdot\vec{\sigma})$, with $\hat{n}\in S^2$ the direction of the rotation and $0\leq\psi\leq\pi$ its class angle, then
\be
\calF_{\operatorname{red}} := \left\{\left(a=\exp(\pm i\psi_a \hat{n}_{h}\cdot\vec{\sigma}), b=\exp(\pm i\psi_b \hat{n}_{h}\cdot\vec{\sigma}), h=\exp(i\psi_h \hat{n}_{h}\cdot\vec{\sigma})\right)\right\},
\ee
with $0\leq\psi_{a},\psi_{b},\psi_{h}<\pi$ and $\hat{n}_{h}\in S^{2}$.
\end{itemize}
Quite obviously, $\calF_{\operatorname{irred}}$ is of dimension $6$, while $\calF_{\operatorname{red}}$ is $5$-dimensional.\footnote{Recall that, by definition, these are the dimensions of the kernel of $dH_\phi$ on non-singular flat connections.} It follows that 
\beq
b^2_{\operatorname{irred}}(\Gamma,\SU(2)) = 3, \qquad\text{and}\qquad b^2_{\operatorname{red}}(\Gamma,\SU(2)) = 2.
\ee
Applying Laplace's method as in sec. \ref{firstresults} therefore suggests that 
\be
\mathcal{A}(\Gamma,\SU(2))\underset{\tau\rightarrow0}{\sim}\calZ_{\tau}(\Gamma,\SU(2))_{\operatorname{irred}}+\calZ_{\tau}(\Gamma,\SU(2))_{\operatorname{red}}
\ee
with
\begin{align}
\calZ_{\tau}(\Gamma,\SU(2))_{\operatorname{irred}}&\underset{\tau\rightarrow0}{\sim}\L_{\tau}^{3}\mathcal{Z}'_{\textrm{irred}}(\Gamma,\SU(2))\\
\calZ_{\tau}(\Gamma,\SU(2))_{\operatorname{red}}&\underset{\tau\rightarrow0}{\sim}\L_{\tau}^{2}\mathcal{Z}'_{\textrm{red}}(\Gamma,\SU(2)).
\end{align}
In other words, it suggests that the divergence of the amplitude is controlled by \emph{irreducible} connections, and $\Omega(\Gamma,\SU(2)) = 3$. However, it so happens that \emph{reducible} connections actually spoil this result, because $\mathcal{A}'_{\textrm{red}}(\Gamma,\SU(2))=\infty$.  This was showed in \cite{Bonzom:2010uq} by direct computation of the contribution of $\calF_{\textrm{red}}$ to the Laplace asymptotics of
\be\label{reminderint}
\mathcal{Z}_{\tau}\big(\G,\SU(2)\big)=\int_{\SU(2)^{3}}dadbdh\, K_{\tau}([a,h])K_{\tau}([b,h]).
\ee
Explicitly, we showed in this reference that
\beq
\mathcal{Z}_{\textrm{red}}\big(\Gamma,\SU(2)\big)\underset{\tau\rightarrow0}{\sim}\f 14\L_{\tau}^{2}\ \Vol(S^2)\ \int_{[0,\pi]^3} \f{d\psi_a\ d\psi_b\ d\psi_h}{\sin^2\psi_h}.
\ee
The factor $\Vol(S^2)$ corresponds to the integral over gauge orbits, which here correspond to rotations of the common direction $\hat{n}_h$ of the three group elements, the class angles remaining fixed; the class angles $\psi_a, \psi_b$ and $\psi_h$ on the other hand parametrize the moduli space. This integral is manifestly \emph{divergent}. This is strictly analogous to the wild singularity discussed in sec. \ref{tamesingularities}, and shows that the divergence degree of $\Gamma$ cannot be estimated by Laplace's method.

\chapter{Applications of homological powercounting}\label{applications}

We found in the previous chapter that, except for a caveat related to possibly wild singularities in the space of flat connections $\calF(\G,G)$, the divergence degree of a foam $\G$ is the second twisted Betti number of $\G$, evaluated on the non-singular elements of $\calF(\G,G)$. We commented on the interpretation of this result as the actual realization of the idea that the divergences of $\calZ(\G,G)$ can be traced back to the presence of ``bubbles''. 

In this chapter, we discuss another aspect of this result: the clarification, unification and generalization of \emph{all} the earlier results on divergences in the Ponzano-Regge and Ooguri models. 

\section{Sorting out topology from cell structure}

One such clarification concerns the extent to which the divergence degree of a foam is determined by the topology of spacetime. We now derive a decomposition of $\Omega(\G,G)$ which completely settles this question.

\subsection{A useful decomposition}

Part of the interest of the powercounting result obtained in the previous chapter is that it holds for \emph{arbitrary} foams. Suppose, however, that $\Gamma$ is actually the $2$-skeleton of a cell decomposition $K_M$ of a $n$-pseudomanifold $M$.\footnote{This includes the manifolds usually considered in general relativity, but also the more singular spaces arising as Feynman diagrams of a group field theory.} Then the result $\Omega(\G,G)=b_{0}^{2}$ can be sharpened considerably. 

Consider indeed the twisted Euler-Poincar\'e formula for the $2$-skeleton $\Gamma$,
\be
\chi_\phi=b_\phi^0-b_\phi^1+b_\phi^2.
\ee
From the discussion in sec. \ref{twistedinflat}, we know that $b^0_\phi$ is the dimension of the stabilizer $\zeta_{\phi}$, and that $b_\phi^1$ is the local dimension of the moduli space $\mathcal{M}(\G,G)$. Combining this with \eqref{characteristic}, we get
\be
\Omega(\Gamma,G)=\dim\mathcal{M}(\Gamma,G)-\dim\zeta_{0}+(\dim G)\chi(\Gamma),
\ee
where $\zeta_{0}$ is the common stabilizer of non-singular flat connections $\phi\in\calF_{0}(\G,G)$. Moreover, by definition of the Euler characteristic, we have
\be
\chi(\Gamma)=V-E+F=\chi(K_M)-\sum_{i=3}^n(-1)^ic_i(K_M),
\ee
where $c_{i}(K_{M})$ is the number of $i$-cells in $K_{M}$. Hence,
\be\label{resultat}
\Omega(\Gamma,G)=\dim\mathcal{M}(\Gamma,G)-\dim\zeta_{0}+(\dim G)\chi(K_M)-(\dim G)\sum_{i=3}^n(-1)^ic_i(K_M).
\ee
Thus, we have a decomposition
\be\label{decomp}
\Omega(\Gamma,G)=I(M,G)+\omega(K_M,G),
\ee
where
\be\label{topological}
I(M,G):=\dim\mathcal{M}(\Gamma,G)-\dim\zeta_{0}+(\dim G)\chi(M).
\ee
is the `topological' part of the divergence degree, depending only on the homeomorphism (and in fact homotopy) class of $M$, and
\be\label{cellular}
\omega(K_M,G):=-(\dim G)\sum_{i=3}^n(-1)^ic_i(K_M)
\ee
is its `cellular' part, depending on the given cell decomposition $K_M$ of $M$.\footnote{Note that this decomposition actually holds for any $n$-dimensional cell complex. The restriction to pseudo-manifolds is made for comparison with the literature and as a natural choice in group field theory.} This decomposition ``sorts out topology from cell structure" in the divergence degree \cite{springerlink:10.1007/s00023-011-0127-y}.

\subsection{On the topological part of $\Omega(\G,G)$}\label{computational}

The above decomposition should be contrasted with the result obtained in sec. \ref{veryfirstresult}, where $\Omega(\G,G)$ was written as the sum of a term proportional to $(F-E)$ (the number of ``redundant'' delta functions in the definition of the flat spinfoam model), and of $\dim\calF_{0}(\G,G)$ (the dimension of the space of flat connections). 

Indeed, unlike $\dim\calF_{0}(\G,G)$ in \eqref{firstresult}, the topological part $I(M,G)$ in \eqref{decomp} only depends on the fundamental group $\pi_{1}(M)$ (and the homotopy class of $M$ via the Euler characteristic $\chi(M)$). 

This fact makes the computation of the divergence degree much easier in the form \eqref{decomp}: in practice, it reduces the problem of computing $I(M,G)$ for a given (pseudo)manifold $M$ to that of identifying the representation variety of its fundamental group $\Hom\bigl(\pi_1(M),G\bigr)$ and computing the the orbits of $\phi\in\Hom\bigl(\pi_1(M),G\bigr)$ under the conjugation action. Since these only depend on $\pi_{1}(M)$, \emph{any} presentation of $\pi_{1}(M)$ can be used to that effect; in particular, from the standard presentation \eqref{standardpresentation} obtained by retracting a tree $T$ in $\G$, as in sec. \ref{reducing}.

%

One more comment about the computation of $I(M,G)$. It may happen that the character variety $\calM(\G,G)=\Hom\bigl(\pi_1(\Gamma),G\bigr)/G$ has several irreducible components. These lead to \emph{different} values of $I(M,G)$; in $\tau\arr0$ limit, the leading behaviour of the amplitude $\calZ_\tau(\Gamma,G)$ is of course given by the largest one. This phenomenon will be illustrated in sec. \ref{3dexamples}.

\section{Comparison with previous results}

Prior to our introduction of twisted cohomology in the discussion of powercounting in the flat model, several results had been presented in the literature, mostly within the context of the Boulatov and Ooguri models, where foams are ``simplicial'': in dimension $d$, exactly $(d+1)$ edges (resp. $d$ faces) are incident on each vertex (resp. edge) of $\Gamma$, corresponding to the number of faces of a $d$-simplex (resp. $(d-1)$-simplex).

These results fall in four classes:

\medskip\begin{itemize}
\item
\emph{Powercounting from Pachner moves}. If one assumes that $\Gamma$ arises from a triangulation $T_M$ of a closed $n$-manifold $M$, as the $2$-skeleton of its dual cell complex, one can see that Pachner moves on $T_M$ generate divergences in the partition function. Ponzano and Regge \cite{PR}, Boulatov \cite{Boulatov:1992vp} and Ooguri \cite{Ooguri:1992eb} relied on this observation to relate the divergence degree to the combinatorics of $\Gamma$.
\item
\emph{Powercounting from bubble counting}. A different approach to the divergences of the flat spinfoam model was initiated by Perez and Rovelli \cite{Perez:2000fs}, who realized that they are related to the presence of ``bubbles" within $\Gamma$. Freidel and Louapre \cite{Freidel2002,Freidel2004}, and later Freidel, Gurau and Oriti \cite{Freidel:2009hd}, pushed this intuition further in three-dimensions and obtained a powercounting estimate for certain special complexes, coined ``type 1". Within Gurau's colored tensor models, this result was then extended to higher dimensions by Ben Geloun \emph{et al.} \cite{Geloun:2010vn}.

\item
\emph{Powercounting from vertex counting}. Yet another approach relies on the field-theoretic notion that the divergence degree of a Feynman diagram can be bounded by the number of its vertices. Together with Magnen, Noui and Rivasseau, I obtained such bounds in the Boulatov model \cite{Magnen:2009dq}; these were adapted to the colored tensor models in \cite{BenGeloun2010}.

\item
\emph{Powercounting from jackets}. The notion of ``jacket" for a (colored) Boulatov-Ooguri complex was introduced by Ben Geloun \emph{et al.} in \cite{Geloun:2010vn}, and used by Gurau and Rivasseau \cite{Gurau2011a,Gurau2011b} to obtain an upper bound on the divergence degree improving the one obtained by vertex counting mentioned above.
\end{itemize}

In this section, we explain how our main formula \eqref{decomp} encompasses all the results quoted above. One important should be stressed: except for the perturbative bounds obtained from vertex counting, these results are \emph{all} based on the (usually implicit) assumption that $\Omega(\Gamma,G)$ is an integer multiple of $\dim G$. An immediate consequence of \eqref{topological} and \eqref{cellular}, however, is that this is not true in general, because neither $\dim\mathcal{M}(\Gamma,G)$ nor $\dim\zeta(\Gamma,G)$ in the topological part $I(M,G)$ are multiples of $\dim G$.

\subsection{Powercounting from Pachner moves}

When Ponzano and Regge introduced their model in 1968, they considered a $3$-manifold $M$ equipped with a triangulation $T_M$, $G=\SU(2)$, and conjectured that the divergence degree was given by three (i.e. $\dim \SU(2)$) times the number of vertices $V(T)$ in the triangulation. They were guided in making this conjecture by the following consideration: because of the Biedenharn-Elliot identity for $\{6j\}$-symbols, the formal amplitudes of two triangulations $T_M$ and $P_{14}T_M$ of $M$ related by a $1-4$ Pachner move satisfy
\be
\mathcal{Z}_{\textrm{PR}}\big(P_{14}T_M\big)=\Big(\sum_{j=0}^{\infty}(2j+1)^2\Big)\mathcal{Z}_{\textrm{PR}}\big(T_M\big).
\ee
Since $\sum_{j=0}^{\Lambda}(2j+1)^2$ scales as $\Lambda^3$ when $\Lambda\rightarrow\infty$, and since the $1-4$ move introduces a single new vertex in the triangulation, the conclusion that each vertex contributes a factor of $\Lambda^3$ in the amplitude appears tantalizing. Indeed, Ponzano and Regge proposed to cut off the sums in their state-sum to a maximal value $\Lambda$ to obtain a finite value $\mathcal{Z}_{\textrm{PR},\Lambda}\big(T\big)$ for the amplitude, and tentatively defined a regularized amplitude by
\be
\mathcal{Z}'_{\textrm{PR}}(T_M)=\underset{\Lambda\rightarrow\infty}{\lim}\ \Lambda^{-3V(T_M)}\mathcal{Z}_{\textrm{PR},\Lambda}\big(T_M\big).
\ee

The origin of the intuition that divergences in the Ponzano-Regge model are related to the vertices of $T_M$, but also the reason why this naive regularization is bound to fail, is completely elucidated by our decomposition \eqref{decomp}. Indeed, taking $K_M$ as the dual cell complex to a triangulation $T_M$, and noting that in three dimensions, a vertex is dual to a $3$-cell, we see that
\be
\omega(T_M,G)=3c_3(K_M)=3V(T_M).
\ee
That is, the relationship between divergences of $\mathcal{A}(T_M)$ and vertices of $T_M$ conjectured by Ponzano and Regge is exact for the non-topological part of $\Omega(T_M,G)$, but misses completely its topological part $I(M,G)$. This is because the argument based on the $1-4$ Pachner move actually estimates not the divergence degree itself, but its \emph{variation} in the move, that is
\be
\Omega(P_{14}T_M)-\Omega(T_M)=\omega(P_{14}T_M)-\omega(T_M)=3\big(V(P_{14}T_M)-V(T_M)\big).
\ee
In this variation, $I(M)$ \emph{cancels}.


A similar observation was made in the four-dimensional case by Ooguri in \cite{Ooguri:1992eb}. He considered separately the $1-5$ move (in which the number of vertices and edges increase by one and five respectively) and the $3-3$ move (in which the number of vertices is left unchanged while the number of edge is increased by one), and found that $\mathcal{Z}_{O}(T_M)$ depends on the triangulation $T_M$ only through a divergent factor measured by $3\big(E(T_M)-V(T_M)\big)$, where $V(T_M)$ and $E(T_M)$ are the number of vertices and edges of $T_M$. Although he could not show this result rigorously (because he could not make sense of the amplitude as a finite number), we can interpret his argument along the same lines as in three dimensions: it provides the correct non-topological divergent degree $\omega(T_M)$, but misses the topological part $I(M)$. Indeed, in the dual complex $K_M$ to $T_M$, there is one $4$-cell per vertex, and one $3$-cell per edge in $T_M$, so that
\be
\omega(K_M)=3\big(c_3(K_M)-c_4(K_M)\big)=3\big(E(T_M)-V(T_M)\big).
\ee

\subsection{Powercounting from bubble counting}

Freidel, Gurau and Oriti defined in \cite{Freidel:2009hd} a restricted class of foams $\Gamma$, which they called ``type $1$", for which they could show that the divergence degree is simply related to the number $B(\Gamma)$ of ``bubbles" in $\Gamma$,
\be\label{FGObubbles}
\Omega(\Gamma)=3\big(B(\Gamma)-1\big).
\ee
With the view that $\Gamma$ is the $2$-skeleton of a cell complex $K_M$ decomposing a $3$-pseudomanifold $M$, the ``bubbles" of $\Gamma$ in the sense of Freidel \emph{et al.} correspond to the $3$-cells of $K_M$, hence \eqref{FGObubbles} reads
\be\label{FGO}
\Omega(K_M)=3(c_3(K_M)-1).
\ee
Taking their cues from the Turaev-Viro model, they speculated that the ``type $1$" condition should correspond to ``topologically trivial manifolds", i.e. to the $3$-sphere. Given the decomposition \eqref{decomp}, the identity \eqref{FGO} is equivalent to
\be
I(M)=\dim\mathcal{M}(\Gamma,G)-\dim\zeta(\Gamma,G)=-3.
\ee
In other words, the ``type 1" complexes are, for a given number of $3$-cells in the corresponding complex $K_M$, are \emph{minimally} divergent. Moreover, they are such that their flat connections up to gauge are isolated and completely reducible. We will see in the next section that this last condition alone does not single out the $3$-sphere (it is also satisfied \emph{e.g.} by the real projective space $\mathbb{R}P^3$).

In \cite{Geloun:2010vn}, Ben Geloun \emph{et al.} considered Gurau's \emph{colored model}, and used the corresponding ``bubble homology" to obtain the following formula for the divergence degree in dimension $d\geq3$ for the \emph{Abelian} structure groups $G=\mathbb{R}$:\footnote{That $\mathbb{R}$ is non-compact introduces another source of divergences, which can be tamed with a second cutoff.}

\be\label{linearized}
\Omega(\mathcal{G})=\sum_{k=3}^{d+1}(-1)^{k-1}c_k(\mathcal{G})+\sum_{k=2}^{d-1}(-1)^kb_k(\mathcal{G}).
\ee
Here, $\mathcal{G}$ is a $(d+1)$-colored graph, $c_k(\mathcal{G})$ is for $0\leq n\leq d$ the number of $k$-bubbles in $\mathcal{G}$ , $c_{d+1}(\mathcal{G}):=1$ and $b_k(\mathcal{G})$ is the $k$-th ``bubble" Betti number of $\mathcal{G}$. All the details are given in the appendix.

The relationship with our result is based on the following correspondence. A $(d+1)$-colored graph $\mathcal{G}$ naturally defines a $d$-dimensional CW complex $\Delta_{\mathcal{G}}$ with the following property:
\medskip\begin{itemize}
\item
the underlying graph of $\mathcal{G}$ is the $1$-skeleton of $\Delta_{\mathcal{G}}$,
\item
the number of $k$-bubbles of $\mathcal{G}$ equals the number of $k$-cells of $\Delta_{\mathcal{G}}$,
\item
the ``bubble" homology of $\mathcal{G}$ coincides with the cellular homology of $\Delta_{\mathcal{G}}$.
\end{itemize}
With this correspondence, the relationship between \eqref{linearized} and our result \eqref{resultat} is straightforward. First, note that what is denoted $c_{d+1}(\mathcal{G})$ in \eqref{linearized} is really $b_d(\mathcal{G})=1$, so that it can (and should) be rewritten

\be
\Omega(\mathcal{G})=\sum_{k=3}^{d}(-1)^{k-1}c_k(\mathcal{G})+\sum_{k=2}^{d}(-1)^kb_k(\mathcal{G}).
\ee
Then, use the identities $c_k(\mathcal{G})=c_k(\Delta_{\mathcal{G}})$ and $b_k(\mathcal{G})=b_k(\Delta_{\mathcal{G}})$ and the Euler-Poincar\'e relation
\be
\chi(\Delta_{\mathcal{G}})=\sum_{k=0}^d(-1)^kb_k(\Delta_{\mathcal{G}})
\ee
to get
\be
\Omega(\mathcal{G})=\sum_{k=3}^{d}(-1)^{k-1}c_k(\Delta_{\mathcal{G}})+\chi(\Delta_{\mathcal{G}})-b_0(\Delta_{\mathcal{G}})+b_1(\Delta_{\mathcal{G}}).
\ee
This is the same as \eqref{resultat} in this particular case. Our result can therefore be described as the generalization of \eqref{linearized} to general cell complexes, and non-Abelian groups.

\subsection{Powercounting from vertex and jacket counting} \label{jacket-counting}

Using tools from perturbative quantum field theory, and notably Cauchy-Schwarz inequalities, an upper bound on the divergence degree in the Boulatov model was obtained in \cite{Magnen:2009dq}. It is formulated in terms of the number of vertices rather than $3$-cells in the complex, and reads for a Boulatov complex $\Gamma$ without generalized tadpole \cite{BenGeloun2010}
\be
\Omega(\Gamma)\leq \f{3}{2}c_0(\Gamma)+6.
\ee
Within the colored model, this bound was then generalized to higher dimensions in \cite{BenGeloun2010}. In dimension $d$, it becomes for a colored graph $\mathcal{G}$ with $V(\mathcal{G})$ vertices
\be\label{bound}
\Omega(\mathcal{G})\leq \f{3(d-1)(d-2)}{4}V(\mathcal{G})+3(d-1)
\ee
Using the notion of ``jacket" introduced in \cite{Geloun:2010vn}, Gurau and Rivasseau further improved this bound in \cite{Gurau2011a,Gurau2011b}, obtaining
\be\label{jacketbound}
\Omega(\mathcal{G})\leq \f{3(d-1)(d-2)}{4}V(\mathcal{G})+3(d-1)-\f{6(d-2)}{d!}\sum_{\mathcal{J}}g(\mathcal{J}),
\ee
where $g(\mathcal{J})$ is the genus of (orientable surface dual to) the jacket $\mathcal{J}$.

Remarkably, the jacket bound \eqref{jacketbound} follows easily from our exact result \eqref{resultat}. Indeed, denoting $E(\mathcal{G})$ the number of edges and $F(\mathcal{G})$ the number of faces of $\mathcal{G}$, the formula \eqref{resultat} reads
\be
\Omega(\mathcal{G})=\dim\mathcal{M}(\Gamma,G)-\dim\zeta(\Gamma,G)+3\big(V(\mathcal{G})-E(\mathcal{G})+F(\mathcal{G})\big).
\ee
Since a colored graph has no tadpole, we have $2E(\mathcal{G})=(d+1)V(\mathcal{G})$, hence
\be\label{intermediate}
\Omega(\mathcal{G})\leq\dim\mathcal{M}(\Gamma,G)-\f{3(d-1)}{2}V(\mathcal{G})+3F(\mathcal{G}).
\ee
Moreover, since
\be
\dim\mathcal{M}(\Gamma,G)\leq\dim\textrm{Hom}\big(\pi_1(\mathcal{G}),\SU(2)\big),
\ee
and $\pi_1(\mathcal{G})$ is a subgroup of $\pi_1(\mathcal{J})$ for each jacket $\mathcal{J}$ of $\mathcal{G}$, we have
\be
\dim\mathcal{M}(\Gamma,G)\leq\f{1}{J(\mathcal{G})}\sum_{\mathcal{J}}\dim\textrm{Hom}\big(\pi_1(\mathcal{J}),\SU(2)\big),
\ee
where
\be
J(\mathcal{G})=\f{d!}{2}
\ee
is the number of jackets of $\mathcal{G}$. Now it is well-known that the dimension of the $\SU(2)$ representation variety of a genus $g$ surface is $6g-3$. Thus,
\be\label{moduligenus}
\dim\mathcal{M}(\Gamma,G)\leq\f{2}{d!}\sum_{\mathcal{J}}\big(6g(\mathcal{J})-3\big).
\ee
Using the Euler relation
\be
\chi(\mathcal{J})=2-2g(\mathcal{J})=V(\mathcal{J})-E(\mathcal{J})+F(\mathcal{J}),
\ee
the combinatorial facts that $V(\mathcal{J})-E(\mathcal{J})=\f{d-1}{2}V(\mathcal{G})$ and that each face of $\mathcal{G}$ belongs to $(d-1)!$ jackets, we have \cite{Gurau2011c}
\be
F(\mathcal{G})=\f{d(d-1)}{4}V(\mathcal{G})+d-\f{2}{(d-1)!}\sum_{\mathcal{J}}g(\mathcal{J}).
\ee
Using this relation in \eqref{moduligenus} gives a bound on $\dim\mathcal{M}(\Gamma,G)$ which, when inserted in \eqref{intermediate}, immediately gives the jacket bound \eqref{jacketbound}.

\section{Three-dimensional examples}\label{3dexamples}

In this section, we illustrate how the divergence degree, and more specifically its topological part $I(M,G)$, can be computed for certain three-dimensional manifolds $M$, always with $G=\SU(2)$. (Hence, hereafter we will drop the reference to $G$ in the notation.) Most of them cannot be handled by the previous methods and do not saturate the corresponding bounds \cite{Freidel2004,Barrett:2009ys,Freidel:2009hd,Geloun:2010vn}.

Closed $3$-manifolds have $\chi(M)=0$, so $I(M)=\dim\mathcal{M}-\dim\zeta$ and it is completely determined by the fundamental group. It is also well-known that $\calM$ is a finite set of points, and hence $\dim\calM=0$, when the fundamental group $\pi_1(M)$ is finite (but we will re-derive this feature in specific examples for illustrative purpose).

The method to compute $I(M)$ for a cellular pseudomanifold (and in fact for any cell complex) has been presented in the section \ref{computational}. The steps are:
\begin{itemize}
\item Choose a presentation of $\pi_1(M)$ (e.g. from a deformation retract of the 2-skeleton $\Gamma$).
\item Identify the representation variety $\Hom\bigl(\pi_1(M),G\bigr)$ using the chosen presentation. In general, it has several irreducible components.
\item Identify the orbit of the $G$-action on each irreducible component.
\end{itemize}


\subsection{The $3$-torus}

The case of the $3$-torus $T^3$ was discussed in Appendix C of \cite{Freidel2004}. Its fundamental group has the presentation
\be
\pi_1(T^3)=\bra a,b,c\ \vert\ [a,b]=[a,c]=[b,c]=1\ket.
\ee
The representations of this group in $\SU(2)$ are of the form
\be
\phi=\Big(\exp(\psi_a\hat{n}.\vec{\tau}),\ \exp(\pm\psi_b\hat{n}.\vec{\tau}),\ \exp(\pm\psi_c\hat{n}.\vec{\tau})\Big)\in\SU(2)^3,
\ee
with $\hat{n}\in S^2$ their common direction of rotation, $\psi_{a,b,c}\in[0,\pi]$ three class angles, and $\vec{\tau}$ the 3-vector formed by a set of (anti-Hermitian) generators of the algebra $\su(2)$. They form a $5$-dimensional manifold $\calF$.

The group action by conjugation rotates the direction $\hat{n}$. For each representation there is a stabilizer $\zeta$ isomorphic to $\textrm{U}(1)$ (and larger if the three rotations are in the center of $\SU(2)$) which leaves it invariant: the subgroup generated by the direction $\hat{n}$. Hence, the dimension of the stabilizer is $\dim \zeta=1$. Thus
\begin{eqnarray}
\dim\mathcal{M}&=&5-2=3,\nonumber\\ \dim\zeta&=&1.
\end{eqnarray}
Hence,
\be
I(T^3)=2.
\ee
Note that this implies that the divergence degree of a cell decomposition of $T^3$ cannot be a multiple of $\dim\SU(2)=3$, as the procedure of \cite{Freidel2004} assumed implicitly.

\subsection{Lens spaces} \label{sec:lens-prism}

Lens spaces $L_{p,q}$ are standard spherical manifolds, with $\Z_p$ as their fundamental group (we exclude the case of $L(0,1)= S^1\times S^2$ which can be understood separately). They include as a particular case the real projective space $\mathbbm{R}P^3$, which is the first example in the appendix of \cite{Gurau2011d}.

The standard presentation of $\Z_p$ is of course
\be
\Z_p=\bra a\ \vert\ a^p=1\ket.
\ee
If $0<r<\pi$, let $S^2_r$ be the subset of $\SU(2)$ defined by
\be \label{2-sphere}
S^2_r=\Bigl\{\exp\bigl(r\,\hat{n}.\vec{\tau}\bigr)\ ; \ \hat{n}\in S^2\Bigr\},
\ee
consisting in those rotations of fixed angle $r$. In the topological picture induced by the identification $\SU(2)\simeq B^3/\pp B^3$, $S^2_r$ is a sphere of radius $r$ centered on the origin. (In this picture,  the central elements $\pm\unit$ of $\SU(2)$ are the center and boundary of $B^3$ respectively.)

Hence,
\begin{itemize}
\item If $p=1$ (the $3$-sphere), there is only one representation, the trivial one.
\item If $p=2$ (the real projective space), the representations of $\Z_2$ send its generator to an element of the center of $\SU(2)$, i.e. $\{\pm \unit\}$.
\item If $p\geq 3$, the set of representations decomposes as $\Hom (\Z_p, \SU(2)) = \calF_p^{(0)}\sqcup\calF_p^{(2)}$, with
\begin{eqnarray} \calF_p^{(0)} & = & \left\{ \begin{array}{ll}
\{\unit\}  & \qquad\textrm{$p$ odd}\\ \zeta(\SU(2))=\{\pm\unit\}  & \qquad\textrm{$p$ even}\end{array} \right. \\ \calF_p^{(2)} & = & \bigsqcup_{1\leq k< p/2}S^2_{2\pi k/p}  \end{eqnarray}
This means that the class angles admit a finite number of values, $r= 2k\pi/p$, for $k=0,\dotsc,\lfloor \f{p}2\rfloor$. The set $\calF_p^{(0)}$ consists of points, while $\calF_p^{(2)}$ is a union of $2$-spheres.
\end{itemize}

\medskip

The next step consists in finding the orbits generated by conjugation in $\SU(2)$. It turns out that all of the above representations are reducible.

\begin{itemize}
\item
In the cases $p=1,2$, and on $\calF_p^{(0)}$ when $p\geq3$, the representations are left invariant by the group action (since they commute with the whole group): they are central. Thus, the centralizer is $\zeta = \SU(2)$, which is 3-dimensional.
\item
On $\calF_p^{(2)}$ for $p\geq 3$, group conjugation corresponds to rotation of the axis $\hat{n}$. Similarly to the 3-torus case, the stabilizer is then $\zeta=\textrm{U}(1)$.
\end{itemize}

In all cases, the moduli space $\calM$ consists in $p$ distinct points, one for each connected component of $\Hom(\Z_p,\SU(2))$, hence $\dim\calM=0$. However, when $p\geq 3$, one can compute two different values of the topological part of the divergence degree, one on $\calF_p^{(0)}$ and another on $\calF_p^{(2)}$. The relevant value is obviously that which gives the most divergent contribution to the amplitude and thus the greatest value of $I= \dim \calM-\dim\zeta$. We get $I = -3$ on $\calF_p^{(0)}$, to be compared with $I=-1$ on $\calF_p^{(2)}$. This means that the relevant value is the one computed on the less reducible representations, i.e. those with the smallest stabilizer.

In conclusion
\be
I\big(L_{p,q}\big)= \begin{cases} -3  & \textrm{if $p=1,2$}\\
 -1  & \textrm{if $p\geq3$.}\end{cases}
\ee

\subsection{Prism spaces}

Prism manifolds $P_{m,n}$, with $m\geq 1,n\geq 2$, form a different class of spherical manifolds. They are characterized by their fundamental group
\be
\pi_1(P_{m,n})=\bra x,y\ \vert\ xyx^{-1}=y\mone,\ x^{2m}=y^{n}\ket.
\ee
Let us first give the irreducible representations $\phi$ in the case of $m$ even. The first relation imposes the direction of the rotation $\phi(x)$ to be orthogonal to that of $\phi(y)$, and its class angle to be $\f{\pi}{2}$. Then the second relation reduces to $y^n=\unit$, which only constrains the class angle of $\phi(y)$ to be $\psi_y = 2k\pi/n$ for $k=1,\dotsc,\lfloor \f{n}{2}\rfloor$ (the case $k=0$ gives a reducible representation). This way the set of irreducible representations is identified as
\begin{align}
\calF_{m,n}^{\rm irr} = \Bigl\{&\left( \phi(x) = \exp\bigl(\f{\pi}{2} \hat{n}_x\cdot\vec{\tau}\bigr),\ \phi(y) = \exp\bigl(\f{2k\pi}{n} \hat{n}_y\cdot\vec{\tau}\bigr)\right);\\ &k=1,\dotsc,\lfloor \f{n}{2}\rfloor,\ (\hat{n}_x, \hat{n}_y)\in (S^2)^2, \hat{n}_x\cdot \hat{n}_y = 0\Bigr\}
\end{align}
Clearly this space is of dimension 3. Moreover, simultaneous conjugation of $\phi(x), \phi(y)$ by some $\SU(2)$ element induces a simultaneous rotation on $\hat{n}_x, \hat{n}_y$. Thus, the orbit is 3-dimensional, isomorphic to $\SU(2)/\{\pm \unit\}=\textrm{SO}(3)$ and the centralizer is just the center of $\SU(2)$. This leads to $\dim \calM = 0, \dim\zeta=0$, and hence:
\beq
I(P_{m,n}) = 0,
\ee
when $m$ is even, and evaluated on irreducible representations. The situation is similar for $m$ odd, only the specific values of the class angle of $\phi(y)$ are changed.

As for reducible representations, they are obtained by taking $\phi(y)=\pm\unit\in\zeta(\SU(2))$. Then, the image of the generator $x$ lives on a sphere of the type $S^2_r$ \eqref{2-sphere} or is $\pm\unit$. One easily sees that $\dim \calM =0$ again, but $\dim\zeta = 1$ or $\dim\zeta=3$. This produces a topological part for the divergence degree which is negative. Thus, the highly divergent contribution to the amplitude comes from the irreducible representations of the fundamental group.

%% file: part3.tex
\part{On the spinfoam continuum limit}
\vspace*{\fill}

The purpose of the third part of this thesis is to present two preliminary results on the \emph{continuum limit} of spinfoam models, viz. the problem of describing a quantum spacetime with \emph{infinitely many} dynamical degrees of freedom from the \emph{truncated} amplitudes $\mathcal{A}(\Gamma)$.

\bigskip

In chap. \ref{spincont}, we address a long-standing question in the spinfoam community: to construct this continuum limit, should one \emph{sum} over all foams $\Gamma$, or infinitely \emph{refine} a foam? We give a mathematical definition for each alternative, and an argument to the effect that, as it happens, they could be two facets of the same solution. 

\bigskip

In chap. \ref{resummation}, we consider a specific framework for the sum over all foams, in three dimensions: Boulatov's \emph{group field theory}. Using a tool of constructive field theory -- Magnen's and Rivasseau's \emph{cactus expansion} -- we improve Freidel and Louapre's proof of Borel summability of the Boulatov series with a high-spin cutoff, and establish a remarkable \emph{scaling limit}. 

\bigskip

The two sections are completely independent. 

\vspace*{\fill}

\chapter{Spinfoams: refining or summing?}\label{spincont}

What is the mathematical definition of the operation of \emph{refining} a foam to increase the number of gravitational degrees of freedom? Does this process have a well-defined limit? How is it related to the \emph{sum} over foams sometimes advocated in the spinfaom community?

These questions were addressed in collaboration with Carlo Rovelli \cite{Rovelli:2010fk}. Some aspects of the arguments presented there, such as the emphasis on the structure of \emph{directed sets} and the r\^ole of zero-spin colorings, were already discussed in Zapata's earlier work.\footnote{We came to know Zapata's work on the continuum limit of spinfoams on the occasion of the Loops 2011 conference, after the paper \cite{Rovelli:2010fk} was completed; the similarities between his approach and ours are thus only fortuitous.}

\section{The refinement limit of spinfoams}

We sketched in the introduction the Ashtekar-Lewandoswki construction of the Hilbert space of loop quantum gravity as the \emph{inductive limit} of truncated spaces $\mathcal{H}_\gamma$ labelled by graphs. This defines the quantum kinematics of the gravitational field as a \emph{continuum limit} of lattice gauge fields.

It turns out that little work has been done to study the analogue of this construction for spinfoams, viz. at the \emph{dynamical} level. (The notable exception is Zapata's papers \cite{Manrique2005,Zapata2002}.) This is surprising: such a hierarchical scheme, whereby more and more degrees of freedom of the field are taken into account according to some notion of \emph{scale}, lies at the core of our modern understanding of quantum field theory, in the form of the \emph{renormalization group}.

The obvious stumbling block to apply Wilsonian techniques to study the dynamics of quantum gravity is the absence of a background metric, which makes the notion the scale difficult to assess.\footnote{Of course, one can run the functional renormalization equation with respect to a fixed, arbitrary, background metric: this is the strategy underlying Weinberg's and Reuter's \emph{asymptotic safety scenario} \cite{Niedermaier2006}. The problem with this approach is the difficulty to extract physics from the truncated flow equation.} But there is a way out of this dead end: to think of the foams themselves as \emph{generalized scales}. Let us elaborate on this idea. 
\subsection{The directed set of generalized scales}

Lattice field theory \cite{Rothe2005} is based on the discretization of (Euclidean) spacetime by means of \emph{regular} lattices: lattices where the distance between neighboring vertices is constant, given by the \emph{lattice spacing} $a$. To \emph{refine} the lattice, one introduces new vertices at the midpoints of edges and plaquettes, and constructs the corresponding regular lattice with lattice spacing $a/2$. Note that both the choice of the initial lattice and the refinement step rely on the metric structure of Euclidean space. (For instance, the midpoint of an edge is singled out by the fact that it is located at a distance $a/2$ from the endpoints of the edge.)

The prominent feature of this metric notion of scale is its \emph{one-dimensional} character, which leads to \emph{sequences} of discretizations. This can be expressed mathematically by the statement that the set of distance scales $\mathcal{S}=\mathbb{R}^+$ is a \emph{totally ordered set}: for any two scales $s_1$ and $s_2$, we have either $s_1\leq s_2$ or  $s_2\leq s_1$, viz. $s_1$ and $s_2$ are mutually comparable.

Non-metric discretizations generally do not share this one-dimensional character. Consider for instance the set $\mathcal{T}_M$ of triangulations of a given manifold $M$. There is a natural order relation on this set, defined by $T_1\preceq T_2$ when all the simplices of $T_2$ are subsets of the simplices of $T_1$; this corresponds to the intuitive notion that $T_2$ is \emph{finer} than $T_1$. It is obvious that not all triangulations are comparable with respect to this relation: $\mathcal{T}_M$ is only a \emph{partially ordered set}, or \emph{poset}.

The limit of a family of objects indexed by a poset is ill-defined, because non-unique. Yet, in the case of triangulations of a manifold, we have the intuition that there is only \emph{one} continuum limit: roughly speaking, it corresponds to a ``triangulation" with infinitely many simplices. The reason why this intuition is meaningful is because the set of triangulations is \emph{directed}: for any two triangulations $T_1,T_2$, there is a triangulation $T_3$ which is finer than both $T_1$ and $T_2$, viz. $T_1\preceq T_3$ and $T_2\preceq T_3$. A poset satisfying this condition is called a \emph{directed set}. Roughly speaking, a directed order relation has a single `asymptotic direction': here, \emph{the} continuum limit.

This observation provides a sound definition of the notion of \emph{generalized scale}: it is simply an element of some directed set, be it totally ordered or not. In this sense, a triangulation is indeed a generalized scale.

\subsection{Kinematical and dynamical continuum limits}


Generally speaking, a quantum system can be described by an algebra of observables $\calA$ acting on a Hilbert space $\calH$, its \emph{kinematics}, and a state over this algebra $\omega:\calA\rightarrow\mathbb{C}$, its \emph{dynamics}. In particular, a truncation of the degrees of freedom by means of a directed set $\mathcal{S}$ associates to each generalized scale $s\in\mathcal{S}$ an algebra $\calA_s$ and a state $\omega_s$, corresponding respectively to the truncated kinematics and dynamics of the system.

Thus, to obtain the continuum limit of a truncated system, we need \emph{two} limits: an algebraic one, for the family of algebras $(\calA_s)_{s\in\mathcal{S}}$ and Hilbert spaces $\calH_\gamma$ (kinematics), and a topological one, for the family of states $(\omega_s)_{s\in\mathcal{S}}$ (dynamics). The corresponding mathematical notions are that of \emph{inductive limits} and \emph{limit of nets}. The definitions are as follows.

\begin{itemize}
\item
Let $\mathcal{S}$ be a directed set, and $(A_s)_{s\in\mathcal{S}}$ a family of algebraic objects indexed by $\mathcal{S}$. Suppose that, for each $s_1\preceq s_2$, we have a homomorphism $f_{s_1s_2}:A_{s_1}\rightarrow A_{s_2}$, such that \begin{itemize}
\item
$f_{ss}$ is the identity of $A_s$
\item
$f_{s_1s_3}=f_{s_2s_3}\circ f_{s_1s_2}$ for all $s_1\preceq s_2\preceq s_3$
\end{itemize}
These conditions ensure that the identification of $a_{s_1}\in A_{s_1}$ with its images $f_{s_1s_2}(a_{s_1})$ for all $s_2\succeq s_1$ defines an equivalence relation over the disjoint union $\bigsqcup_{s\in\mathcal{S}}A_s$. The \emph{inductive limit} $A_\infty$ of $(A_s)_{s\in\mathcal{S}}$ is then the set of equivalence classes for this relation. It naturally comes equipped with the same algebraic structure as the $A_s'$. For instance, if the latter are inner product spaces, then the maps $f_{s_1s_2}$ are isometric embeddings, and the inner product on $A_\infty$ is given by $\langle a_1,a_2\rangle_{A_\infty}=\langle f_{s_1s_3}(a_{s_1}),\vert f_{s_2s_3}(a_{s_2})\rangle_{A_3}$, where $a_{s_1}\in A_{s_1}$ and $a_{s_2}\in A_{s_2}$ are representatives of $a_1$ and $a_2$ respectively, and $s_3\succeq s_1,s_2$.

\item
Let $X$ be a topological space, and $N:\mathcal{S}\rightarrow X$ a function over a directed set $\mathcal{S}$ valued in $X$, viz. a \emph{net} over $\mathcal{S}$. Then $N$ converges to $N_\infty\in X$, or that $N_\infty$ is the limit of $N$, if for any neighborhood $U$ of $N_\infty$ there exists $s_U\in\mathcal{S}$ such that $N(s)\in U$ for any $s\succeq s_U$. It is easy to check that, thanks to the directedness of $\mathcal{S}$, the limit of a net over $\mathcal{S}$, when it exists, is unique (assuming $X$ is Hausdorff).\footnote{Standard examples of nets are provided by Riemann sums. Let $\mathcal{P}$ denote the set of partitions of the unit interval $[0,1]$ directed by inclusion, and $f:[0,1]\rightarrow\mathbb{C}$ a (continuous) function. The map $N_f:\mathcal{P}\rightarrow\mathbb{C}$ associating to each partition $p$ the corresponding Riemann sum of $f$ is a net over $\mathcal{P}$. Its limit is the Riemann integral $\int_0^1dtf(t)$.}
\end{itemize}

These definitions clarify the general notion of \emph{refinement}: in general terms, a set of generalized scales is given in the form a directed set; kinematics (resp. dynamics) defines an inductive system (resp. a net) over the set of generalized scales; the continuum limit is the corresponding limit. No metric is involved in this process.

Although little known outside the mathematical community, the structure of directed set plays a fundamental r\^ole in the conceptual architecture of physics. Indeed, it captures in its general form the concept of \emph{approximation scheme}: a hierarchical structure directed towards a unique limit. Since physics is the science of approximations, we could say (paraphrasing Rivasseau on the renormalization group) that \emph{directed sets are the soul of physics}.

\subsection{Loop quantization as a continuum limit}

The kinematical truncation scheme used in loop quantum gravity is precisely of the above type, with the notion of generalized scale provided by \emph{graphs}.\footnote{Along the lines of the canonical formulation of general relativity, these are usually defined as \emph{embedded} in a three-dimensional Cauchy surface in some globally hyperbolic spacetime. Rovelli has recently considered a purely combinatorial version of this construction, but it is not clear that the inductive limit procedure is meaningful in this case.} For each graph $\gamma$, one considers the algebra $\calA_\gamma$ of gauge-invariant functions on the cotangent bundle $T^*\su^{E(\gamma)}$ and its standard quantization over the Hilbert space $\calH_\gamma:=L^2(\su^{E(\gamma)}/\su^{V(\gamma)})$. This defines the quantum kinematics of a system of discrete connections on $\gamma$.

The kinematics of the quantum gravitational field is then constructed as the continuum limit of this system of discrete connections. To this effect, one uses the natural order relation on the set of graphs $\mathcal{G}$ defined by $\gamma_1\preceq\gamma_2$ if $\gamma_1$ is a subgraph of $\gamma_2$. The latter clearly equips $\mathcal{G}$ with a directed set structure: given any two graphs $\gamma_1$ and $\gamma_2$, it is easy to construct $\gamma_3\succeq\gamma_1,\gamma_2$ by adding all the vertices and edges of $\gamma_1$ to $\gamma_2$. Moreover, each embedding $\gamma_1\hookrightarrow\gamma_2$ naturally induces maps $\calA_{\gamma_1}\rightarrow\calA_{\gamma_2}$ and $\calH_{\gamma_1}\rightarrow\calH_{\gamma_2}$ which turn the family of algebras and Hilbert spaces into inductive systems. Their inductive limits $\calA_\infty$ and $\calH_\infty$ form the kinematical arena of loop quantum gravity, where the action of (some extension of) the diffeomorphism group can be implemented.

Let us be more specific about the mappings $i_{\gamma_1\gamma_2}:\calH_{\gamma_1}\rightarrow\calH_{\gamma_2}$ induced by the inclusion of $\gamma_1$ into $\gamma_2$. Let $\psi\in\calH_{\gamma_1}$ and denote $e^{\gamma_1}_i$ (resp. $e^{\gamma_2}_j$) the edges of $\gamma_1$ (resp. $\gamma_2$). Then 
\be
(i_{\gamma_1\gamma_2}\psi)\big(g_{e^{\gamma_2}_1},\dots,g_{e^{\gamma_2}_{E(\gamma_2)}}\big):=\psi\big(g_{e^{\gamma_1}_1},\dots,g_{e^{\gamma_1}_{E(\gamma_1)}}\big).
\ee
In other words, the image of $i_{\gamma_1\gamma_2}$ in $\calH_{\gamma_2}$ is the set of gauge-invariant functions on $\su^{E(\gamma_2)}$ which are constant on those edges of $\gamma_2$ that are not edges of $\gamma_1$; equivalently, it is the subspace of $\calH_{\gamma_2}$ spanned by spin-network functions with zero spins on the edges of $\gamma_2\setminus\gamma_1$. It follows from this definition that the inductive limit $\calH_\infty$ can be written as
\be
\calH_\infty=\bigoplus^{\perp}_{\gamma}\calH_\gamma^*,
\ee
where $\calH_\gamma^*$ denotes the subspace of $\calH_\gamma$ spanned by spin-network functions with non-zero spins. This is an interesting observation: at the kinematical level, the continuum limit of loop quantum gravity consists of superselection sectors with non-zero spin colorings. Pictorially, we can say that ``refining the graphs'' is the same as ``summing over the graphs with zero-spins excluded''. 


\subsection{The spinfoam continuum limit}

To discuss the analogue construction for spinfoam models, we must begin by identifying the relevant directed set. One possibility (explored by Zapata in \cite{Zapata2002}) is to work with embedded foams and their diffeomorphisms classes. Let us use instead the combinatorial framework sketched in sec. \ref{deffoams}, in which a foam $\Gamma$ is defined as a triple of sets $(V_\Gamma,E_\Gamma,F_\Gamma)$ enumerating its vertices, edges and faces. 


In a given spinfoam model, to each foam $\Gamma$ with boundary $\gamma$ is associated a vector $ \calZ(\Gamma)$ in the boundary Hilbert space $\mathcal{H}_\gamma$. Now, this map $\Gamma\mapsto \calZ(\Gamma)$  defines a \emph{net} over the set $\mathcal{F}_\gamma$ of foams with fixed boundary $\gamma$. 

Consider indeed the natural order relation defined by $\G_1\preceq\G_2$ if there is an \emph{embedding} $\iota:\G_1\hookrightarrow\G_2$, namely a triple of injective maps $\iota_v:V_{\G_1}\rightarrow V_{\G_2}$, $\iota_e:E_{\G_1}\rightarrow E_{\G_2}$, $\iota_f:F_{\G_1}\rightarrow F_{\G_2}$ preserving the relations between vertices, edges and faces. It is easy to see that any pair of foams with the same boundary has an upper bound $\G\succeq\G_1,\G_2$ with respect to this relation, given essentially by attaching the vertices, edges and faces of $\G_2$ to $\G_1$. Thus, a spinfoam model is a \emph{net} over the set of proper foams with a given boundary. 


%

Hence, the set $\mathcal{F}_\gamma$ is naturally a directed set, and $\Gamma\mapsto  \calZ(\Gamma)$ a net over it. We can therefore define formally, for each boundary $\gamma$, the \emph{spinfoam continuum limit} as the limit $W^\infty_\gamma$ of this net:
\begin{equation}
\calZ^\infty_\gamma:\doteq\underset{\Gamma\in\mathcal{F}_\gamma}{\lim}\  \calZ(\Gamma).
\end{equation}

\section{Refining = summing?}

The spinfoam continuum limit defined above, if compelling, is at variance with Baez's, Rovelli's and Reisenberger's original idea that foams are ``quantum histories'' of the gravitational field, like Feynman diagrams are ``quantum histories'' of systems of particles. Indeed, this latter conception suggests that the proper way to unfreeze the truncated degrees of freedom is not by refining a foam, as in lattice field theory, but rather by \emph{summing} over the set of all foams. Is there a relationship between refinement and sum over foams, similar to kinematical relation between refinement and sum over graphs?

\subsection{Spinfoam cylindrical consistency}

From now, let us assume that the foam amplitudes $\calZ(\Gamma)$ are given as sums over colorings of the faces of $\Gamma$ by irreducible representations of a compact Lie group $G$ (typically $G=\su$). More specifically, we assume that for each coloring $\sigma$ of $\Gamma$, we have a \emph{spinfoam amplitude} $\calZ(\G,\sigma)$. This assignement defines \emph{two} spinfoam models, depending whether the trivial representations are included or not:
\be
\label{net}
\calZ(\Gamma):\doteq\!\!\!\sum_{\sigma\in\Col(\Gamma)} \calZ(\Gamma,\sigma),
 \ \ \ {\rm and} \ \ \  
\calZ^*(\Gamma):\doteq\!\!\!\sum_{\sigma\in\Col^*(\Gamma)} \calZ(\Gamma,\sigma),
\ee
Here, for a fixed foam $\Gamma$, we denoted $\Col(\Gamma)$ the set of all colorings $\sigma$, and $\Col^*(\Gamma)$ the subset consisting of colorings by \emph{non-trivial} ($j_f\neq0$) representations.

Since trivial representations play no r\^ole in the kinematics of loop quantum gravity, it is natural to impose a \emph{cylindrical consistency condition} relating $\calZ(\Gamma)$ to $\calZ^*(\Gamma)$. To state it, we must first observe that each coloring $\sigma$ of a foam $\Gamma$ comes with a multiplicity, related to the symmetries of $\Gamma$. The \emph{multiplicity} $\vert\sigma\vert_{\Gamma}$ of a coloring $\sigma\in\Col(\Gamma)$ is the number of colorings $\sigma'$ such that $\sigma=\sigma'\circ\phi$, with $\phi$ an automorphism of $\G$, i.e. a foam embedding of $\Gamma$ into itself. Then we say that the amplitude
\be
A_\Gamma(\sigma) := \vert\sigma\vert_{\Gamma}\  \calZ(\Gamma,\sigma),
\label{AZ}
\ee
is \emph{cylindrically consistent} if $A(\Gamma,\sigma)=A(\Gamma',\sigma')$ when $(\Gamma',\sigma')$ is a \emph{trivial extension} of $(\Gamma,\sigma)$, that is when $\Gamma$ is a subfoam of $\Gamma'$, $\sigma$ and $\sigma'$ coincide on the faces of $\Gamma$ and $\sigma'$ is trivial on the other faces of $\Gamma'$.

\subsection{Summing is refining}

Under this cylindrical consistency condition, the relationship between the continuum limit $\calZ_{\g}^\infty$ and the sum of $\calZ(\Gamma)$ over all the foams $\Gamma$ with fixed boundary is the following:
\be\label{result}\calZ(\Gamma)=\sum_{\substack{\Gamma'\preceq\Gamma\\ \pp\Gamma'=\gamma}}\calZ^{*}(\Gamma).
\ee
If we assume furthermore that the net $\calZ$ is convergent, then we have by passing to the limit in the identity above
\be
\calZ^{\infty}_{\gamma} \doteq\sum_{\pp\Gamma=\gamma} \calZ^{*}(\Gamma).
\label{main}
\ee
Thus, \emph{refining foams in the model $\calZ$ is the same as summing over foams in the model $W^*$}.   Note that, due to the peculiar directed set structure, \emph{all} foams appear in \eqref{main} the following sense: every foam $\Gamma$ with boundary $\pp\Gamma=\gamma$ appears in one finite sum \eqref{result} whose value can be chosen arbitrarily close to $W^{\infty}_{\gamma}$.

The proof of (\ref{result}) is easy. First, observe that the subfoams of $\Gamma$ index a partition of $\Col(\Gamma)$, in which each each class is made of the trivial extensions of a given subfoam $\Gamma'\subset\Gamma$:
 \be\Col(\Gamma)=\bigsqcup_{\Gamma'\subset\Gamma}\Col^*({\Gamma'}).\ee This implies that \be \calZ(\Gamma)=\sum_{\Gamma'\subset\Gamma}\left(\sum_{\sigma'\in\Col^*({\Gamma'})}\vert\sigma'\vert_{\Gamma}^{-1}A_\Gamma(\sigma')\right).\ee Second, check that we have \be\vert\sigma'\vert_{\Gamma}=\vert\sigma'\vert_{\Gamma'}N_{\Gamma',\Gamma},\ee with $N_{\Gamma',\Gamma}$ the number of subfoams of $\Gamma$ isomorphic to $\Gamma'$. Third, use cylindrical consistency to get \be
\calZ(\Gamma)=\sum_{\Gamma'\subset\Gamma}N_{\Gamma',\Gamma}^{-1}\calZ^*(\Gamma')
\ee
and conclude to (\ref{result}).

\subsection{Discussion}

But is there any rationale for assuming the cylindrical consistency condition and expecting it to be part of the quantum theory of gravity?
There is.  The spinfoam sum is meant to be an implementation of the Misner-Hawking sum over geometries
\be
Z=\int_{\rm Metrics/Diff}\ Dg_{\mu\nu}\ e^{\frac{i}{\hbar}S[g_{\mu\nu}]}.
\label{sog}
\ee
Here the integral is not over metrics, but over equivalence classes of metrics under diffeomorphisms. In the truncation induced by the choice of a foam, the diffeomorphisms are reduced to the automorphisms $\phi$ of the foam.  Therefore the colorings $\sigma$ and $\sigma\circ\phi$ have a natural interpretation as the discrete residual equivalent of diffeomorphism-related metrics. This interpretation is reinforced by the fact that the amplitude is in fact invariant. If we want to integrate over geometries, then, the contributions of $\sigma$ and $\sigma\circ\phi$ represent an overcounting, and we must divide by the number of them, namely by $\vert\sigma\vert_{\G}$. The same conclusion can be reached by interpreting colored spinfoams as histories of nontrivial spin networks. Then colorings related by foam automorphisms clearly represent the same history. For both these reasons, it is  physically interesting to consider the amplitudes modified as in \eqref{AZ} and the cylindrical consistency condition.  Then the sum over foams is equal to the continuum limit. 
\chapter{Resummation of the Boulatov series}\label{resummation}

In this section, we address the issue of the continuum limit from the angle of \emph{group field theory} (GFT). In this scheme, the sum over foams is nothing but a Feynman series, and the issue of convergence is a problem of \emph{constructive field theory}. Is the sum over foams the asymptotic expansion of a well-defined GFT correlation functions?

The first results in this direction were obtained in the Boulatov model by Freidel and Louapre \cite{Freidel:2002tg}. Here, using more advanced constructive techniques, we improve their proof of Borel summability, notably by establishing the constructive scaling behaviour of the correlation functions in the large cutoff limit.  These results were obtained under the lead of Rivasseau, in collaboration with Magnen and Noui \cite{Magnen:2009dq}.

\section{Constructive field theory}

\emph{Constructive field theory} \cite{Riv} is the name of a mathematical physics program launched around 1970 by Wightman. It aims at defining \emph{rigorously} certain field theory models, in increasing order of difficulty, by checking that they obey the ``Wightman axioms" \cite{Streater:1989vi}. The strategy is clear-cut: introduce as many cutoffs as necessary for the
correlation functions of these models to be well-defined, and then develop 
the necessary methods to lift these cutoffs. 

\subsection{Current status of constructive field theory}

Early constructive field theory succeeded in building rigorously super-renormalizable
field theories such as the emblematic $\phi^4$ model in two and three dimensions. It also elucidated
their relationship to perturbation theory: the Schwinger functions of 
these models are the Borel sum of their perturbation theory \cite{EMS,MS}. But four-dimensional $\phi^4$ theory itself could not be built, because its coupling constant does not remain small in the ultraviolet regime.
Being asymptotically free, non-Abelian gauge theories do not have this problem.
Nevertheless, although some partial results were obtained, they could not be built in the full constructive sense either, due to technical difficulties such as Gribov ambiguities; neither could the interesting infrared confining regime of the theory be understood rigorously.
Probably the first four-dimensional field theory to be built 
completely through constructive methods will be the Grosse-Wulkenhaar model \cite{Grosse:2004yu},
a non-commutative field theory which, ironically, should not satisfy
the Wightman axioms of the initial constructive program.


The constructive program is largely unknown in the quantum gravity community, with the notable exception
of \cite{Freidel:2002tg}. However, it can be argued that constructive theory embodies a deeper point of view on quantum field theory than the textbook one.
A modern constructive technique such as the \emph{cactus expansion} used hereafter allows to resum perturbation theory by reorganizing it in a precise, explicit manner.

\subsection{Power series and Borel sums}

Renormalized perturbation theory writes field-theoretic observables ${O}(\lambda)$, notably correlation functions, as \emph{formal power series} in the coupling constant(s) $\lambda$:
\be\label{powerseries}
 {O}(\lambda)\doteq\sum_{n=0}^{\infty}{O}_{n} \lambda^n.
 \ee
The coefficients ${O}_{n}$ are, in turn, given by a sum of Feynman amplitudes $O_G$, where $G$ spans the set of Feynman graphs with $V(G)=n$ vertices:
\be
O_n=\sum_{V(G)=n}O_G.
\ee 

It was early realized by Dyson \cite{PhysRev.85.631} that the construction of observables by power series in $\lambda$ is bound to remain heuristic, because \eqref{powerseries} \emph{cannot converge}. Indeed, if it did, it would define an analytic function of $\lambda$ in a disk around $\lambda$, and hence would be equally well-defined for $\lambda\geq0$ and $\lambda\leq0$. But this is physically absurd. Think of $\lambda$ as the fine structure constant of quantum electrodynamics: $\lambda\leq0$ means that charge of the same sign attract each other. But the vacuum is clearly unstable in this regime: to decrease the energy of the field, create many electrons and as many positrons and let them collapse on each other. The same argument can be run for $\lambda\phi^4$ theory. 

The consequence of the non-analyticity of ${O}(\lambda)$ at $\lambda=0$ is irrevocable: the perturbative series \eqref{powerseries} is, at best, an \emph{asymptotic expansion}\footnote{A formal power series $\sum_{n\geq0}f_n\lambda^n$ is an \emph{asymptotic expansion} of a function $f(\lambda)$ as $\lambda\rightarrow0$ if, for any $N>0$, $f(\lambda)=\sum_{n=0}^Nf_n\lambda^n+\mathcal{O}(\lambda^{N+1})$. This does not imply that $\sum_{n\geq0}f_n\lambda^n$ converges.} of ${O}(\lambda)$ in the $\lambda\rightarrow0$ limit. But it is a classic theorem of Borel that there is an \emph{infinite} number of smooth functions asymptotic to {\it any} power series: the sequence of coefficients $(O_n)_{n\geq0}$ does not uniquely defines the function $O(\lambda)$, as it would if $O(\lambda)$ were analytic in a disk centered on $\lambda=0$.

\begin{figure}[h]
\begin{center}

\includegraphics[scale=0.2]{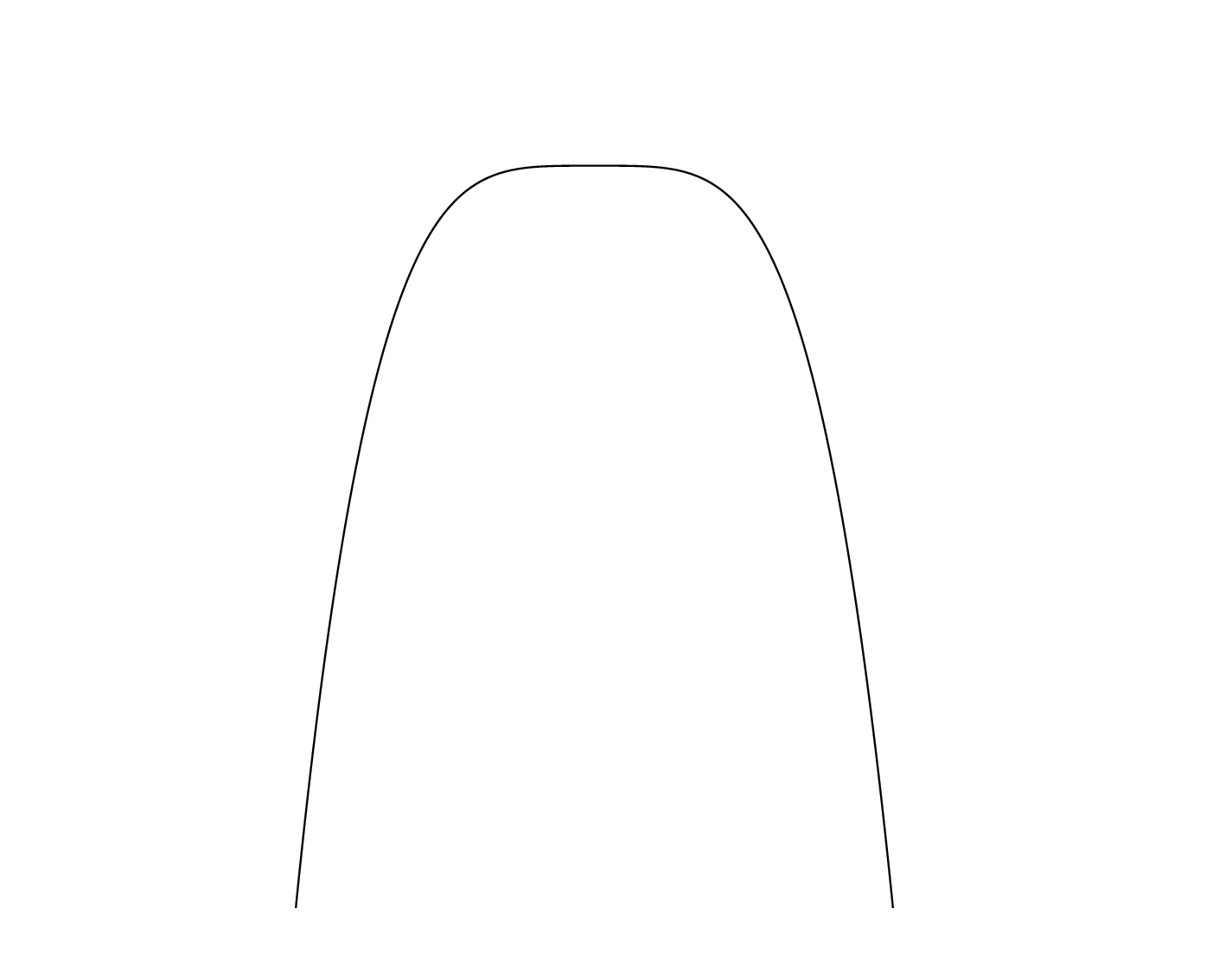}
\includegraphics[scale=0.3]{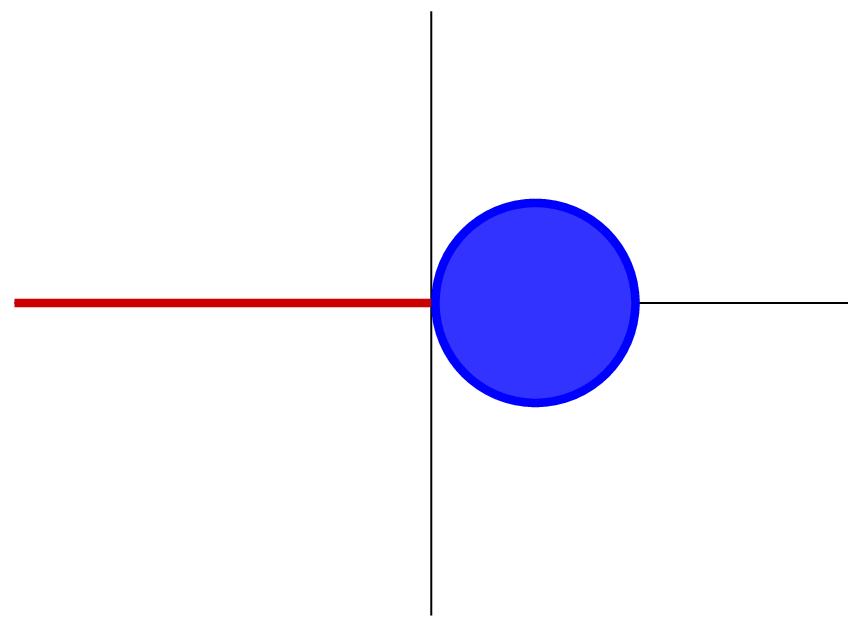}
\includegraphics[scale=0.2]{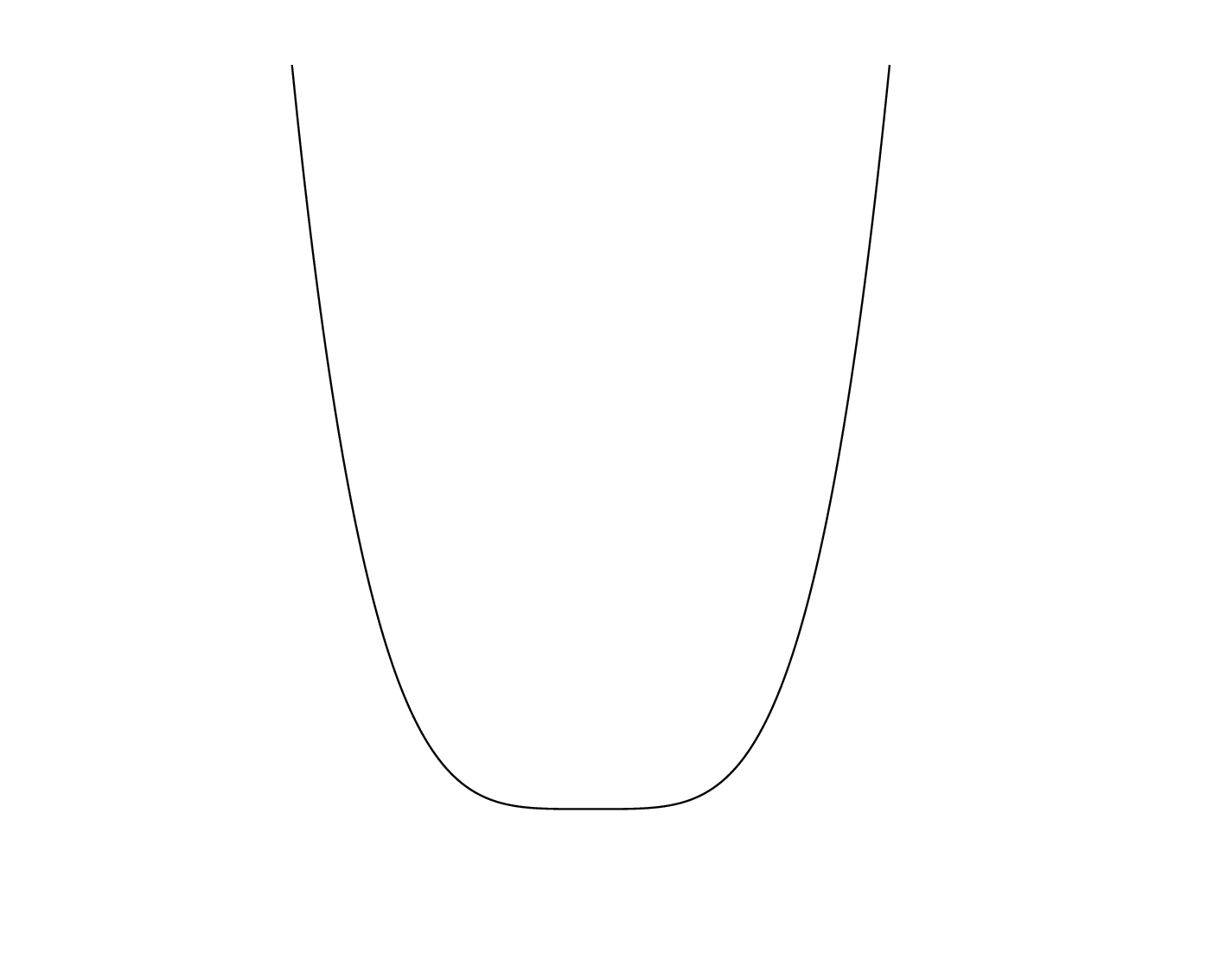}
\end{center}
\caption{Typical analyticity domain of a Borel summable function $O(\lambda)$: it is analytic in a disk where $\textrm{Re}\ \lambda>0$, corresponding to a stable potential (right), and has a branch cut along the $\lambda\leq0$ axis, corresponding to an unstable potential (left).}
\label{analyticborel}
\end{figure}

But what if $O(\lambda)$ is analytic in a disk \emph{tangent} to $\lambda=0$ in the $\textrm{Re}\ \lambda>0$ half-space, as in Fig. \ref{analyticborel}? This corresponds to the case mentioned above, where $O(\lambda)$ has singularities on the $\lambda\leq0$ axis (a branch cut), where the vacuum is unstable. 

\emph{Borel summability} states that, under certain conditions formulated by Nevanlinna \cite{Nevanlinna} and rediscovered by Sokal \cite{Sokal:1980ey}, the function $O(\lambda)$ \emph{can} be reconstructed from the coefficients $(O_n)_{n\geq0}$. These conditions are that\medskip\begin{itemize}
\item
$O$ is analytic in the disk $C_R := \{ \lambda\in\mathbb{C}\ \vert\ {\rm Re}\, \lambda^{-1} > 1/R\}$
\item
there are constants $A,B$ such that the bound 
\be\label{bor2}
\Big| O(\lambda)-\sum_{n=0}^NO_n\lambda^n\Big| \le A B^{N+1} (N+1)! \vert \lambda \vert^{N+1} 
\ee 
holds uniformly in $r$ and $\lambda$.
\end{itemize}\medskip
Then the power series 
\be\label{boreltransform}
B(t):=\sum_{n=0}^{\infty} O_n {
t^n \over n!}
\ee converges for $\vert t \vert < {1 \over B}$, and admits an analytic continuation in the strip 
\be
S_{B} := \{t\in\mathbb{C}\ \vert\ {\rm \ dist \ } (t, {\mathbb R}^+) < {1 \over B}\},
\ee
satisfying the bound
\be \label{bor3}
\vert B(t)  \vert \le { \rm C} e^{t/R} \qquad {\rm for \ } t\geq0  
\ee
for some constant $C$. Moreover, $O$ is represented in $C_R$ by the absolutely convergent integral
\be  \label{bor4}
O(\lambda) = {1 \over \lambda} \int_{0}^{\infty}dt\ e^{-{t/\lambda}}  B(t).
\ee
$B$ is then called the \emph{Borel transform} of $O$, and the complex $t$ plane is called the \emph{Borel plane}. 

There is a reciprocal to this theorem: consider the power series $\sum_{n\geq0} O_n \lambda^n$. If the power series $\sum_{n\geq0} O_n t^n/n!$ converges in a disk $\{\vert t \vert < {1 \over B}\}$, admits an analytic continuation 
$B(t)$ in the strip $S_{B}$ and satisfies the bound (\ref{bor3}) in this 
strip, then the function $O$ defined by the integral representation (\ref{bor4}) is analytic in 
$C_R$, has $\sum_{n=0}^{\infty} O_n \lambda^n$ as Taylor series at the origin and satisfies the 
uniform remainder estimates \eqref{bor2}. In this case we say that the series $\sum_{n\geq0} O_n \lambda^n$ is \emph{Borel summable}, and call the series $\sum_{n\geq0} O_n t^n/n!$ its \emph{Borel transform} and the function $O$ its \emph{Borel sum.} 

In other words, Borel summable series and Borel summable functions are in one-to-one
correspondence just like are ordinary series and germs of analytic functions.

\subsection{Proliferating and non-proliferating species}

Even when the perturbative series \eqref{powerseries} is actually Borel summable, it is very difficult to obtain the relevant information about its Borel transform, and in particular to show that \eqref{boreltransform} converges. The main reason for this is purely combinatorial: there are \emph{too many graphs} at a fixed order $n$. Indeed, the typical bound for a Feynman amplitude is of the type \cite{Riv}
\be
\vert O_G\vert\leq K^{V(G)}
\ee
for some constant $K$. Hence, one usually has to deal with the series
\be
\sum_{n\geq0}a_nK^n\lambda^n,
\ee
where $a_n$ is the number of Feynman graphs with $n$ vertices. Its Borel summability is in turn subject to the convergence of 
\be
\sum_{n\geq0}\f{a_n\lambda^n}{n!},
\ee
which is known as the \emph{generating series of Feynman graphs}. Unfortunately, it so happens that the latter has \emph{zero} radius of convergence, because $a_n$ grows too fast (typically like $n!^2$). Rivasseau has called \emph{proliferating} \cite{CFT0d} a combinatorial species whose generating series has zero radius of convergence. 

Thus, to get started with the constructive program, one must abandon the Feynman expansion and try to replace it by a series labelled by a non-proliferating species, such as trees.\footnote{Cayley's theorem states that there are $n^{n-2}$ labelled trees over $n$ vertices. The corresponding generating series clearly has a non-zero radius of convergence.} Each such tree should resum an infinite number of Feynman amplitudes \cite{Rivasseau2010}. This is precisely what the Magnen-Rivasseau cactus expansion does.





\section{A constructive tool: the cactus expansion}

Although it was initially designed to deal with \emph{matrix} $\phi^4$ theory \cite{Rivasseau:2007fr}, the cactus expansion applies to scalar models \cite{Magnen:2007uy,Rivasseau2011a} and higher-order interactions \cite{Rivasseau2010} as well. 

An important mathematical prerequisite for constructive field theory is the notion of \emph{degenerate Gaussian measure}, which we now review.


\subsection{Degenerate Gaussian measures}

If $A$ is a $n\times n$ positive-definite matrix, the Gaussian measure $d\mu^A$ associated to $A$ is usually defined by
\be\label{gaussianpositive}
d\mu_A(x)=\f{\sqrt{\det A}}{(2\pi)^{n/2}}e^{-\f{x.Ax}{2}}dx
\ee
where is $x\in\mathbb{R}^n$ and $dx$ is the standard $n$-dimensional Lebesgue measure. The inverse matrix $C=A^{-1}$ is the \emph{covariance} of the measure:
\be
\int_{\mathbb{R}^n}d\mu_A(x)\ x_ix_j=C_{ij}. 
\ee
When $A=C$ is the unit matrix, we call $d\mu_A$ the \emph{standard} Gaussian measure on $\mathbb{R}^n$. 

The key property of Gaussian measures is expressed by \emph{Wick's theorem}\footnote{This is a slight generalization of the usual Wick theorem, which is stated for monomials only.}: for any smooth, summable function $f$, we have
\be\label{expquadr}  \int d\mu_C (x)\  f(x)  = \exp\Big({\frac{1}{2}\frac{\partial}{\partial x}  C \frac{\partial}{\partial x }  }\Big) f(x)_{\vert_{x =0}}.
\ee
This implies that $d\mu_A$ is completely determined by its covariance $C$. 

Now, observe that unlike \eqref{gaussianpositive}, \eqref{expquadr} still makes sense if $C$ is \emph{non-invertible}, i.e. only positive-semidefinite. For any such matrix $C$, we can thus define the \emph{degenerate Gaussian measure} $d\mu_C$ with covariance $C$ by \eqref{expquadr}. When $C$ actually has zero modes, it is not be absolutely continuous with respect to the Lebesgue measure $dx$: it is a Dirac delta along each zero mode.

One important observation about (degenerate) Gaussian measures is that they have a well-defined \emph{functional} counterpart -- unlike the Lebesgue measure. The covariance $C$ is then a positive semi-definite \emph{operator}, and by Minlos' theorem $d\mu_C$ is supported on a suitable space of distributions. (In this setup the zero modes of $C$ are interpreted as \emph{gauge modes}.)
 
\subsection{An absolutely convergent expansion}

We are now ready to discuss the \emph{cactus expansion}. Consider a quartic field theory $\lambda\phi^4$, with \emph{partition function}
\be
\mathcal{Z}(\lambda):\doteq\int d\mu(\phi)\ e^{-\lambda\int\phi^4/8},
\ee
Here, $d\mu$ is some Gaussian measure, and we leave the number of components of $\phi$ (scalar, vector, matrix...) unspecified. The cactus expansion constructs $\mathcal{Z}(\lambda)$ as an \emph{absolutely convergent} series
\be\label{cactus}
\mathcal{Z}(\lambda)=\sum_F\mathcal{Z}_F(\lambda),
\ee
where $F$ ranges over the set of \emph{forests}, i.e. graphs without loops. The amplitudes $\mathcal{Z}_F(\lambda)$ have the following  properties:
\medskip\begin{enumerate}
\item
They factorize over the connected components of $F$. It follows that the \emph{free energy} $\mathcal{F}(\lambda):=\log\mathcal{Z}(\lambda)$ is given by the very same expansion \eqref{cactus}, with the sum restricted to \emph{connected} forests, viz. trees. 
\item
They are analytic functions of $\lambda$ in a Nevanlinna-Sokal disk $C_R$, hence so are $\mathcal{Z}(\lambda)$ and $\mathcal{F}(\lambda)$.
\item
They are easily bounded, and it is straightforward to check that the Taylor remainders of \eqref{cactus} about $\lambda=0$ satisfy the Nevanlinna-Sokal bound \eqref{bor2}. 
\end{enumerate}\medskip
Thus, in addition to giving a rigorous definition of (single-scale, i.e. with a ultraviolet cutoff) $\phi^4$ theory, the cactus expansion naturally shows that it is the Borel sum of its perturbative series. 

\subsection{Four ingredients for a cactus expansion}\label{recipe}

The cactus expansion is a recipe with four ingredients: (1) an intermediate field representation, (3) a replica trick, (3) a forest formula and (4) a resolvent bound.
\medskip

(1) The \emph{intermediate field representation} is a very useful trick to represent a quartic
interaction $\int\phi^4$ in terms of a cubic one $\int\phi\sigma\phi$. Schematically, it reads
\be
e^{-\lambda\int\phi^4/8}\doteq\int d\nu(\sigma)\ e^{-\f{1}{2}i\sqrt{\lambda}\int\phi \sigma\phi},
\ee
where $\sigma$ is a field with Gaussian ultralocal\footnote{To say that a Gaussian measure is \emph{ultralocal} means that, as a differential operator, its covariance is of zeroth-order, viz. contains no derivatives.} measure $d\nu$ (which in zero dimension
would simply be the standard Gaussian measure over $\mathbb{R}$). Performing the Gaussian integration over $\phi$, 
one can then express the original path integral $Z(\lambda)$ in terms of $\sigma$ only. For instance, in zero spacetime dimension, viz. for the numerical integral
\be\label{zerodim}
z(\lambda):=\int_{\mathbb{R}} d\mu(\phi)\ e^{-\lambda\phi^4/8},
\ee
we have
\begin{align}
z(\lambda)=\int_{\mathbb{R}} d\nu(\sigma) \int_{\mathbb{R}} d\mu(\phi)\ e^{-\f{1}{2}i\sqrt{\lambda}\int_{\mathbb{R}} \phi \sigma\phi}=\int_{\mathbb{R}} d\nu(\sigma) \det(1+i\sqrt{\lambda}\sigma)^{-1/2}
\end{align}
hence
\be
z(\lambda)=\int_{\mathbb{R}} d\nu(\sigma)\ e^{-\f{1}{2} \textrm{Log}(1+i\sqrt{\lambda}\sigma)}.
\ee
The new interaction vertex $V_\lambda(\sigma):=-\f{1}{2} \textrm{Log}(1+i\sqrt{\lambda}\sigma)$ was called a \emph{loop vertex} in \cite{Rivasseau:2007fr}. In more than zero dimension, i.e. for a genuine field theory, a propagator $C^{1/2}$ would typically sandwich the $\sigma$ field on both sides, and the loop vertex would be given by the trace of operator $\textrm{Log}(1+i\sqrt{\lambda}C^{1/2}\sigma C^{1/2})$. This trick can be generalized to any correlation functions and to more complicated models \cite{Rivasseau2010a}.

\medskip
(2) The \emph{replica trick} is a property of degenerate Gaussian measures. Let  $d\nu(\sigma)$ denote the standard Gaussian measure on $\mathbb{R}$, and $V\in L^n(\mathbb{R})$. Now, `replicate' $n$ times the variable $\sigma$, and consider the degenerate Gaussian measure $d\nu_n$ on $\mathbb{R}^n$ with covariance $C_{ij}=\la\sigma_i\sigma_j\ra=1$. The replica trick is then the statement that
\be
\int_{\mathbb{R}} d\nu(\sigma)\ V(\sigma)^n=\int_{\mathbb{R}^n} d\nu_n(\sigma_1,\dots,\sigma_n)\ \prod_{v=1}^nV(\sigma_v).
\ee
Let us emphasize that this is not a form of Fubini's theorem, which expresses an $n$-dimensional integral as a product of $n$ integrals. Here, we replace {\it one} $1$-dimensional integral by an $n$-dimensional one, with $(n-1)$ delta functions. 

\medskip
(3) Consider a smooth function $H$ of $\f{n(n-1)}{2}$ variables $\boldsymbol{h}=(h_{l})$, living on the lines $l$ of the complete graph over $n$ vertices. The so-called Brydges-Kennedy Taylor \emph{forest formula} \cite{Abdesselam:1994ap,BK} is a Taylor interpolation of $H$ with integral remainders indexed by labeled forests over $n$ vertices:

\be\label{forest}
H(\boldsymbol{1})=\sum_{F\in\mathcal{F}_n}\left(\prod_{l\in F}\int_0^1dh_l\right)\left(\prod_{l\in F}\f{\partial}{\partial h_l}\right)H(\boldsymbol{h}^F).
\ee
In this expression, $\cF_n$ denotes the set of forests over $n$ vertices, the products are over lines $l$ of each forest $F$, and $\boldsymbol{h}^F$ is the $\f{n(n-1)}{2}$-uple defined by $h^F_l:=\min_ph_p$, where $p$ runs over the unique path in $F$ connecting the source and target vertices of $l$. (If they are not connected by $F$, then $h^F_l:=0$.) 

\begin{figure}[t]
\begin{center}
\scalebox{1} 
{
\begin{pspicture}(0,-1.37)(10.56,1.39)
\psdots[dotsize=0.12](0.08,-1.29)
\psdots[dotsize=0.12](0.48,-0.69)
\psdots[dotsize=0.12](0.88,-1.29)
\psdots[dotsize=0.12](1.68,-1.29)
\psdots[dotsize=0.12](2.08,-0.69)
\psdots[dotsize=0.12](2.48,-1.29)
\psline[linewidth=0.04cm](2.08,-0.69)(1.68,-1.29)
\psdots[dotsize=0.12](3.28,-1.29)
\psdots[dotsize=0.12](3.68,-0.69)
\psdots[dotsize=0.12](4.08,-1.29)
\psdots[dotsize=0.12](4.88,-1.29)
\psdots[dotsize=0.12](5.28,-0.69)
\psdots[dotsize=0.12](5.68,-1.29)
\psdots[dotsize=0.12](6.48,-1.29)
\psdots[dotsize=0.12](6.88,-0.69)
\psdots[dotsize=0.12](7.28,-1.29)
\psdots[dotsize=0.12](8.08,-1.29)
\psdots[dotsize=0.12](8.88,-1.29)
\psdots[dotsize=0.12](8.48,-0.69)
\psdots[dotsize=0.12](9.68,-1.29)
\psdots[dotsize=0.12](10.08,-0.69)
\psdots[dotsize=0.12](10.48,-1.29)
\psline[linewidth=0.04cm](3.68,-0.69)(4.08,-1.29)
\psline[linewidth=0.04cm](4.88,-1.29)(5.68,-1.29)
\psline[linewidth=0.04cm](6.48,-1.29)(6.88,-0.69)
\psline[linewidth=0.04cm](6.88,-0.69)(7.28,-1.29)
\psline[linewidth=0.04cm](8.48,-0.69)(8.08,-1.29)
\psline[linewidth=0.04cm](8.08,-1.29)(8.88,-1.29)
\psline[linewidth=0.04cm](10.08,-0.69)(10.48,-1.29)
\psline[linewidth=0.04cm](9.68,-1.29)(10.48,-1.29)
\psdots[dotsize=0.12](4.68,0.31)
\psdots[dotsize=0.12](5.88,0.31)
\psdots[dotsize=0.12](5.28,1.31)
\psline[linewidth=0.04cm](5.28,1.31)(5.88,0.31)
\psline[linewidth=0.04cm](5.88,0.31)(4.68,0.31)
\psline[linewidth=0.04cm](4.68,0.31)(5.28,1.31)
\usefont{T1}{ptm}{m}{n}
\rput(4.73,0.855){\small 1}
\usefont{T1}{ptm}{m}{n}
\rput(5.8,0.875){\small 2}
\usefont{T1}{ptm}{m}{n}
\rput(5.27,0.055){\small 3}
\end{pspicture} 
}
\end{center}
\caption{The complete graph over 3 vertices, and its 7 forests, matching the $7$ terms in (\ref{exforest}).}
\label{forests}
\end{figure}
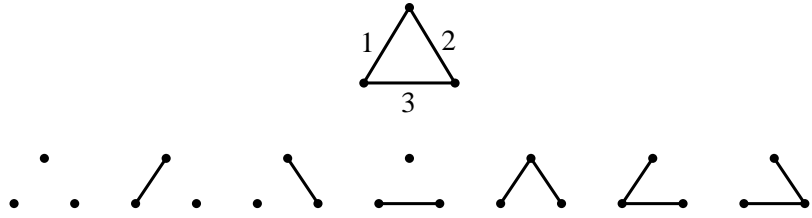

One can easily check that for $n=2$, this is nothing but the fundamental theorem of calculus: 
\be
H(1)=H(0)+\int_{0}^1dh\ H'(h).
\ee 
For higher values of $n$, on the other hand, the outcome of (\ref{forest}) is genuinely non-trivial, as the case $n=3$ already demonstrates (Fig. \ref{forests}):

\begin{align}\label{exforest}
	\begin{split}
H(1,1,1)&=H(0,0,0)\\+&\int_0^1 dh_1\ \partial_1H(h_1,0,0)+\int_0^1 dh_2\ \partial_2H(0,h_2,0)+\int_0^1 dh_3\ \partial_3H(0,0,h_3)\\
 &+ \int_0^1dh_1\int_0^1dh_2\ \partial^2_{12}H(h_1,h_2,\min(h_1,h_2))\\&+\int_0^1dh_1\int_0^1dh_3\ \partial^2_{13}H(h_1,\min(h_1,h_3),h_3)
\\ 
 &+ \int_0^1dh_2\int_0^1dh_3\ \partial^2_{23}H(\min(h_2,h_3),h_2,h_3).
\end{split}
\end{align}

\medskip
(4) If $\lambda$ is a complex number with positive real part, and $\sigma$ a real number, one has
\be\label{ineq}
\vert1+i\sqrt{\lambda}\sigma\vert^{-1}\leq\sqrt{2}.
\ee
Let $\Sigma$ be a Hermitian matrix. Taking the supremum over the spectrum of $\Sigma$ in (\ref{ineq}) gives the \emph{resolvent bound}

\be
\Vert(1+i\sqrt{\lambda}\Sigma)^{-1}\Vert\leq\sqrt{2},
\ee
where $\Vert\cdot\Vert$ denotes the operator norm, $\Vert A\Vert:=\sup_{\vert x\vert=1}\vert Ax\vert$. 

\subsection{A toy example: $\phi^4$ theory in $0$ dimensions}\label{zerodimensions}

To illustrate how these four ingredients can be combined to yield the cactus expansion, we now consider the $\phi^4$ field theory in $0$ dimension,\footnote{For a historical perspective on constructive methods applied to this $0$-dimensional model, see \cite{CFT0d}.} i.e. the integral \eqref{zerodim}. The functions $z(\lambda)$ and $f(\lambda):=\log z(\lambda)$ it defines are analytic in the cut plane $\mathbb{C}\setminus\mathbb{R}_-$, and hence admit, at best, a Borel expansion about $\lambda=0$. Using the four ingredients presented above, we now show that this is indeed the case.

We start from the intermediate field representation
\be\label{intermediatefield0dim}
z(\lambda)=\int_{\mathbb{R}} d\nu(\sigma)\ e^{V_{\lambda}(\sigma)},
\ee
with loop vertex $V_{\lambda}(\sigma):=-\f{1}{2}\textrm{Log}(1+i\sqrt{\lambda}\sigma)$. 

First, expand the exponential in \eqref{intermediatefield0dim} in powers of $V_\lambda(\sigma)$, swap integration and summation\footnote{Of course, it is precisely such an interchange between integration and summation that yields the divergent perturbative series. Note that here, however, the process is licit because $\int d\mu(\sigma)\ e^{1/2\vert\mathrm{Log}(1+i\sqrt{\lambda}\sigma)\vert}<\infty$, and so Lebesgue's dominated convergence theorem applies.} and apply the replica trick to the order-$n$ term to obtain
\be\label{sum}
Z(\lambda)=\sum_{n=0}^{\infty}\f{1}{n!}\int_{\mathbb{R}^n} d\nu_n(\sigma_1,\dots,\sigma_n)\ \prod_{v=1}^nV_{\lambda}(\sigma_v).
\ee

Next, consider the matrix $C^{\boldsymbol{h}}$, parametrized by an $\f{n(n-1)}{2}$-uple $\boldsymbol{h}$ and defined by $C^{\boldsymbol{h}}_{ii}:=1$ and $C^{\boldsymbol{h}}_l:=h_l$, where $l=\{ij\}$ ($i\neq j$), and let $d\nu^{\boldsymbol{h}}_n(\sigma_1,\dots,\sigma_n)$ be the Gaussian measure with covariance $C^{\boldsymbol{h}}$. Then apply the forest formula to the function

\be
H(\boldsymbol{h}):=\int d\nu_n^{\boldsymbol{h}}(\sigma_1,\dots,\sigma_n)\ \prod_{v=1}^nV_{\lambda}(\sigma_v).
\ee
This gives
\be\label{toy loop vertex}
z(\lambda)=\sum_{n=0}^{\infty}\f{1}{n!}\sum_{F\in\mathcal{F}_n}\left(\prod_{l\in F}\int_0^1dh_l\right)\left(\prod_{l\in F}\f{\partial}{\partial h_l}\right)\int_{\mathbb{R}^n} d\nu^{\boldsymbol{h^F}}_n(\sigma_1,\dots,\sigma_n)\prod_{v=1}^nV_{\lambda}(\sigma_v),
\ee
or equivalently
\be
z(\lambda)=\sum_{F}z_F(\lambda)
\ee
with $F$ ranging over the set of all forests, and
\be
z_F(\lambda)=\f{1}{n(F)!}\left(\prod_{l\in F}\int_0^1dh_l\right)\left(\prod_{l\in F}\f{\partial}{\partial h_l}\right)\int_{\mathbb{R}^n} d\nu^{\boldsymbol{h^F}}_n(\sigma_1,\dots,\sigma_n)\prod_{v=1}^nV_{\lambda}(\sigma_v).
\ee

As announced, the summand factorizes along connected components of each forest. This implies that 
\be\label{toytrees}
f(\lambda)=\sum_{T}z_T(\lambda)
\ee
where now $T$ ranges over the set of all \emph{trees}. It is these trees $T$ over loop vertices $V_{\lambda}(\sigma_{v})$ which we coin \emph{cacti}.

Since the dependence of the covariance of $d\nu^{\boldsymbol{h^T}}_n(\sigma_1,\dots,\sigma_n)$
in the $h$ variables is linear, applying the derivative $\partial/\partial h_l$ in \eqref{toytrees} is easy using (\ref{expquadr}):
it amounts to an additional insertions of $\partial^2/\partial\sigma_{s(l)}\partial\sigma_{t(l)}$ in the integral, where $s(l)$ and $t(l)$
are respectively the starting and ending vertices of the line $l$. Hence
\begin{multline}\label{derivatexp}
\left(\prod_{l\in T}\f{\partial}{\partial h_l}\right)\int_{\mathbb{R}^n} d\nu^{\boldsymbol{h^T}}_n(\sigma_1,\dots,\sigma_n)\prod_{v=1}^nV_{\lambda}(\sigma_v)\\=\int_{\mathbb{R}^n} d\nu^{\boldsymbol{h^T}}_n(\sigma_1,\dots,\sigma_n)\left(\prod_{l\in T}\f{\partial^2}{\partial\sigma_{s(l)}\partial\sigma_{t(l)}}\right)\prod_{v=1}^nV_{\lambda}(\sigma_v).
\end{multline}

Consider the loop vertex $V_{\lambda}(\sigma_v)$, with coordination $k_v$ in the tree $T$. Thanks to the resolvent bound, we have

\be\label{bound}
\Big| \f{\partial^{k_v}}{\partial\sigma_v^{k_v}}V_{\lambda}(\sigma_v)\Big|=(k_v-1)!\vert\lambda\vert^{k_v/2}\vert1+i\sqrt{\lambda}\sigma_v\vert^{-k_v}\leq2^{k_v/2}(k_v-1)!\vert\lambda\vert^{k_v/2},
\ee
and thus, since there are $n-1$ lines in a tree over $n$ vertices,
\be
\Big|\left(\prod_{l\in T}\f{\partial^2}{\partial\sigma_{s(l)}\partial\sigma_{t(l)}}\right)\prod_{v=1}^nV_{\lambda}(\sigma_v)\Big|\leq2^{n-1}\vert\lambda\vert^{n-1}\prod_{v=1}^n(k_v-1)!
\ee
This bound goes through the normalized integrals over the $\sigma$'s and the $h$'s. Using Cayley's formula for the number of trees over $n$ labeled vertices with fixed coordinations $k_v$
\be
\f{\prod_{v=1}^n(k_v-1)!}{n!},
\ee
we find that
\be
z_T(\lambda)\leq 2^{n-1}\vert\lambda\vert^{n-1}.
\ee
This shows that the cactus expansion (\ref{toytrees}) of $f$ converges uniformly in a half-disk $D_{R}=\{\lambda\in\mathbb{C},\Re\lambda\geq0,\vert\lambda\vert\leq R\}$, with $R<\f{1}{2}$, which obviously contains the Nevanlinna-Sokal disk $C_{R/2}$.

Furthermore, the order-$r$ Taylor-Lagrange remainder 
\be
T_rf(\lambda):= f(\lambda)- \sum_{p=0}^{r-1} 
\frac{\lambda^p}{p!} f^{(p)} (0)
\ee
can easily be shown to satisfy the Nevanlinna-Sokal criterion (\ref{bor2}). Indeed, consider a cactus 
amplitude $z_T(\lambda)$ with $n$ loop vertices. By (\ref{derivatexp})-(\ref{bound}), each such amplitude is made of an explicit factor $\lambda^{n-1}$ times an integral over $d\nu^{\boldsymbol{h^T}}_n(\sigma_1,\dots,\sigma_n)$ of
a product of $2n-2$ resolvents
\be
R_{l_{v}} (\lambda, \sigma_{v})  :=(1+i\sqrt{\lambda}\sigma_{v})^{-1},
\ee
where $l_{v}$ denote a half-line hooked to a vertex $v$. Hence for $r \le n-1$, $T_r  z_T(\lambda) =  z_T(\lambda) $, and for $r\ge n$
\be  T_r z_T(\lambda)   =\f{1}{n!}  \lambda^{n-1} T_{r-n+1}
 \left(\prod_{l\in T}\int_0^1dh_l\right)\int_{\mathbb{R}^n} d\nu^{\boldsymbol{h^T}}_n(\sigma_1,\dots,\sigma_n) \prod_{l_{v}=1}^{2n-2} R_{l_{v}} (\lambda, \sigma_{v}) .
\ee
But by (\ref{expquadr}) and since $F=  \prod_{l_{v}=1}^{2n-2} R_{l_{v}}$ is solely a function of $\sqrt{\lambda} \sigma $
we have
\be
\int_{\mathbb{R}^n} d\nu^{\boldsymbol{h^T}}_n(\sigma_1,\dots,\sigma_n)  \prod_{l_{v}=1}^{2n-2} R_{l_{v}} (\lambda, \sigma_{v}) = 
e^{\frac{1}{2}\lambda \frac{\partial}{\partial \sigma}  C^{\boldsymbol{h}^T} \frac{\partial}{\partial \sigma }  }  
 \prod_{l_{v}=1}^{2n-2}  R_{l_{v}} (1, \sigma)  \; \vert_{\sigma_{v} =0}.
\ee 
Hence the Taylor-Lagrange formula applies to the exponential:
\be  T_k e^{\lambda H}  = \int_0^1 dt\ \frac{(1-t)^{k-1}}{(k-1)!}   \lambda^{k}  H^{k} e^{\lambda t H} 
\ee
and the operator $H^k$ creates exactly $k$ additional insertions of 
lines $\f{\partial^2}{\partial\sigma_{s(l')}\partial\sigma_{t(l')}}$. The combinatorics of $2k$ derivations 
on a product of $2n-2$ resolvents costs a factor $\frac{(2r-1))!}{(2n-1)!}$
times a new Gaussian integral with $2n-2+2k$ resolvents of the same type $R_{l_{v}'}$. The $\lambda$
factor can be then transferred back to a $\sqrt\lambda$ factor in the resolvents:
\be  e^{\frac{1}{2}\lambda t \frac{\partial}{\partial \sigma}  C^{\boldsymbol{h}^T} \frac{\partial}{\partial \sigma }  }   \prod_{l_{v}'=1}^{2n-2+2k} 
R_{l_{v}'} (1, \sigma_{v}) \; \vert_{\sigma_{v} =0} =  e^{\frac{1}{2}t \frac{\partial}{\partial \sigma}  C^{\boldsymbol{h}^T} \frac{\partial}{\partial \sigma }  }   
\prod_{l_{v}'=1}^{2n-2+2k}  R_{l_{v}'} (\lambda, \sigma_{v}) \; \vert_{\sigma_{v} =0}  .
\ee
Hence, since $k = r-n+1$ the convergent series $\sum_{T}T_r z_T(\lambda)$ can be bounded exactly as before,
except for two facts: each term contains a factor $\lambda^{r}$ and we have also to add to the bounds a factor 
\be
\int_0^1 dt \frac{(1-t)^{r-n}}{(r-n)!} \frac{(2r-1))!}{(2n-1)!}=\frac{(2r-1))!}{(r-n+1)!(2n-1)!}.
\ee
This last factor is maximal for the trivial tree with $n=1$,  and certainly bounded by $2^r\, r!$, from 
which it follows that for some constant $K$
\be
\vert T_r  f(\lambda)\vert\leq K^r  r!\vert\lambda\vert^r.
\ee

To summarize, the cactus expansion allows not only to trade the {\it asymptotic} perturbative series
\be\label{asymptotictaylor}
f(\lambda)\simeq\sum_{n=0}^{\infty}\f{1}{n!}f^{(n)}(0)
\ee
 for the {\it convergent} expression (\ref{toytrees}), but also to check the Sokal-Nevanlinna criteria, proving Borel summability of \eqref{asymptotictaylor}. Of course, in this toy example, Borel summability of $z$ is obvious and Borel summability of $f=\log z$ could be shown by more elementary methods; the power of the cactus expansion becomes manifest when it comes to the constructive analysis of the $\phi^4$ \emph{field} theory \cite{Magnen:2007uy}, and of the matrix $\phi^4$ model \cite{Rivasseau:2007fr}. And of Boulatov's group field theory. 

\section{Borel summability and scaling behavior}

Indeed, the fact that Boulatov's group field theory has a \emph{quartic} interaction allows to use the cactus expansion to study its Borel summability. The great advantage of this method over Freidel and Louapre's  \cite{Freidel:2002tg} lies in the control it provides over the high-spin cutoff: by this token, we can unravel the constructive \emph{scaling limit} of this group field theory. 

\subsection{A twofold regularization}

We introduced Boulatov's model in the first chapter of this thesis. In the language of degenerate Gaussian measures, the Boulatov partition function is defined formally by 
\be\label{formal}
\mathcal{Z}_B(\lambda):\doteq\int d\mu[\phi]\ e^{-\lambda T[\phi]/8},
\ee
where 
\be
\int_{\textrm{SU}(2)^6}\prod_{i=1}^{6}dg_i\ \phi(g_1,g_2,g_3)\phi(g_3,g_4,g_5)\phi(g_5,g_2,g_6)\phi(g_6,g_4,g_1),
\ee
and $d\mu[\phi]$ is the Gaussian measure with covariance $C_{\phi}$ defined by
\be
(C_{\phi}\phi)(g_1,g_2,g_3):=\f{1}{3}\sum_c\int_{\textrm{SU}(2)} dh\ \phi(g_{c(1)}h,g_{c(2)}h,g_{c(3)}h).
\ee
Here, $c$ ranges over the cyclic permutations of three elements. This measure is clearly degenerate: the covariance $C_\phi$ is the \emph{orthogonal projector} onto the subspace of $L^2(\textrm{SU}(2)^3)$ invariant under $\textrm{SU}(2)$ averaging and cyclic permutations. 

Of course, the partition function (\ref{formal}) is only formal and needs regularizations to become mathematically well-defined. The problem is twofold: \medskip\begin{itemize}
\item
 the Fourier space of the field $\phi$ is non-compact, although discrete, and hence `ultraviolet' divergences arise,
 \item
 Boulatov's quartic interaction $T[\phi]$ is not positive, hence unstable. This can be seen from its Fourier space formulation where the interaction term reduces to an oscillatory $\{6j\}$ symbol.
\end{itemize}\medskip

To cure the first problem, we follow \cite{Freidel:2002tg} and introduce a cutoff $\La$ truncating the Peter-Weyl (or Fourier) decomposition of the field:
\be
\phi(g_1,g_2,g_3)=\sum_{{j}_1,j_2,j_3}^{\La} \textrm{tr} \left(\Phi_{j_1,j_2,j_3} D^{j_1}(g_1) D^{j_2}(g_2) D^{j_3}(g_3)\right).
\ee
In this formula, the sum runs over the spins $j_1,j_2,j_3$ up to $\La$; $D^j(g)$ denotes the $(2j+1)$-dimensional matrix representation of $g$; $\Phi_{j_1,j_2,j_3}$ are the Fourier modes of the field $\phi$
viewed as complex-valued tensors and $\textrm{tr}$ denotes the trace in the space carrying the tensor product representation associated to the spins $j_{1}$, $j_{2}$, $j_{3}$. In the following, we denote $\cH^{(\La)}$ the subspace of $L^2(\su^3)$ resulting from this truncation, and $\cH_{0}^{(\La)}:=\cH^{(\La)}\cap\im C_{\phi}$. The number of degrees of freedom left is thus given by

\be
\dim\cH_{0}^{(\La)}=\mathcal{O}(\La^6).
\ee

\begin{figure}[h]
\begin{center}
\scalebox{0.8} 
{
\begin{pspicture}(0,-2.4)(15.696783,2.4)
\psbezier[linewidth=0.024](7.496783,1.92)(7.496783,1.12)(7.496783,0.72)(6.2967825,0.72)
\psbezier[linewidth=0.024](8.2967825,-1.28)(8.2967825,-0.48)(8.2967825,-0.08)(9.496782,-0.08)
\psbezier[linewidth=0.024](9.496782,0.72)(8.696783,0.72)(8.2967825,0.72)(8.2967825,1.92)
\psbezier[linewidth=0.024](6.2967825,-0.08)(7.0967827,-0.08)(7.496783,-0.08)(7.496783,-1.28)
\psline[linewidth=0.024cm](7.8967824,1.92)(7.8967824,-1.28)
\psline[linewidth=0.024cm](6.2967825,0.32)(7.6967826,0.32)
\psline[linewidth=0.024cm](8.096783,0.32)(9.496782,0.32)
\psbezier[linewidth=0.024](13.096783,1.92)(13.096783,1.12)(13.096783,0.72)(11.896783,0.72)
\psbezier[linewidth=0.024](15.096783,0.72)(14.2967825,0.72)(13.896783,0.72)(13.896783,1.92)
\psbezier[linewidth=0.024](11.896783,-0.08)(12.696783,-0.08)(13.096783,-0.08)(13.096783,-1.28)
\psbezier[linewidth=0.024](13.896783,-1.28)(13.896783,-0.48)(13.896783,-0.08)(15.096783,-0.08)
\psbezier[linewidth=0.024](11.896783,0.32)(13.696783,0.32)(13.496782,1.12)(13.496782,1.92)
\psbezier[linewidth=0.024](13.496782,-1.28)(13.496782,0.52)(14.096783,0.32)(15.096783,0.32)
\usefont{T1}{ptm}{m}{n}
\rput(7.976783,-2.235){Tetrahedron}
\usefont{T1}{ptm}{m}{n}
\rput(13.5467825,-2.215){Pillow}
\usefont{T1}{ptm}{m}{n}
\rput(7.5167828,2.16){\footnotesize $1$}
\usefont{T1}{ptm}{m}{n}
\rput(7.9167824,2.16){\footnotesize $2$}
\usefont{T1}{ptm}{m}{n}
\rput(8.316783,2.16){\footnotesize $3$}
\usefont{T1}{ptm}{m}{n}
\rput(9.736783,0.72){\footnotesize $3$}
\usefont{T1}{ptm}{m}{n}
\rput(9.736783,0.32){\footnotesize $4$}
\usefont{T1}{ptm}{m}{n}
\rput(9.736783,-0.04){\footnotesize $5$}
\usefont{T1}{ptm}{m}{n}
\rput(8.2967825,-1.5){\footnotesize $5$}
\usefont{T1}{ptm}{m}{n}
\rput(7.8767824,-1.5){\footnotesize $2$}
\usefont{T1}{ptm}{m}{n}
\rput(7.5167828,-1.5){\footnotesize $6$}
\usefont{T1}{ptm}{m}{n}
\rput(6.0767827,0.7){\footnotesize $1$}
\usefont{T1}{ptm}{m}{n}
\rput(6.0767827,0.3){\footnotesize $4$}
\usefont{T1}{ptm}{m}{n}
\rput(6.0767827,-0.06){\footnotesize $6$}
\usefont{T1}{ptm}{m}{n}
\rput(13.116782,2.22){\footnotesize $1$}
\usefont{T1}{ptm}{m}{n}
\rput(13.516783,2.22){\footnotesize $2$}
\usefont{T1}{ptm}{m}{n}
\rput(13.916782,2.22){\footnotesize $3$}
\usefont{T1}{ptm}{m}{n}
\rput(15.416782,0.7){\footnotesize $3$}
\usefont{T1}{ptm}{m}{n}
\rput(15.416782,0.3){\footnotesize $4$}
\usefont{T1}{ptm}{m}{n}
\rput(15.416782,-0.06){\footnotesize $5$}
\usefont{T1}{ptm}{m}{n}
\rput(11.656782,0.68){\footnotesize $1$}
\usefont{T1}{ptm}{m}{n}
\rput(11.656782,0.28){\footnotesize $2$}
\usefont{T1}{ptm}{m}{n}
\rput(11.656782,-0.08){\footnotesize $6$}
\usefont{T1}{ptm}{m}{n}
\rput(13.936783,-1.5){\footnotesize $5$}
\usefont{T1}{ptm}{m}{n}
\rput(13.516783,-1.5){\footnotesize $4$}
\usefont{T1}{ptm}{m}{n}
\rput(13.156782,-1.5){\footnotesize $6$}
\psline[linewidth=0.024cm](0.012,0.8)(1.5967826,0.8)
\psline[linewidth=0.024cm](0.012,0.4)(1.5967826,0.4)
\psline[linewidth=0.024cm](0.012,0.0)(1.5967826,0.0)
\psframe[linewidth=0.024,dimen=outer](2.1967826,1.0)(1.5967826,-0.2)
\psline[linewidth=0.024cm](2.1967826,0.8)(3.7967825,0.8)
\psline[linewidth=0.024cm](2.1967826,0.4)(3.7967825,0.4)
\psline[linewidth=0.024cm](2.1967826,0.0)(3.7967825,0.0)
\usefont{T1}{ptm}{m}{n}
\rput(1.9167826,-2.235){Covariance}
\end{pspicture} 
}
\caption{The covariance $C_{\phi}$, the Boulatov tetrahedral vertex $T[\phi]$, and the Freidel-Louapre pillow $P[\phi]$. The labels on the vertices match the ordering of the group elements in the integrand of $T[\phi]$ and $P[\phi]$.}
\label{tetrahedron}
\end{center}
\end{figure}
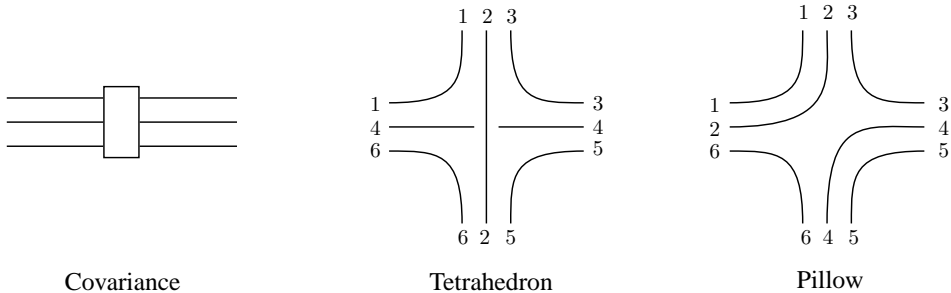

Now for the second problem: the fact that Boulatov's interaction $T[\phi]$ is not positive. 
To fix this shortcoming, Freidel and Louapre proposed to add the following `pillow' term\footnote{The word `pillow' refers to the geometric interpretation of the GFT vertex: if Boulatov's $T[\phi]$ is a tetrahedron, then Freidel and Louapre's $P[\phi]$ are two tetrahedra glued along two triangles -- a pillow.} to the action \cite{Freidel:2002tg}, see Fig. \ref{tetrahedron}
\be
P[\phi]:=\int_{\textrm{SU}(2)^6}\prod_{i=1}^{6}dg_i\ \phi(g_1,g_2,g_3)\phi(g_3,g_4,g_5)\phi(g_5,g_4,g_6)\phi(g_6,g_2,g_1).
\ee
Indeed, they showed that when $\vert\delta\vert\leq1$,  $I_{\delta}[\phi]:=P[\phi]+\delta T[\phi]$ is positive. To this aim, they introduce the `squaring' operator $S$ mapping $\phi$ to the function $S\phi$ on $\su^4$ defined by
 \be
 S\phi(g_1,g_2,g_3,g_4):=\int dg\ \phi(g_1,g_2,g)\phi(g,g_3,g_4).
 \ee
In terms of this new field, the modified interaction $I_{\delta}$ reads
\be
I_{\delta}[\phi]=\la S\phi\vert(1+\delta\mathcal{T}) S\phi\ra_4,
\ee
where $\mathcal{T}$ is the involution transposing the central arguments of $S\phi$:
\be
\mathcal{T}S\phi(g_1,g_2,g_3,g_4):=S\phi(g_1,g_3,g_2,g_4),
\ee 
and $\la\cdot\vert\cdot\ra_{4}$ is the standard inner product in $L^2(\su^4)$. Since $(1+\delta\mathcal{T})$ is a positive operator, the modified quartic interaction $I_{\delta}$ is indeed positive. 

Combining the cutoff on spins $\La$ and the Boulatov-Freidel-Louapre (BFL) interaction, we get the (now well-defined) regularized partition function
\be\label{regularizedBFL}
\mathcal{Z}^{(\La)}_{\textrm{BFL}}(\lambda):=\int d\mu^{(\La)}[\phi]\ e^{-\lambda I_{\delta}[\phi]/8}.
\ee


The following bound on this model's Feynman amplitudes, proved in \cite{Magnen:2009dq}, will useful later on. Let $G$ be a (vacuum, i.e. without external legs) Feynman graph of \eqref{regularizedBFL}. The corresponding Feynman amplitude $\mathcal{A}_{\textrm{BFL}}(G)$ satisfies\footnote{Rivasseau and Magnen later found a caveat in this theorem, see \cite{BenGeloun2010}.}
\be\label{perturbativebound}
\vert\mathcal{A}_{\textrm{BFL}}(G)\vert \leq (\lambda K_\delta)^{V(G)} \Lambda^{6 + 3V(G)}
\ee
for some $\delta$-dependent constant $K_\delta$. 

\subsection{Intermediate field representation}

\begin{figure}[h]
\begin{center}
\scalebox{0.9} 
{
\begin{pspicture}(0,-2.056)(9.024,2.016)
\psbezier[linewidth=0.024](5.4166675,1.1479173)(6.212,0.604)(6.212,0.0039999997)(5.432209,-0.54906785)
\psbezier[linewidth=0.024](6.412,-0.196)(6.212,-0.196)(5.812,-0.596)(5.432209,-0.9490679)
\psbezier[linewidth=0.024](6.412,-0.596)(6.012,-0.596)(5.612,-1.196)(5.432209,-1.3490678)
\psbezier[linewidth=0.024](6.412,1.004)(6.012,1.204)(5.812,1.604)(5.412,2.004)
\psbezier[linewidth=0.024](6.412,0.604)(6.212,0.604)(5.612,1.404)(5.412,1.604)
\psline[linewidth=0.04cm,linestyle=dashed,dash=0.16cm 0.16cm](6.412,1.004)(8.012,1.004)
\psline[linewidth=0.04cm,linestyle=dashed,dash=0.16cm 0.16cm](6.412,0.604)(8.012,0.604)
\psline[linewidth=0.04cm,linestyle=dashed,dash=0.16cm 0.16cm](6.412,-0.196)(8.012,-0.196)
\psline[linewidth=0.04cm,linestyle=dashed,dash=0.16cm 0.16cm](6.412,-0.596)(8.012,-0.596)
\psbezier[linewidth=0.024](9.007333,1.1479173)(8.212,0.604)(8.212,0.0039999997)(8.991791,-0.54906785)
\psbezier[linewidth=0.024](8.012,-0.196)(8.212,-0.196)(8.612,-0.596)(8.991791,-0.9490679)
\psbezier[linewidth=0.024](8.012,-0.596)(8.412,-0.596)(8.812,-1.196)(8.991791,-1.3490678)
\psbezier[linewidth=0.024](8.012,1.004)(8.412,1.204)(8.612,1.604)(9.012,2.004)
\psbezier[linewidth=0.024](8.012,0.604)(8.212,0.604)(8.812,1.404)(9.012,1.604)
\psbezier[linewidth=0.024](0.016667435,1.1479173)(0.812,0.604)(0.812,0.0039999997)(0.03220895,-0.54906785)
\psbezier[linewidth=0.024](1.012,-0.196)(0.612,-0.396)(0.212,-0.796)(0.03220895,-0.9490679)
\psbezier[linewidth=0.024](1.012,-0.596)(0.612,-0.596)(0.212,-1.196)(0.03220895,-1.3490678)
\psbezier[linewidth=0.024](1.012,1.004)(0.612,1.204)(0.412,1.604)(0.012,2.004)
\psbezier[linewidth=0.024](1.012,0.604)(0.612,0.804)(0.212,1.404)(0.012,1.604)
\psline[linewidth=0.04cm,linestyle=dashed,dash=0.16cm 0.16cm](1.012,1.004)(2.612,1.004)
\psline[linewidth=0.04cm,linestyle=dashed,dash=0.16cm 0.16cm](1.012,0.604)(2.612,-0.196)
\psline[linewidth=0.04cm,linestyle=dashed,dash=0.16cm 0.16cm](1.012,-0.196)(2.612,0.604)
\psline[linewidth=0.04cm,linestyle=dashed,dash=0.16cm 0.16cm](1.012,-0.596)(2.612,-0.596)
\psbezier[linewidth=0.024](3.6073325,1.1479173)(2.812,0.604)(2.812,0.0039999997)(3.5917912,-0.54906785)
\psbezier[linewidth=0.024](2.612,-0.196)(3.012,-0.396)(3.412,-0.796)(3.5917912,-0.9490679)
\psbezier[linewidth=0.024](2.612,-0.596)(3.012,-0.596)(3.412,-1.196)(3.5917912,-1.3490678)
\psbezier[linewidth=0.024](2.612,1.004)(3.012,1.204)(3.212,1.604)(3.612,2.004)
\psbezier[linewidth=0.024](2.612,0.604)(3.012,0.804)(3.412,1.404)(3.612,1.604)
\usefont{T1}{ptm}{m}{n}
\rput(7.282,-1.891){Pillow}
\usefont{T1}{ptm}{m}{n}
\rput(1.892,-1.891){Tetrahedron}
\end{pspicture} 
}
\end{center}
\caption{Slicing the BFL vertices with an intermediate field $\sigma$ over $\su^4$: the dashed lines are combined in the covariance $C_{\sigma}$.}
\label{split}
\end{figure}
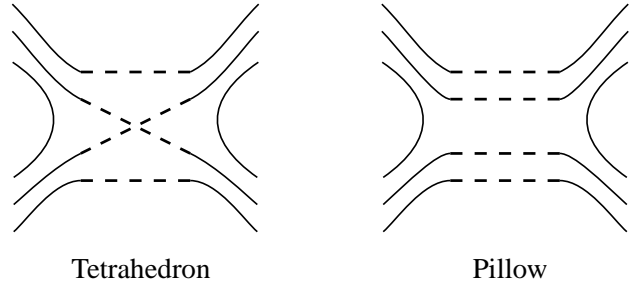

Let us now construct the cactus expansion of the BFL model. Following the recipe explained in sec. \ref{recipe}, we start by introducing a ultralocal intermediate field $\sigma$ on $\su^4$ slicing the BFL $\phi^4$ vertices into two $\phi^2\sigma$ vertices as in Fig. \ref{split}:
\be
e^{-\lambda I_{\delta}[\phi]/8}=\int d\nu^{(\La)}_{\delta}[\sigma]\ e^{-\f{i}{2}\sqrt{\lambda}\la S\phi\vert\sigma\ra_4}.
\ee
In this intermediate field picture, the tetrahedral and pillow interactions are encapsulated in the ultralocal Gaussian measure $d\nu^{(\La)}_{\delta}$ through its covariance $C_{\sigma}:=(1+\delta\mathcal{T})$. 

Let us then define the operator $\Sigma$ coupling $\phi$ to $\sigma$
\be
\Sigma\phi(g_1,g_2,g_3):=\int dg_4dg_5\ \sigma(g_1,g_2,g_4,g_5)\phi(g_3,g_4,g_5)
\ee
in such a way that
\be
\la S\phi\vert\sigma\ra_4=\la\phi\vert\Sigma\phi\ra_3,
\ee
where $\la\cdot\vert\cdot\ra_{3}$ is the standard inner product in $L^2(\su^3)$. After integration over the original field $\phi$, we obtain
\be
\mathcal{Z}_{\textrm{BFL}}^{(\La)}(\lambda)=\int d\nu^{(\La)}_{\delta}[\sigma]\ e^{V_{\lambda}[\sigma]},
\ee
where the loop vertex is given by 
\be
V_{\lambda}[\sigma]:=-\f{1}{2} \textrm{Tr}\ \textrm{Log}(1+i\sqrt{\lambda}C_{\phi}\Sigma C_{\phi}).
\ee
One easily checks that $\widetilde{\Sigma}:=C_{\phi}\Sigma C_{\phi}$ is a Hermitian operator and therefore that the resolvent bound applies to the derivatives of this loop vertex just like in the toy example.

\subsection{Cactus expansion}

Following the same steps as in sec. \ref{zerodimensions} yields the cactus expansion of the BFL free energy:

\begin{align}\label{cactusBFL}
\mathcal{F}^{(\La)}_{\textrm{BFL}}(\lambda)=&\sum_{n=1}^{\infty}\f{1}{n!}\sum_{T\in\mathcal{T}_n}\left(\prod_{l\in T}\int_0^1dh_l\right)\int d\nu^{\boldsymbol{h^T}}_n(\sigma_1,\dots,\sigma_n)\nonumber\\
&\left(\prod_{l\in T}\int d^4g^{s(l)}d^4g^{t(l)}\ C_{\sigma}(g^{s(l)};g^{t(l)})\f{\delta^2}{\delta\sigma_{s(l)}(g^{s(l)})\delta\sigma_{t(l)}(g^{t(l)})}\right)\prod_{v=1}^nV_{\lambda}(\sigma_v).
\end{align}
At this stage, the only difference with the 0-dimensional case is the insertion of a covariance $C_{\sigma}(g_{s(l)};g_{t(l)})$ on each line $l$ of the tree, and the integration with respect to the corresponding $4$-uples of group elements $g^{l_v}:=(g_i^{l_v})_{i=1}^4$, attached to the half-lines $l_{v}$.

Computing the effect of the $k_v$ derivatives on the loop vertex $V_{\lambda}[\sigma_v]$, labeled by the half-lines $l_v$ connecting it to the tree, we obtain
\be
\left(\prod_{l_v=1}^{k_v}\f{\delta}{\delta\sigma_{v}(g^{l_v})}\right)V_{\lambda}(\sigma_v)=\f{(i\sqrt{\lambda})^{k_v}}{2}\textrm{Tr}\left(\prod_{l_v=1}^{k_v}(1+i\sqrt{\lambda}\widetilde{\Sigma}(\sigma_v))^{-1}\f{\delta\widetilde{\Sigma}}{\delta\sigma_v(g^{l_v})}\right).
\ee
Hence, the cactus amplitude $F_T(\lambda)$ is given by the following product of traces connected by ultralocal covariances:
\begin{align}\label{bigprod}
F_T(\lambda):=&\f{1}{n(T)!}\left(\prod_{l\in T}\int_0^1dh_l\right)\int d\nu^{\boldsymbol{h^T}}_n(\sigma_1,\dots,\sigma_n)\nonumber\\ &\prod_{l\in T}\int d^4g^{s(l)}d^4g^{t(l)}\ C_{\sigma}(g^{s(l)};g^{t(l)})\prod_{v\in T}\textrm{Tr}\left(\prod_{l_v=1}^{k_v}(1+i\sqrt{\lambda}\widetilde{\Sigma}(\sigma_v))^{-1}\f{\delta\widetilde{\Sigma}}{\delta\sigma_v(g^{l_v})}\right).
\end{align}

\subsection{Cauchy-Schwarz inequalities}

\begin{figure}[h]
\begin{center}
\scalebox{1.2} 
{
\begin{pspicture}(0,-3.036)(7.681,3.043)
\psdots[dotsize=0.12](3.672,2.636)
\psdots[dotsize=0.12](3.672,1.836)
\psdots[dotsize=0.12](2.872,1.836)
\psdots[dotsize=0.12](4.272,1.836)
\psdots[dotsize=0.12](4.672,2.236)
\psdots[dotsize=0.12](4.672,1.236)
\psdots[dotsize=0.12](3.672,1.036)
\psdots[dotsize=0.12](3.672,0.236)
\psline[linewidth=0.024cm](3.672,2.636)(3.672,1.836)
\psline[linewidth=0.024cm](3.672,1.836)(2.872,1.836)
\psline[linewidth=0.024cm](3.672,1.836)(3.672,1.036)
\psline[linewidth=0.024cm](3.672,1.836)(4.272,1.836)
\psline[linewidth=0.024cm](4.272,1.836)(4.672,2.236)
\psline[linewidth=0.024cm](4.272,1.836)(4.672,1.236)
\psline[linewidth=0.024cm](3.672,1.036)(3.672,0.236)
\psellipse[linewidth=0.0139999995,dimen=middle](3.772,1.536)(1.3,1.5)
\psbezier[linewidth=0.018,linestyle=dashed,dash=0.16cm 0.16cm](4.512,2.756)(4.192,2.3256774)(4.397714,1.616)(5.012,2.0263226)
\psbezier[linewidth=0.018,linestyle=dashed,dash=0.16cm 0.16cm](4.052,0.096)(3.932,0.416)(3.552,0.776)(3.252,0.176)
\psbezier[linewidth=0.018,linestyle=dashed,dash=0.16cm 0.16cm](2.8334925,2.556)(3.332,2.096)(3.232,1.316)(2.572,1.296)
\psbezier[linewidth=0.018,linestyle=dashed,dash=0.16cm 0.16cm](3.252,2.916)(3.457,2.236)(3.872,2.236)(4.072,2.976)
\psbezier[linewidth=0.018,linestyle=dashed,dash=0.16cm 0.16cm](2.712,0.696)(3.512,1.516)(3.932,1.616)(4.352,0.216)
\psbezier[linewidth=0.018,linestyle=dashed,dash=0.16cm 0.16cm](4.292,2.916)(3.932,2.156)(3.912,1.376)(4.672,0.476)
\psbezier[linewidth=0.018,linestyle=dashed,dash=0.16cm 0.16cm](5.072,1.636)(4.672,1.636)(4.012,1.456)(4.852,0.676)
\psellipse[linewidth=0.0139999995,dimen=middle](6.372,-1.661001)(1.3,1.2970009)
\psbezier[linewidth=0.018,linestyle=dashed,dash=0.16cm 0.16cm](7.112,-0.6061068)(6.792,-0.9781927)(6.9977145,-1.5918275)(7.612,-1.237035)
\psbezier[linewidth=0.018,linestyle=dashed,dash=0.16cm 0.16cm](6.652,-2.9061217)(6.532,-2.6294281)(6.152,-2.3181481)(5.852,-2.8369484)
\psbezier[linewidth=0.018,linestyle=dashed,dash=0.16cm 0.16cm](5.4334927,-0.7790403)(5.932,-1.1767873)(5.832,-1.8512278)(5.172,-1.8685211)
\psbezier[linewidth=0.018,linestyle=dashed,dash=0.16cm 0.16cm](5.852,-0.4677601)(6.057,-1.0557338)(6.472,-1.0557338)(6.672,-0.41588002)
\psbezier[linewidth=0.018,linestyle=dashed,dash=0.16cm 0.16cm](5.312,-2.3873215)(5.992,-1.6437075)(6.632,-1.9420543)(6.932,-2.824)
\psbezier[linewidth=0.018,linestyle=dashed,dash=0.16cm 0.16cm](6.892,-0.4677601)(6.532,-1.1249073)(6.512,-1.7712277)(7.272,-2.604)
\psbezier[linewidth=0.018,linestyle=dashed,dash=0.16cm 0.16cm](7.672,-1.5745342)(7.272,-1.5745342)(6.672,-1.7474676)(7.472,-2.364)
\psdots[dotsize=0.12](0.872,-0.564)
\psdots[dotsize=0.12](0.872,-1.364)
\psdots[dotsize=0.12](0.072,-1.364)
\psdots[dotsize=0.12](1.472,-1.364)
\psdots[dotsize=0.12](1.872,-0.964)
\psdots[dotsize=0.12](1.872,-1.964)
\psdots[dotsize=0.12](0.872,-2.164)
\psdots[dotsize=0.12](0.872,-2.964)
\psline[linewidth=0.024cm](0.872,-0.564)(0.872,-1.364)
\psline[linewidth=0.024cm](0.872,-1.364)(0.072,-1.364)
\psline[linewidth=0.024cm](0.872,-1.364)(0.872,-2.164)
\psline[linewidth=0.024cm](0.872,-1.364)(1.472,-1.364)
\psline[linewidth=0.024cm](1.472,-1.364)(1.872,-0.964)
\psline[linewidth=0.024cm](1.472,-1.364)(1.872,-1.964)
\psline[linewidth=0.024cm](0.872,-2.164)(0.872,-2.964)
\usefont{T1}{ptm}{m}{n}
\rput(3.652,-1.679){\Large $\longmapsto$}
\end{pspicture} 
}
\end{center}
\caption{The planar representation of a tree.}
\label{disk}
\end{figure}
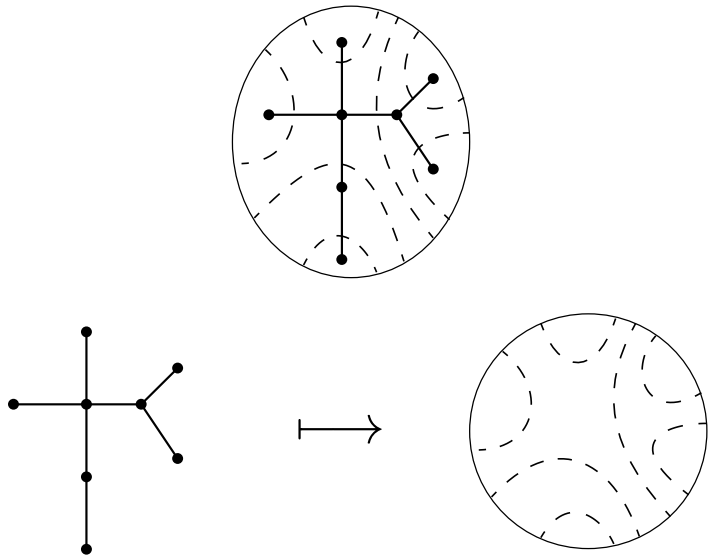

To get some insight into this cactus amplitude, it is handy to introduce a `dual' representation of a tree $T$, as a planar partition of the disk. The boundary of the disk is obtained by turning around $T$, while the dotted lines partitioning it cross the boundary twice and each line of $T$ exactly once, without crossing each other, see Fig. \ref{disk}.

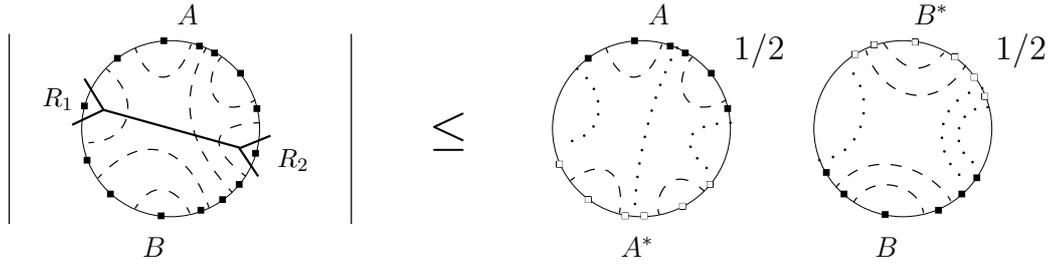
\begin{figure}[h]
\begin{center}
\scalebox{0.9} 
{
\begin{pspicture}(0,-1.9)(15.32,1.9)
\psellipse[linewidth=0.0139999995,dimen=middle](2.36,0.022999091)(1.3,1.2970009)
\psbezier[linewidth=0.018,linestyle=dashed,dash=0.16cm 0.16cm](3.1,1.0778931)(2.78,0.7058073)(2.9857142,0.09217247)(3.6,0.44696498)
\psbezier[linewidth=0.018,linestyle=dashed,dash=0.16cm 0.16cm](2.64,-1.2221218)(2.52,-0.94542825)(2.14,-0.63414806)(1.84,-1.1529484)
\psbezier[linewidth=0.018,linestyle=dashed,dash=0.16cm 0.16cm](1.4214926,0.90495974)(1.92,0.50721276)(1.82,-0.16722772)(1.16,-0.18452105)
\psbezier[linewidth=0.018,linestyle=dashed,dash=0.16cm 0.16cm](1.84,1.2162399)(2.045,0.62826616)(2.46,0.62826616)(2.66,1.2681199)
\psbezier[linewidth=0.018,linestyle=dashed,dash=0.16cm 0.16cm](1.3,-0.7033214)(1.98,0.040292434)(2.62,-0.25805432)(2.92,-1.14)
\psbezier[linewidth=0.018,linestyle=dashed,dash=0.16cm 0.16cm](2.88,1.2162399)(2.52,0.5590928)(2.5,-0.08722771)(3.26,-0.92)
\psbezier[linewidth=0.018,linestyle=dashed,dash=0.16cm 0.16cm](3.66,0.109465815)(3.26,0.109465815)(2.66,-0.06346764)(3.46,-0.68)
\psline[linewidth=0.032cm](1.38,0.3)(3.38,-0.26)
\psline[linewidth=0.032cm](1.1,0.76)(1.38,0.3)
\psline[linewidth=0.032cm](1.38,0.3)(0.92,0.08)
\psline[linewidth=0.032cm](3.38,-0.26)(3.82,-0.08)
\psline[linewidth=0.032cm](3.36,-0.26)(3.64,-0.68)
\psdots[dotsize=0.11,dotstyle=square*](2.26,1.32)
\psdots[dotsize=0.11,dotstyle=square*](2.78,1.24)
\psdots[dotsize=0.11,dotstyle=square*](3.0,1.14)
\psdots[dotsize=0.11,dotstyle=square*](3.38,0.84)
\psdots[dotsize=0.11,dotstyle=square*](3.62,0.32)
\psdots[dotsize=0.11,dotstyle=square*](3.6,-0.34)
\psdots[dotsize=0.11,dotstyle=square*](3.36,-0.8)
\psdots[dotsize=0.11,dotstyle=square*](3.12,-1.02)
\psdots[dotsize=0.11,dotstyle=square*](2.8,-1.18)
\psdots[dotsize=0.11,dotstyle=square*](2.22,-1.26)
\psdots[dotsize=0.11,dotstyle=square*](1.48,-0.94)
\psdots[dotsize=0.11,dotstyle=square*](1.14,-0.44)
\psdots[dotsize=0.11,dotstyle=square*](1.58,1.06)
\psdots[dotsize=0.11,dotstyle=square*](1.1,0.36)
\usefont{T1}{ptm}{m}{n}
\rput(2.12,-1.715){$B$}
\usefont{T1}{ptm}{m}{n}
\rput(0.71,0.465){\small $R_1$}
\usefont{T1}{ptm}{m}{n}
\rput(2.61,1.705){$A$}
\usefont{T1}{ptm}{m}{n}
\rput(4.15,-0.435){\small $R_2$}
\psline[linewidth=0.018cm](0.0,1.42)(0.0,-1.38)
\psline[linewidth=0.018cm](5.0,1.42)(5.0,-1.38)
\usefont{T1}{ptm}{m}{n}
\rput(6.42,-0.025){\LARGE $\leq$}
\psellipse[linewidth=0.0139999995,dimen=middle](9.24,0.022999091)(1.3,1.2970009)
\psbezier[linewidth=0.018,linestyle=dashed,dash=0.16cm 0.16cm](9.98,1.0778931)(9.66,0.7058073)(9.865714,0.09217247)(10.48,0.44696498)
\psbezier[linewidth=0.04,linestyle=dotted,dotsep=0.16cm](8.301493,0.90495974)(8.8,0.50721276)(8.7,-0.16722772)(8.04,-0.18452105)
\psbezier[linewidth=0.018,linestyle=dashed,dash=0.16cm 0.16cm](8.72,1.2162399)(8.925,0.62826616)(9.34,0.62826616)(9.54,1.2681199)
\psbezier[linewidth=0.04,linestyle=dotted,dotsep=0.16cm](9.76,1.2162399)(9.4,0.5590928)(9.14,-0.5)(9.14,-1.28)
\psbezier[linewidth=0.04,linestyle=dotted,dotsep=0.16cm](10.54,0.109465815)(10.14,0.109465815)(9.54,-0.06346764)(10.34,-0.68)
\psdots[dotsize=0.11,dotstyle=square*](9.14,1.32)
\psdots[dotsize=0.11,dotstyle=square*](9.66,1.24)
\psdots[dotsize=0.11,dotstyle=square*](9.88,1.14)
\psdots[dotsize=0.11,dotstyle=square*](10.26,0.84)
\psdots[dotsize=0.11,dotstyle=square*](10.5,0.32)
\psdots[dotsize=0.11,dotstyle=square*](8.46,1.06)
\usefont{T1}{ptm}{m}{n}
\rput(9.49,1.705){$A$}
\usefont{T1}{ptm}{m}{n}
\rput(9.18,-1.715){$A^*$}
\psbezier[linewidth=0.018,linestyle=dashed,dash=0.16cm 0.16cm](10.16,-0.9)(9.68,-0.44)(9.4,-0.98)(9.5,-1.24)
\psbezier[linewidth=0.018,linestyle=dashed,dash=0.16cm 0.16cm](8.2,-0.74)(8.78,-0.32)(9.0,-0.74)(8.92,-1.22)
\psdots[dotsize=0.11,fillstyle=solid,dotstyle=square](8.04,-0.5)
\psdots[dotsize=0.11,fillstyle=solid,dotstyle=square](8.46,-1.02)
\psdots[dotsize=0.11,fillstyle=solid,dotstyle=square](9.0,-1.26)
\psdots[dotsize=0.11,fillstyle=solid,dotstyle=square](9.28,-1.26)
\psdots[dotsize=0.11,fillstyle=solid,dotstyle=square](9.84,-1.12)
\psdots[dotsize=0.11,fillstyle=solid,dotstyle=square](10.24,-0.8)
\usefont{T1}{ptm}{m}{n}
\rput(10.96,1.17){\large $1/2$}
\psellipse[linewidth=0.0139999995,dimen=middle](13.06,0.022999091)(1.3,1.2970009)
\usefont{T1}{ptm}{m}{n}
\rput(13.47,1.705){$B^*$}
\usefont{T1}{ptm}{m}{n}
\rput(12.82,-1.715){$B$}
\usefont{T1}{ptm}{m}{n}
\rput(14.78,1.17){\large $1/2$}
\psbezier[linewidth=0.018,linestyle=dashed,dash=0.16cm 0.16cm](12.34,-1.06)(12.62,-0.6)(13.32,-0.82)(13.36,-1.22)
\psbezier[linewidth=0.018,linestyle=dashed,dash=0.16cm 0.16cm](12.08,-0.8423529)(12.6,-0.2)(13.5,-0.58)(13.76,-1.04)
\psbezier[linewidth=0.04,linestyle=dotted,dotsep=0.16cm](11.84,-0.44)(12.48,-0.4)(12.74,0.54)(12.2,0.98)
\psbezier[linewidth=0.04,linestyle=dotted,dotsep=0.16cm](14.22,0.62)(13.52,0.46)(13.42,-0.38)(14.04,-0.78)
\psbezier[linewidth=0.04,linestyle=dotted,dotsep=0.16cm](14.3,0.34)(13.8,0.36)(13.68,-0.34)(14.24,-0.56)
\psbezier[linewidth=0.018,linestyle=dashed,dash=0.16cm 0.16cm](12.52,1.18)(12.72,0.5)(13.52,0.2)(13.98,0.96)
\psbezier[linewidth=0.018,linestyle=dashed,dash=0.16cm 0.16cm](12.78,1.28)(13.1,0.82)(13.36,0.82)(13.66,1.14)
\psdots[dotsize=0.11,fillstyle=solid,dotstyle=square](12.36,1.12)
\psdots[dotsize=0.11,fillstyle=solid,dotstyle=square](12.64,1.26)
\psdots[dotsize=0.11,fillstyle=solid,dotstyle=square](13.24,1.3)
\psdots[dotsize=0.11,fillstyle=solid,dotstyle=square](13.82,1.08)
\psdots[dotsize=0.11,fillstyle=solid,dotstyle=square](14.1,0.78)
\psdots[dotsize=0.11,fillstyle=solid,dotstyle=square](14.26,0.5)
\psdots[dotsize=0.11,dotstyle=square*](11.94,-0.62)
\psdots[dotsize=0.11,dotstyle=square*](12.2,-0.94)
\psdots[dotsize=0.11,dotstyle=square*](12.8,-1.24)
\psdots[dotsize=0.11,dotstyle=square*](13.58,-1.18)
\psdots[dotsize=0.11,dotstyle=square*](13.92,-0.94)
\psdots[dotsize=0.11,dotstyle=square*](14.14,-0.7)
\end{pspicture} 
}
\end{center}
\caption{Splitting the disk in two parts to apply the Cauchy-Schwarz inequality. On the LHS, black squares are resolvents, dashed lines are covariances $C_{\sigma}$, and the thick line expresses the amplitude (\ref{bigprod}) as the inner product between the upper and lower parts $A$ and $B$. On the right, white squares are Hermitian conjugates of resolvents and dotted lines are covariances $C_{\sigma}'$.}
\label{schwarz1}
\end{figure}

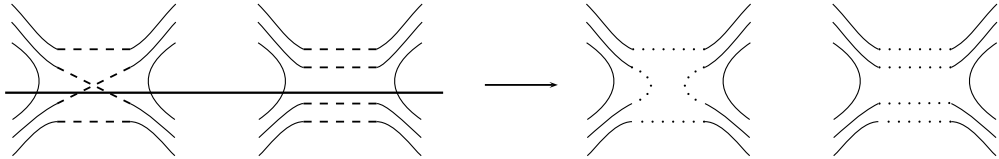
\begin{figure}[h]
\begin{center}
\scalebox{0.6} 
{
\begin{pspicture}(0,-1.6885338)(21.752,1.6885339)
\psbezier[linewidth=0.024](5.5446677,0.82045114)(6.34,0.27653393)(6.34,-0.32346618)(5.560209,-0.876534)
\psbezier[linewidth=0.024](6.54,-0.52346605)(6.34,-0.52346605)(5.94,-0.9234661)(5.560209,-1.276534)
\psbezier[linewidth=0.024](6.54,-0.9234661)(6.14,-0.9234661)(5.74,-1.523466)(5.560209,-1.6765339)
\psbezier[linewidth=0.024](6.54,0.6765338)(6.14,0.8765338)(5.94,1.276534)(5.54,1.6765338)
\psbezier[linewidth=0.024](6.54,0.27653393)(6.34,0.27653393)(5.74,1.0765339)(5.54,1.276534)
\psline[linewidth=0.04cm,linestyle=dashed,dash=0.16cm 0.16cm](6.54,0.6765338)(8.14,0.6765338)
\psline[linewidth=0.04cm,linestyle=dashed,dash=0.16cm 0.16cm](6.54,0.27653393)(8.14,0.27653393)
\psline[linewidth=0.04cm,linestyle=dashed,dash=0.16cm 0.16cm](6.54,-0.52346605)(8.14,-0.52346605)
\psline[linewidth=0.04cm,linestyle=dashed,dash=0.16cm 0.16cm](6.54,-0.9234661)(8.14,-0.9234661)
\psbezier[linewidth=0.024](9.135333,0.82045114)(8.34,0.27653393)(8.34,-0.32346618)(9.119791,-0.876534)
\psbezier[linewidth=0.024](8.14,-0.52346605)(8.34,-0.52346605)(8.74,-0.9234661)(9.119791,-1.276534)
\psbezier[linewidth=0.024](8.14,-0.9234661)(8.54,-0.9234661)(8.94,-1.523466)(9.119791,-1.6765339)
\psbezier[linewidth=0.024](8.14,0.6765338)(8.54,0.8765338)(8.74,1.276534)(9.14,1.6765338)
\psbezier[linewidth=0.024](8.14,0.27653393)(8.34,0.27653393)(8.94,1.0765339)(9.14,1.276534)
\psbezier[linewidth=0.024](0.14466743,0.82045114)(0.94,0.27653393)(0.94,-0.32346618)(0.16020894,-0.876534)
\psbezier[linewidth=0.024](1.14,-0.52346605)(0.74,-0.7234661)(0.34,-1.123466)(0.16020894,-1.276534)
\psbezier[linewidth=0.024](1.14,-0.9234661)(0.74,-0.9234661)(0.34,-1.523466)(0.16020894,-1.6765339)
\psbezier[linewidth=0.024](1.14,0.6765338)(0.74,0.8765338)(0.54,1.276534)(0.14,1.6765338)
\psbezier[linewidth=0.024](1.14,0.27653393)(0.74,0.47653413)(0.34,1.0765339)(0.14,1.276534)
\psline[linewidth=0.04cm,linestyle=dashed,dash=0.16cm 0.16cm](1.14,0.6765338)(2.74,0.6765338)
\psline[linewidth=0.04cm,linestyle=dashed,dash=0.16cm 0.16cm](1.14,0.27653393)(2.74,-0.52346605)
\psline[linewidth=0.04cm,linestyle=dashed,dash=0.16cm 0.16cm](1.14,-0.52346605)(2.74,0.27653393)
\psline[linewidth=0.04cm,linestyle=dashed,dash=0.16cm 0.16cm](1.14,-0.9234661)(2.74,-0.9234661)
\psbezier[linewidth=0.024](3.7353327,0.82045114)(2.94,0.27653393)(2.94,-0.32346618)(3.719791,-0.876534)
\psbezier[linewidth=0.024](2.74,-0.52346605)(3.14,-0.7234661)(3.54,-1.123466)(3.719791,-1.276534)
\psbezier[linewidth=0.024](2.74,-0.9234661)(3.14,-0.9234661)(3.54,-1.523466)(3.719791,-1.6765339)
\psbezier[linewidth=0.024](2.74,0.6765338)(3.14,0.8765338)(3.34,1.276534)(3.74,1.6765338)
\psbezier[linewidth=0.024](2.74,0.27653393)(3.14,0.47653413)(3.54,1.0765339)(3.74,1.276534)
\psline[linewidth=0.05cm](0.0,-0.28346616)(9.6,-0.28346616)
\psline[linewidth=0.04cm,arrowsize=0.05291667cm 2.0,arrowlength=1.4,arrowinset=0.4]{<-}(12.115,-0.11146606)(10.515,-0.11146606)
\psbezier[linewidth=0.024](12.744667,0.82045114)(13.54,0.27653393)(13.54,-0.32346618)(12.760209,-0.87653387)
\psbezier[linewidth=0.024](13.74,0.67653394)(13.34,0.87653404)(13.14,1.276534)(12.74,1.6765339)
\psbezier[linewidth=0.024](13.74,0.27653393)(13.34,0.47653413)(12.94,1.0765339)(12.74,1.276534)
\psbezier[linewidth=0.024](13.74,-0.5234659)(13.34,-0.7234661)(12.94,-1.1234661)(12.760209,-1.2765338)
\psbezier[linewidth=0.024](13.74,-0.9234661)(13.34,-0.9234661)(12.94,-1.5234659)(12.760209,-1.6765338)
\psbezier[linewidth=0.024](16.335333,0.82045114)(15.54,0.27653393)(15.54,-0.32346618)(16.319792,-0.87653387)
\psbezier[linewidth=0.024](15.34,0.67653394)(15.74,0.87653404)(15.94,1.276534)(16.34,1.6765339)
\psbezier[linewidth=0.024](15.34,0.27653393)(15.74,0.47653413)(16.14,1.0765339)(16.34,1.276534)
\psbezier[linewidth=0.024](15.34,-0.5234659)(15.74,-0.7234661)(16.14,-1.1234661)(16.319792,-1.2765338)
\psbezier[linewidth=0.024](15.34,-0.9234661)(15.74,-0.9234661)(16.14,-1.5234659)(16.319792,-1.6765338)
\psbezier[linewidth=0.024](18.144669,0.82045114)(18.94,0.27653393)(18.94,-0.32346618)(18.16021,-0.87653387)
\psbezier[linewidth=0.024](19.14,0.67653394)(18.74,0.87653404)(18.54,1.276534)(18.14,1.6765339)
\psbezier[linewidth=0.024](19.14,0.27653393)(18.74,0.47653413)(18.34,1.0765339)(18.14,1.276534)
\psbezier[linewidth=0.024](19.14,-0.5234659)(18.74,-0.7234661)(18.34,-1.1234661)(18.16021,-1.2765338)
\psbezier[linewidth=0.024](19.14,-0.9234661)(18.74,-0.9234661)(18.34,-1.5234659)(18.16021,-1.6765338)
\psbezier[linewidth=0.024](21.735334,0.82045114)(20.94,0.27653393)(20.94,-0.32346618)(21.71979,-0.87653387)
\psbezier[linewidth=0.024](20.74,0.67653394)(21.14,0.87653404)(21.34,1.276534)(21.74,1.6765339)
\psbezier[linewidth=0.024](20.74,0.27653393)(21.14,0.47653413)(21.54,1.0765339)(21.74,1.276534)
\psbezier[linewidth=0.024](20.74,-0.5234659)(21.14,-0.7234661)(21.54,-1.1234661)(21.71979,-1.2765338)
\psbezier[linewidth=0.024](20.74,-0.9234661)(21.14,-0.9234661)(21.54,-1.5234659)(21.71979,-1.6765338)
\psline[linewidth=0.05cm,linestyle=dotted,dotsep=0.16cm](13.74,0.67653394)(15.34,0.67653394)
\psline[linewidth=0.05cm,linestyle=dotted,dotsep=0.16cm](13.94,-0.9234661)(15.34,-0.9234661)
\psbezier[linewidth=0.05,linestyle=dotted,dotsep=0.16cm](13.74,0.27653393)(14.295,0.048533965)(14.295,-0.31146604)(13.74,-0.5234659)
\psline[linewidth=0.05cm,linestyle=dotted,dotsep=0.16cm](19.34,-0.9234661)(20.74,-0.9234661)
\psline[linewidth=0.05cm,linestyle=dotted,dotsep=0.16cm](19.14,0.67653394)(20.74,0.67653394)
\psline[linewidth=0.05cm,linestyle=dotted,dotsep=0.16cm](19.14,0.27653393)(20.74,0.27653393)
\psline[linewidth=0.05cm,linestyle=dotted,dotsep=0.16cm](19.14,-0.52346605)(20.715,-0.511466)
\psbezier[linewidth=0.05,linestyle=dotted,dotsep=0.16cm](15.315,0.27653393)(14.76,0.048533965)(14.76,-0.31146604)(15.315,-0.5234659)
\end{pspicture} 
}
\end{center}
\caption{Covariances $C_{\sigma}$ (dashed lines) sandwiched in the inner product in the Cauchy-Schwarz inequality (thick line) are replaced by modified covariances $C'_{\sigma}$ (dotted lines).}
\label{covariances}
\end{figure}

In such a picture, the resolvents $(1+i\sqrt{\lambda}\widetilde{\Sigma})^{-1}$ are attached to the arcs on the boundary of the disk, while the covariances $C_{\sigma}$ are attached to the dotted lines. To bound (\ref{bigprod}), we can apply the Cauchy-Schwarz inequality along a line splitting the disk in two parts with the same number of consecutive resolvents. Indeed, the number of half-lines of a tree being even, it is always possible to pick two arcs with resolvents $R_1$ and $R_2$, and express (\ref{bigprod}) as the inner product $\la A\vert R_1\otimes R_2\vert B\ra$, where $A$ and $B$ contain the same number of arcs, and thus of resolvents (see Fig. \ref{schwarz1}). By the Cauchy-Schwarz inequality, we have

\be
\vert\la A\vert R_1\otimes R_2\vert B\ra\vert\leq\Vert R_1\Vert\vert R_2\Vert\sqrt{\la A\vert A\ra}\sqrt{\la B\vert B\ra}\leq 2\sqrt{\la A\vert A\ra}\sqrt{\la B\vert B\ra}.
\ee
This process has two effects: thanks to the resolvent bound, it trades the original tree for two trees, with the same number of vertices, but with two resolvents replaced by the identity ($\Vert R_1\Vert\Vert R_2\Vert\leq 2$); each covariance $C_{\sigma}$ which is sandwiched in the inner product is replaced by a modified covariance $C_{\sigma}':=1+\delta\cT'$, where $\cT'$ identifies the two central arguments of $\sigma$ instead of twisting them (Fig. \ref{covariances}):
\be
\cT'(g_1,\dots,g_4;g'_1,\dots,g'_4):=\delta(g_{1}g_{1}'^{-1})\delta(g_{4}g_{4}'^{-1})\delta(g_{2}g_{3}^{-1})\delta(g_{2}'g_{3}'^{-1}).
\ee

We can iterate the process $(n-1)$ times, until all resolvents are removed, and all covariances $C_{\sigma}$ replaced by $C_{\sigma}'$. We are then left with a BFL \emph{Feynman graph}, hence

\be
\vert F_T(\lambda)\vert\leq 2^{n(T)-1}\ \textrm{sup}_{T'\in\cT_{n}}\ \vert\cA_{\textrm{BFL}}(T')\vert,
\ee
Now, from theorem \eqref{perturbativebound} we have $\textrm{sup}_{T'\in\cT_{n}}\ \vert\cA_{\textrm{BFL}}(T')\vert\le K_\delta^n\Lambda^{6 + 3n}$, hence 
\be
\vert F_T(\lambda)\vert\leq2^{n(T)-1}K_\delta^n \Lambda^{6 + 3n(T)}. 
\ee

Before we can conclude, we should check that this estimate also holds for the  trivial tree with just one vertex ($n=1$): since it contains no half-line, the loop vertex $\textrm{Log}(1+i\sqrt{\lambda}\widetilde{\Sigma})$ is not acted upon by a $\sigma$-derivative, and therefore the resolvent bound does not apply. However, using standard convexity inequalities, we have for any $z\in\mathbb{C}$
\be
\vert\text{Log}(1+z)\vert\le\f{1}{2}+\f{1}{2}(\vert z\vert^2+4\pi^2),
\ee 
hence
\be
\Vert\text{Log}(1+i\sqrt{\lambda}\widetilde{\Sigma})\Vert\le\f{1}{2}+\f{1}{2}(\Vert\widetilde{\Sigma}\Vert^2+4\pi^2)\le\f{1}{2}+\f{1}{2}(\sup\vert\sigma\vert^2+4\pi^2),
\ee
and therefore
\be
\vert\text{Tr}\int d\nu^{(\La)}_{\delta}(\sigma)\ \textrm{Log}(1+i\sqrt{\lambda}\widetilde{\Sigma})\vert\le K\dim\cH^{(\La)}_{0}=\mathcal{O}(\La^6).
\ee
Absorbing the uniform factor $\dim\cH^{(\La)}_{0}=\mathcal{O}(\La^6)$ appearing in all estimates in the definition of a free energy per degree of freedom 
\be
\mathcal{G}_{\textrm{BFL}}^{(\La)}:=\f{1}{\dim\cH^{(\La)}_{0}}\mathcal{F}_{\textrm{BFL}}^{(\La)},
\ee 
we have proved that \emph{the cactus expansion of the BFL free energy per degree of freedom $\mathcal{G}_{\textrm{BFL}}^{(\La)}$ is uniformly convergent in a half-disk }
\be
\{\lambda\in\mathbb{C},\Re\lambda\geq0,\vert\lambda\vert\leq K_\delta\La^{-3}\}.
\ee 
Moreover, the Nevanlinna-Sokal criteria can be checked exactly as in the toy example, hence $\mathcal{G}_{\textrm{BFL}}^{(\La)}$ defines the Borel sum of the BFL perturbative series. 

\subsection{A constructive scaling limit}

The remarkable feature of this result lies in the control it provides over the summability radius: it shows that, as the cutoff $\Lambda$ is sent to infinity, the coupling constant should shrink as $\Lambda^{-3}$. This immediately suggests the definition of a \emph{scaling limit} for the BFL model, analogous to one discovered in matrix models by 't Hooft at the perturbative level \cite{Hooft:1973jz}, and confirmed by Rivasseau at the constructive level \cite{Rivasseau:2007fr}. 

Define indeed a \emph{running coupling constant}
\be\label{constructivescaling}
\lambda'(\Lambda)=\f{\lambda}{\Lambda^3}.
\ee
Then, seen as a function of $\lambda'$, the BFL free energy per degree of freedom $\mathcal{G}_{\textrm{BFL}}^{(\La)}(\lambda')$ has a summability radius which is \emph{independent of $\Lambda$}. This fact suggests the existence of a limit 
\be
\underset{\Lambda\rightarrow\infty}{\lim}\mathcal{G}_{\textrm{BFL}}^{(\La)}(\lambda')
\ee 
where $\lambda'\rightarrow0$. Three comments are in order.

\medskip\begin{itemize}
\item
This scaling behavior was beyond the scope of the Freidel-Louapre analysis \cite{Freidel:2002tg}. It is only thanks to our use of the (much less elementary) cactus expansion that we could establish this result. 
\item
The constructive scaling \eqref{constructivescaling} matches the perturbative scaling \eqref{perturbativebound}. In other words, \emph{the sum over all foams does not introduce further divergences in the high-spin limit}. 
\item
The presence of the pillow term in the BFL action is key to this result. Let us emphasize however that the BFL scaling behavior does \emph{not} coincide with the one of the original Boulatov model. A perturbative estimate established in \cite{Magnen:2009dq} shows indeed that, in the pure Boulatov model, we have
\be
\vert\mathcal{A}_{\textrm{B}}(G)\vert\leq (\lambda K)^{V(G)}\Lambda^{6+3V(G)/2}. 
\ee
\end{itemize}\medskip

Let us conclude by remarking that, since we established in 2009 the perturbative \eqref{perturbativebound} and constructive \eqref{constructivescaling} scaling behavior of three-dimensional group field theory, Gurau and Rivasseau have made significant progress towards the identification of dominant contributions in the Boulatov \cite{Gurau2011a} and Ooguri models \cite{Gurau2011a,Gurau2011b}, analogous to 't Hooft's planar diagrams \cite{Hooft:1973jz}.

%% file: conclusion.tex
\chapter{Conclusion}

\section{Des fondations pour les mousses de spins}

La viabilité du modèle plat en tant que théorie quantique de la géométrie est cruciale pour le programme des mousses de spins. Dans cette thèse, nous avons abordé certains des aspects réputés obscurs de ce modèle, et notamment celui de ses \emph{divergences} : divergences de bulles d'une amplitude donnée, mais aussi divergence de la série perturbative, du moins dans le cadre de la \emph{group field theory} de Boulatov. Nous avons obtenus les résultats suivants :
\begin{itemize}
\item
Le degré de divergence d'une mousse $\G$, pour un groupe de structure compact $G$, est donné par le deuxième nombre de Betti tordu par les $G$-connexions plates non-singulières sur $\G$. Ce résultat confirme l'intuition de Perez-Rovelli et de Freidel-Louapre concernant le r\^ole des \emph{bulles} pour ce comptage de puissance, quoique dans un sens éminemment plus subtil que la notion de ``surface fermée'' ne le laisse entendre \emph{a priori}.
\item
Il existe une prescription combinatoire réalisant, dans les modèles de mousses de spins, l'identité heuristique entre ``limite de raffinement'' et ``somme sur les histoires''. Ce résultat structurel 
suggère qu'en ce qui concerne la limite continue des mousses de spins, raffiner ou sommer, telle n'est pas la question.
\item
La série perturbative du modèle de Boulatov, avec régularisation aux grands spins et terme de ``coussin'', n'est pas seulement sommable au sens de Borel : elle possède une limite d'échelle où la constante de couplage \emph{s'annule}.
\end{itemize}
Certes, ces résultats sont loin d'apporter une réponse, même partielle, au problème de la gravité quantique évoqué dans l'introduction. Ils peuvent toutefois être vu comme des fondements \emph{solides} pour la théorie des mousses de spin.

\section{Le modèle plat, une phase de la gravité quantique ?}

S'il est directement relié à la théorie de Yang-Mills en deux dimensions et à la gravité de Palatini-Cartan en trois dimensions, le modèle plat n'est lié à la gravité quantique en \emph{quatre} dimensions qu'au niveau \emph{heuristique} : modèle auxiliaire sans dynamique gravitationnelle, il permet la construction des modèles du type EPRL ou FK par l'imposition de certaines contraintes dites ``de simplicité''. On pourrait donc -- légitimement -- s'interroger sur la pertinence physique des résultats obtenus dans cette thèse. 

Il s'avère que la relation entre le modèle plat et la gravité quantique quadri-dimensionnelle pourrait être plus étroite que ne le suggère la construction des modèles EPRL-FK. Considérant le comportement d'une action \emph{invariante} \emph{par} \emph{reparamétrisation} sous l'effet d'une discrétisation de l'espace-temps, Rovelli a en effet suggéré récemment l'existence d'une phase ``topologique'' de la gravité, où la géométrie serait \emph{plate} : un mécanisme qu'il a baptisé \emph{Ditt}-invariance \cite{Rovelli2011b}, en hommage au travail de B. Dittrich.

Une idée similaire a également été défendue par Rivasseau \cite{Rivasseau2011}. Inspiré par le comportement du paramètre $\Omega$ de Grosse-Wulkenhaar \cite{Grosse2003} à haute énergie, il conjecture un ``running'' du paramètre d'Immirzi dans le modèle EPRL vers la valeur critique $1$, où l'amplitude de mousse de spins se réduit à celle du modèle plat pour $G=\SU(2)\times\SU(2)$. Ceci correspondrait selon lui à la phase ``pré-géométrique'' de la gravité quantique. 

Il est clair que, si l'un de ces scénarios venait à se réaliser, le modèle plat reviendrait naturellement sur le devant de la scène physique. Les résultats présentés dans cette thèse deviendraient alors \emph{directement} pertinents pour étudier le régime quantique de la gravité.


\section{Quel ``groupe'' de renormalisation ?}

Pour faire sens d'une telle hypothèse, il reste toutefois à formuler une généralisation du \emph{groupe de renormalisation} adaptée au cadre des mousses de spins. Plusieurs pistes semblent envisageables, parmi lesquelles : 

\begin{itemize}
\item
Adapter le groupe de renormalisation de Wilson-Kadanoff au \emph{filet des mousses de spins}. Comme nous l'avons rappelé au chap. \ref{spincont}, celui-ci repose dans sa formulation traditionnelle sur la présence d'un réseau \emph{métrique}. Le caractère \emph{background independent} de la relativité générale nécessite le recours à une notion d'échelle généralisée ; nous postulons que la notion d'ensemble ordonné filtrant (\emph{directed set}) a un role essentiel à jouer dans ce programme. Nous renvoyons le lecteur aux travaux préliminaires de Zapata \cite{Zapata2002}, Oeckl et leurs collaborateurs \cite{Manrique2005} à ce sujet, ainsi qu'aux résultats récents de Dittrich \emph{et al.} \cite{Dittrich2011}.

\item
Une autre piste, poursuivie notamment par Rivasseau et ses collaborateurs, s'appuie sur l'analogie structurelle entre la théorie des champs sur le groupe (\emph{group field theory}) et la théorie quantique des champs usuelle pour tenter de construire un groupe de renormalisation généralisé. Pour mener à bien ce programme, il est crucial de se doter d'une group field theory avec un propagateur non-trivial, donc de mousses de spins non-plates \cite{Rivasseau2011}. Tout reste à faire dans cette approche.
\end{itemize}

Développer ces points de vue, ou d'autres encore, afin de mieux maîtriser la troncation du champ gravitationnel en mousses de spins : c'est à mon sens le défi crucial pour la théorie des boucles. Courage à ceux qui s'y attelleront, ou s'y attellent déjà !




%% file: appendices.tex
\chapter{Appendix A: Cellular homology}\label{appendixcellular}

For the reader's convenience, we recall here some basic definitions concerning cell complexes and their (co)homology (a well-written, pedagogical reference is \cite{hatcher}). A cell complex, or finite CW complex, is a topological space $X$ presented as the disjoint union of finitely many open cells, such that for each open $p$-cell $\sigma^{p}$ there is a continuous map $f:B^p\rightarrow X$ whose restriction to the interior of $B^p$ is a homeomorphism onto $\sigma^p$, and such that the image of the boundary of $B^p$ is contained in the union of lower dimensional cells. The dimension of $X$, $\dim X$, is the maximal dimension of its cells. If $c_{p}(X)$ is the number of $p$-cells of $X$, the Euler characteristic of $X$ is defined as
\be
\chi(X)=\sum_{p=0}^{\dim X}(-1)^pc_{p}(X).
\ee
It is homotopy invariant.

Cellular homology associates to each cell complex $X$ a sequence of homotopy invariant abelian groups, the homology groups of $X$, as follows. For each dimension $p$, consider the set $C_{p}(X)$ of formal linear combination of $p$-cells with integer coefficients (the `free Abelian group' over the $p$-cells of $X$); its elements are called $p$-chains. Define, for each $p$, the boundary map $\pp_{p}:C_{p}(X)\rightarrow C_{p-1}(X)$ by its action on $p$-cells $\sigma_{\alpha}^p$
\be
\pp_{p}\sigma_{\alpha}^p=\sum_{\beta}[\sigma^{p}_{\alpha},\sigma_{\beta}^{p-1}]\sigma_{\beta}^{p-1}
\ee
and linearity. Here, the sum runs over the $(p-1)$-cells on the boundary of $\sigma^{p}_{\alpha}$, and $[\sigma^{p}_{\alpha},\sigma_{\beta}^{p-1}]$ is the incidence number of $\sigma_{\alpha}^p$ on $\sigma^{p-1}_{\beta}$ -- that is, the number of times $\sigma_{\alpha}^p$ wraps around $\sigma^{p-1}_{\beta}$, with relative orientations taken into account (see \cite{hatcher} for a precise definition). Elements of $\ker\pp_{p}$ are called $p$-cycles, and elements of $\textrm{Im}\ \pp_{p+1}$ are called $p$-boundaries. The fundamental property of the boundary maps is that, for each $p$,

\be
\pp_{p}\pp_{p+1}=0.
\ee

This implies that $\textrm{Im}\ \pp_{p+1}\subset\ker\pp_{p}$, and allows to consider the quotients $H_{p}(X)=\ker\pp_{p}/\textrm{Im}\ \pp_{p+1}$ of $p$-cycles modulo $p$-boundaries. The sequence of Abelian groups $C_{p}(X)$, together with the boundaries maps $\pp_{p}$
\be
0\longrightarrow C_{\dim X}(X)\longrightarrow\dots\longrightarrow C_{p+1}(X)\longrightarrow C_{p}(X)\longrightarrow \dots\longrightarrow C_{0}(X)\longrightarrow 0
\ee
forms the cellular chain complex of $X$, and the $H_{p}(X)$'s are the homology groups of $X$. They are homotopy invariant, and in particular so are their ranks $b_{p}(X)$, the Betti numbers of $X$. Intuitively, $b_{p}(X)$ is the number of `independent $p$-holes' of $X$.
The Euler-Poincar\'e theorem states that
\be
\chi(X)=\sum_{p=0}^{\dim X}(-1)^pb_{p}(X).
\ee

Dualization of this construction leads to cellular cohomology. Explicitely, for each p, the cochain group $C^p(X)$ of $X$ is defined as the set of linear maps from $C_{p}(X)$ to $\mathbb{Z}$, and the coboundary operator $\delta^{p}$ as the transpose of $\pp^{p+1}$. One checks that
\be
\delta^{p+1}\delta^{p}=0
\ee
and the resulting complex
\be
0\longrightarrow C_{0}(X)\longrightarrow\dots\longrightarrow C_{p}(X)\longrightarrow C_{p+1}(X)\longrightarrow \dots\longrightarrow C_{\dim X}(X)\longrightarrow 0
\ee
is called the cellular cochain complex of $X$, and the quotients $H^p(X)=\ker\delta_{p}/\textrm{Im}\ \delta_{p-1}$ its cohomology groups. They are homotopy invariant as well.

This construction can be generalized by replacing the integer coefficients in the (co)chains by elements of an arbitrary Abelian group $A$. The corresponding homology and cohomology groups are then denoted $H_{p}(X,A)$ and $H^p(X,A)$ respectively. One shows in particular that, whenever $A$ is actually a vector space, $H_{p}(X,A)$ and $H^p(X,A)$ are dual to each other. Also, in this case (and more generally if $A$ is torsion-free), it holds that $H_{p}(X,A)=H_{p}(X)\otimes A$, and thus $b_{p}(X,A)=b_{p}(X)\dim(A)$.

Eventually, let us mention Poincar\'e duality: when $X$ is the cell decomposition of an oriented, closed $d$-manifold, we have

\be
H^p(X)\simeq H_{d-p}(X).
\ee
\chapter{Appendix B: Colored homology}\label{appendixcolored}

Due to the several simplifications it provides, the recent literature on group field theory has focused on Gurau's colored model \cite{Gurau2011d}. To ease the translation between this framework and ours, based on finite CW complexes, we recall here the basics of cristallization and colored graph theory, and explicit the relationship between colored and cellular homology. Although it seems that Gurau did not know about this work, colored homology was actually introduced in \cite{CAVICCHIOLI1995}.

An $(d + 1)$-colored graph is a pair $\mathcal{G}=(G, c)$, where $G = (V(G), E(G))$ is a connected multigraph (without tadpoles) regular of degree $d + 1$, and $c : E(G)\rightarrow\Delta_d = \{0, 1, . . . , d\}$ is an edge-coloring such that $c(e)\neq c(f )$ for any pair $e$ and $f$ of adjacent edges of $G$. For each proper subset $B$ of $\Delta_n$, let $\mathcal{G}_B$ denote the subgraph of $G$ defined by $(V (G), c^{-1}(B))$; each connected component of $\mathcal{G}_B$ is called a $B$-residue, or $B$-bubble. If $B$ has $\vert B\vert=k$ elements, it is also called a $k$-bubble. The $0$-bubbles of $\mathcal{G}$ are its vertices.

The \emph{colored}, or \emph{bubble}, homology of such a colored graph $\mathcal{G}$ is defined as follows. Let $C_k(\mathcal{G})$ be the free Abelian group generated by the $k$-bubbles of $\mathcal{G}$, and $d_k:C_k(\mathcal{G})\rightarrow C_{k-1}(\mathcal{G})$ the linear map defined by
\be\label{coloredboundary}
d_k(b_k):=\sum_{q\in B}(-1)^{\vert B^<(q)\vert}\sum_{c_{k-1}\in\mathcal{G}_{B\setminus\{q\}}}c_{k-1}
\ee
where $b_k$ is a $B$-bubble $k$ colors, and
\be
B^<(q):=\left\{p\in B\ ,\ p<q\right\}.
\ee
It is not difficult to check that $d_{k}d_{k+1}=0$, and thus that
\be
C_k(\mathcal{G})\xrightarrow{\ d_d\ }
C_{d-1}(\mathcal{G})\xrightarrow{\ d_{d-1}\ }  \dots C_1(\mathcal{G})\xrightarrow{\ d_1\ }C_0(\mathcal{G})\xrightarrow{\ d_0\ }0,
\ee
forms a chain complex. Its homology groups $H_k(\mathcal{G})=\ker d_k/\im d_{k+1}$ are the \emph{colored homology groups} of $\mathcal{G}$. We denote the corresponding Betti numbers $b_k(\mathcal{G})$.

\medskip

Every $(d + 1)$-colored graph $(G, c)$ determines a $d$-dimensional CW complex $K_{\mathcal{G}}$ as follows. For each vertex $v$ of $G$, consider an $d$-simplex $\sigma_d(v)$ and label its vertices by $\Delta^d$. If $v$ and $w$ are joined in $G$ by an $i$-colored edge, $i\in\Delta_d$, then identify the $(d - 1)$-faces of $\sigma_d(v)$ and $\sigma_d(w)$ opposite to the vertex labelled by $i$, so that equally labelled vertices coincide. The quotient space is $K_{\mathcal{G}}$. It is obviously connected.

The CW complex thus defined is special in that every cell $e^k_\alpha$ arises as the projection of a $k$-simplex $\sigma^k_\alpha$ with vertices labelled in a subset $B_\alpha$ of $\Delta^d$. Moreover, the cells on the boundary of $e^k_\alpha$ are the projections of the faces of $\sigma^k_\alpha$. Such a CW complexes is called a $\Delta$-complex by Hatcher \cite{hatcher}, and a pseudo-complex by other authors \cite{CAVICCHIOLI1995}. Thanks to its simplicial character, its boundary operator is readily computed.  For $q\in B_\alpha$, denote $e^{k-1}_{\alpha(q)}$ the $(k-1)$-cell on the boundary of $e^k_\alpha$ arising as the projection of the face of $\sigma^k_\alpha$ opposite to the vertex labelled by $q$. Then $[e^k_\alpha,e^{k-1}_{\alpha(q)}]=(-1)^{\vert B_\alpha^<(q)\vert}$, and thus
\be
\pp_k(e^k_\alpha)=\sum_{q\in B_\alpha}(-1)^{\vert B_\alpha^<(q)\vert}e^{k-1}_{\alpha(q)}.
\ee

Moreover, like a simplicial complex, $K_{\mathcal{G}}$ possesses a dual CW complex $\Delta_{\mathcal{G}}$, whose $k$-cells $f^k_\alpha$ are in one-to-one correspondence with the $(d-k)$-cells $e^{d-k}_\alpha$ of $K_{\mathcal{G}}$ (see \cite{hatcher} for details). Let
\be
E^k_{\alpha,q}=\left\{\beta\ ,\ e^{k}_{\beta(q)}=e^k_\alpha\right\}
\ee
index the set of cofaces of $e^k_\alpha$. Then $\Delta_{\mathcal{G}}$ is such that
\be\label{cellularboundary}
\pp_k(f^k_\alpha)=\sum_{q\in B_\alpha}(-1)^{\vert B_\alpha^<(q)\vert}\sum_{\beta\in E_{\alpha,q}^k} f^{k-1}_\beta,
\ee
i.e. the boundary operator $\pp_k$ of $\Delta_{\mathcal{G}}$ is the coboundary operator $\delta^{d-k}$ of $K_{\mathcal{G}}$.  Hence, \be H_k(\Delta_{\mathcal{G}})=H^{d-k}(K_{\mathcal{G}}),\ee as in Poincar\'e duality.

To relate the colored homology of the colored graph $\mathcal{G}$ to the cellular homology of the corresponding complex, it suffices to note that the $B$-bubbles of $\mathcal{G}$ are in one-to-one correspondence with the cells of $K_{\mathcal{G}}$ arising from simplices with vertices labelled in $\Delta^d\setminus B$. Indeed, let $v$ be a vertex of a $B$-bubble $b$ with $\vert B\vert=k$, and $\sigma_d(v)$ the corresponding simplex, with vertices labelled by $\Delta_d$. Let $f_B(\sigma_d(v))$ be the $(d-k)$-subsimplex of $\sigma_d(v)$ defined by those of its vertices which are not in $B$, and $e^{d-k}_b(v)$ the corresponding $(d-k)$-cell in $K_{\mathcal{G}}$. By definition of the quotient space $K_{\mathcal{G}}$, the cell $e^{d-k}_b(v)$ actually does not depend on $v$. Thus, for each $k$-bubble $b$ of $\mathcal{G}$ there is a $(d-k)$-cell $e^{d-k}_b$ of $K_{\mathcal{G}}$, and therefore a $k$-cell $f^k_b$ of $\Delta_{\mathcal{G}}$. It is not hard to check that this correspondence is actually bijective. Moreover, inspection of \eqref{coloredboundary} and \eqref{cellularboundary} shows that
\be
F_k(d_kb)=\delta^{d-k}F_k(b),
\ee
where $F_k$ the mapping $b\mapsto f^k_b$. In other words, we have an isomorphism of chain complexes $F=(F_k)_{k=0}^d:C_*(\mathcal{G})\rightarrow C^{*}(\Delta_{\mathcal{G}})$, and thus
\be
H_k(\mathcal{G})\simeq H_k(\Delta_{\mathcal{G}})=H^{d-k}(K_{\mathcal{G}}).
\ee
In particular, the colored Betti numbers of $\mathcal{G}$ coincide with the cellular Betti numbers of $\Delta_{\mathcal{G}}$. Note moreover that, since the $0$-th Betti number of a topological space counts the number of its connected components, we have $b_d(\mathcal{G})=b_0(K_{\mathcal{G}})=1$ and $b_0(\mathcal{G})=b_0(\Delta_{\mathcal{G}})=1$.